\documentclass[reqno,12pt,a4paper]{article}

\usepackage{amsfonts,amssymb,amsthm, 
amsmath,amscd, bbm, bm, mathabx, %bbold,
mathrsfs,delarray,subfigure}
\usepackage[dvipsnames]{xcolor}
\usepackage{xr-hyper} 
\usepackage{hyperref,cite}
\usepackage{xypic}
\usepackage{sectsty}

\usepackage{accents}

\sectionfont{\fontsize{12}{15}\selectfont}
\subsectionfont{\fontsize{12}{15}\selectfont}

\usepackage{enumitem}

\DeclareMathOperator{\ad}{ad}

\DeclareMathOperator{\id}{id}

\DeclareMathOperator{\Hom}{Hom}

\DeclareMathOperator{\Der}{Der}

\DeclareMathOperator{\rank}{rank}

\DeclareMathOperator{\Poly}{Poly}

\numberwithin{equation}{subsection} 
\numberwithin{subsection}{section} 

\newcommand{\ceqref}[1]{{\textcolor{blue}{\eqref{#1}}}}
\newcommand{\cref}[1]{{\textcolor{blue}{\ref{#1}}}}
\newcommand{\ccite}[1]{{\textcolor{blue}{\!\cite{#1}}}}

\newcommand{\sss}{{\hbox{$\sum$}}}
\newcommand{\ppp}{{\hbox{$\prod$}}}
\newcommand{\ooo}{{\hbox{$\bigotimes$}}}

\newcommand{\nnn}{{\hbox{$\bigcap$}}}
\newcommand{\ddd}{{\hbox{$\bigoplus$}}}
\newcommand{\ul}[1]{{\underline{#1}}}

\newcommand{\www}{{\hbox{$\text{\footnotesize 
$\bigwedge$}$}}}

\newcommand{\mathsans}[1]{{{\sf #1}}}

\font\euler=eusm10 at 12.8 truept
\font\scripteuler=eusm7
\font\scriptscripteuler=eusm5 
\newtheorem{defi}{{\sf Definition}}[section]
\newtheorem{prop}{{\sf Proposition}}[section]

\newtheorem{lemma}{{\sf Lemma}}[section]
\newtheorem{exa}{{\sf Example}}[section]

\begin{document}

%\hrule\vskip.5cm
%\hbox to 14.5 truecm{October 2016 \hfil DIFA 16}
%\hbox to 14.5 truecm{Version 2 \hfil}
%\vskip.5cm\hrule
\vskip1.5cm
\begin{large}
{\flushleft\textcolor{blue}{\sffamily\bfseries Algebraic formulation of higher gauge theory}}  
\end{large}
\vskip1.3cm
\hrule height 1.5pt
\vskip1.3cm
{\flushleft{\sffamily \bfseries Roberto Zucchini}\\
\it Dipartimento di Fisica ed Astronomia,\\
Universit\`a di Bologna,\\
I.N.F.N., sezione di Bologna,\\
viale Berti Pichat, 6/2\\
Bologna, Italy\\
Email: \textcolor{blue}{\tt \href{mailto:roberto.zucchini@unibo.it}{roberto.zucchini@unibo.it}}, 
\textcolor{blue}{\tt \href{mailto:zucchinir@bo.infn.it}{zucchinir@bo.infn.it}}}

%roberto.zucchini@unibo.it, zucchinir@bo.infn.it}

\vskip.7cm
%\hrule height 1.5pt
\vskip.6cm 
{\flushleft\sc% \sffamily \bfseries 
Abstract:} 
%\par\noindent
In this paper, we present a purely algebraic formulation of higher gauge theory and gauged sigma models 
based on the abstract theory of graded commutative algebras and their morphisms. 
The formulation incorporates naturally BRST symmetry 
and is also suitable for AKSZ type constructions. 
It is also shown that for a full--fledged 
BV formulation including ghost degrees of freedom, higher gauge and gauged sigma model fields must be viewed as
internal smooth functions on the shifted tangent bundle 
of a space time manifold valued in a shifted $L_\infty$--algebroid encoding symmetry. 
The relationship to other formulations where the $L_\infty$--algebroid
arises from a higher Lie groupoid by Lie differentiation is highlighted. 

\vspace{2mm}

\vfil\eject

\tableofcontents

\vfil\eject

\section{\textcolor{blue}{\sffamily Introduction}}\label{sec:intro}

\vspace{-.49mm}
{\it Higher gauge theory} is an generalization of ordinary gauge theory 
where gauge fields and field strengths are higher degree forms. %Presently, %The consensus is that 
Higher gauge theory appears to be the most promising candidate for the description 
of the dynamics of the higher--dimensional extended objects %, strings and branes, 
thought to be the basic constituents of matter and mediators of fundamental interactions. 
In various forms, higher gauge theory is relevant in supergravity,  
in string and brane theory and in the loop and spin foam formulations
of quantum gravity. %\cite{Baez:1999sr,Rovelli:2004tv}. 
Presently, the interest in higher gauge theory rests on the hope that 
it may eventually provide a Lagrangian formulation of the mysterious
$N=(2, 0)$ $6$--dimensional superconformal field theory 
describing the effective dynamics of $M5$--branes.
%\ccite{Fiorenza:2012tb,Fiorenza:2015gla,Palmer:2012ya,Saemann:2013pca,Lavau:2014iva}.
See \ccite{Baez:2010ya} and references therein for a readable introduction to higher gauge theory
and \ccite{Saemann:2016sis} for an updated account of the latest developments. % of the subject. 

\vspace{-.1mm}
Quite early in the history of the subject, it was realized that in the non Abelian case 
the symmetry of higher gauge theory cannot be described by ordinary Lie groups. 
The solution to this problem proposed in \ccite{Baez:1998,Baez:2002jn,Baez:2004in,Baez:2005qu}
and now widely accepted is that higher gauge theory should result from a categorification of
ordinary gauge theory by formalizing higher gauge symmetry through the 
algebraic structures stemming from the categorification of ordinary 
groups, the so--called $2$--groups. These ideas have been developed to an increasing degree of generality  
in the context of $\infty$--Lie theory in refs. \ccite{Sati:2008eg,Fiorenza2011}. 
Related approaches to the problem were followed in refs. \ccite{Ritter:2013wpa,%Ritter:2015ymv,
Ritter:2015zur,Jurco:2014mva,Jurco:2016qwv}.

\vspace{-.1mm}
All these endeavours require rather sophisticated mathematical constructions.
Indeed, higher gauge theory meets many areas of contemporary mathematics: 
higher algebraic structures, such as higher categories, 
higher groups \ccite{Baez5,Baez:2003fs} and strong homotopy Lie or $L_\infty$--algebras 
\ccite{Lada:1992wc,Lada:1994mn}, 
higher geometrical structures, such as gerbes \ccite{Brylinski:1993ab,Breen:2001ie}
and higher topological structures, such as higher knots \ccite{Zucchini:2015wba,Zucchini:2015xba}.
An illustration of these multiple topics and their relationship to
fundamental physics can be found in \ccite{Schreiber2011,Sharpe:2015mja}. 

\vspace{-.1mm}
The expression of the full potential of higher gauge theory has been hindered so far by 
the scarcity of interesting physically motivated examples, though some progress has been made on 
this score. A non exhaustive list of contributions in which models relevant in string theory and %categorified 
higher Chern--Simons theory are worked out is \ccite{Palmer:2013ena,Palmer:2013pka,Lavau:2014iva,
Zucchini:2011aa,Soncini:2014ara,Ritter:2015ymv,Zucchini:2015ohw}.
It is therefore desirable to devise formulations of higher gauge theory with general 
templates allowing for broad classes of interesting examples. This is the point of view 
adopted in this paper.

In this paper, we propose a purely algebraic formulation of higher gauge theory %and gauged sigma models 
based on the abstract theory of graded commutative algebras and their morphisms
%In this paper, we propose a new formulation of higher gauge theory and gauged sigma models 
%with general $L_\infty$--algebra or algebroid symmetry 
that naturally incorporates BRST symmetry and is thus suitable for perturbative quantization. % and gauge fixing. 
We do so by building higher gauge theory as a generalization of or\-dinary gauge theory  
cast in  the language of NQ--manifold and $L_\infty$--algebroid geometry 
\ccite{Roytenberg:0203110} which has a natural graded commutative algebraic reading. 

Our formulation builds on the framework developed 
by Bojowald, Gruetzmann, Kotov and Strobl in refs. 
\ccite{Bojowald:0406445,Kotov:2007nr,Gruetzmann:2014ica}, which is called 
BGKS theory in the following for short and is briefly reviewed in subsect.
\cref{subsec:bgkstheory}. We stress our indebtedness to these authors. 
A related approach is provided %by Fiorenza, Rogers and Schreiber 
in ref. \ccite{Fiorenza:2011jr}.

A basic element of our formulation is BRST theory (see \ccite{Barnich:2000zw} for a review 
and extensive referencing) in the superfield framework pioneered a long time ago by  
Baulieu and Thierry--Mieg in refs. \ccite{Baulieu:1981sb,Baulieu:1984iw,Baulieu:1984ih} 
and Bonora, Pasti and Tonin in refs, \ccite{Bonora:1980pt,Bonora:1980ar,Bonora:1981rw}. 
This is reviewed in subsect \cref{subsec:brsttheory}.

In subsect. \cref{subsec:ztheory}, we explain in broad lines our the basic tenets on which our theory
is based in order to make the content the main body of the paper more easily understandable.
In subsect. \cref{subsec:plan}, we finally outline the plan of the paper.

%Most of the examples of higher gauge theoretic models worked out in the literature so far  
%involve trivial higher principal bundles and connections. This has prevented the investigation of 
%non pertutbative aspects which are the focus of much interest in ordinary gauge theory
%and which are expected to be relevant in higher gauge theory as well. 
%In particular, one would expect categorified analogues of non-abelian 
%monopoles and instantons to exist. 

%Although higher analogues of the twistor descriptions of monopoles 
%and instantons have been constructed [2, 3, 4], 
%the known solutions as in particular those of [5] do not quite fit the picture completely.
%This presents an obstacle to both mathematical and physical progress in 
%the study of higher gauge theory. 

%\vfil\eject

\subsection{\textcolor{blue}{\sffamily %Review of the 
BGKS theory}}\label{subsec:bgkstheory}

BGKS theory is a geometrical formulation of higher gauge theory. In the usual way, 
a higher gauge field is viewed as a higher connection on a background higher principal bundle 
and its gauge field strength as the connection's curvature. Everything is however 
cast in the language of NQ--manifold theory, which renders the generalization of the 
customary notions of ordinary gauge theory to the higher case particularly 
simple and elegant. 

Let $G$ be a compact Lie group with Lie algebra $\mathfrak{g}$, $M$ be a manifold
and $P$ be a trivial principal $G$--bundle over $M$. A connection of $P$ is then simply 
a $\mathfrak{g}$--valued $1$--form $A$. The curvature of $A$ is the $\mathfrak{g}$--valued $2$--form 
%\hphantom{xxxxxxxxxxxxxxxxx}
\begin{equation}
F_A=dA+\tfrac{1}{2}[A,A]. 
\label{FA}
\end{equation}
%A gauge transformations is a $G$-valued map $g$ acting on $A$ as ${}^gA=gAg^{-1}-dg g^{-1}$. 
Gauge transformations are the symmetry transformations of gauge theory. Geometrically, they
are automorphisms of $P$ covering the identity. 
An infinitesimal gauge transformation is simply a $\mathfrak{g}$-valued $0$--form $\epsilon$. It acts on  
a connection $A$ as %given by \hphantom{xxxxxxxxxxxxxxxxx}
\begin{equation}
\delta A=-d\epsilon-[A,\epsilon]. \vphantom{\bigg]}
\label{deltaA}
\end{equation}
The corresponding variation of the curvature $F_A$ of $A$ is 
\begin{equation}
\delta F_A=[\epsilon,F_A]. %\vphantom{\bigg]}
\label{deltaFA}
\end{equation}
Following \ccite{Bojowald:0406445,Kotov:2007nr,Gruetzmann:2014ica}, we reformulate the above 
well--known facts as follows. 

A differential graded commutative algebra
is a graded commutative algebra $C$ endowed with a nilpotent degree $1$ derivation $Q_C$.
A NQ--manifold is a non negatively graded manifold $X$ equipped with a homological
vector field $Q_X$, i. e. a degree 1 nilpotent vector field, so that the smooth function algebra 
$C^\infty(X)$ together with $Q_X$ constitute a differential graded commutative algebra.  

By a standard construction, with the manifold $M$ one can associate the NQ--manifold $(T[1]M,d)$, where
$T[1]M$ is the $1$--shifted tangent bundle of $M$ and $d$ is the homological vector field \hphantom{xxxxxxxx}
\begin{equation}
d=\xi^i\partial_{xi}, %\vphantom{\bigg]}
\label{dt1M}
\end{equation}
$x^i$, $\xi^i$ being degree $0$, $1$ base and fiber coordinates of $T[1]M$. 
Similarly, with the Lie algebra $\mathfrak{g}$ one can associate the NQ--manifold 
$(\mathfrak{g}[1],Q_{\mathfrak{g}})$, where $\mathfrak{g}[1]$ is the $1$--shifted 
vector space $\mathfrak{g}$ and $Q_{\mathfrak{g}}$ is the Chevalley--Eilenberg differential, 
\begin{equation}
Q_{\mathfrak{g}} \pi=-\tfrac{1}{2}[\pi,\pi]  %\vphantom{\bigg]}
%\xi^\alpha=-\tfrac12 f^\alpha_{\beta\gamma}\xi^\beta\xi^\gamma~,
\label{qce}
\end{equation}
with $\pi=\pi^at_a$, $\pi^a$ being degree $1$ coordinates of $\mathfrak{g}[1]$ with respect to
a given basis $t_a$ of $\mathfrak{g}$ assumed of degree $0$.   %of smooth functions of $T[1]M$ 

It is a basic fact that the graded commutative algebra $\Omega^*(M)$ of differential forms 
and the de Rham differential $d_{dR}$ of $M$ can be identified with the graded commutative algebra 
$C^\infty(T[1]M)$ and the homological vector field $d$, respectively. The datum of 
a connection $1$--form $A$ is then equivalent to that of a graded manifold morphism  
$a:T[1]M\rightarrow \mathfrak{g}[1]$ if $A$ and $a$ are related as 
%we stipulate that
%. The relationship of $A$ to $a$ is %former to the latter
%being . $A$ and $a$ are related as In fact, by \ceqref{A=a*pi}, 
\begin{equation}
A=a^*\pi, 
\label{A=a*pi}
\end{equation}
since in this way $A$ is determined by and determines $a$. Correspondingly,
The curvature $2$--form $F_A$ of $A$ can be expressed in terms of $a$ as 
\begin{equation}
F_A=da^*\pi+\tfrac{1}{2}[a^*\pi,a^*\pi]=da^*\pi-a^*Q_{\mathfrak{g}}\pi.
\label{FA=a*pi}
\end{equation}
%Cast in this form, 
$F_A$ thus provides a measure of the failure of $a$ 
to be a NQ--manifold morphism. 

To see how gauge transformation can be accommodated in this framework, one proceeds as follows. 
We replace the map $a$ with its graph $\hat a:T[1]M\rightarrow T[1]M\times\mathfrak{g}[1]$.
Since the datum of $a$ is equivalent to that of $\hat a$, it should be possible to reformulate 
everything in terms of $\hat a$. Indeed, $\hat a$ is also a graded manifold morphism. 
$T[1]M\times\mathfrak{g}[1]$ is a NQ--manifold with homological vector field $d+Q_{\mathfrak{g}}$. 
Furthermore, the datum of a connection $A$ amounts to that of $\hat a$, since \ceqref{A=a*pi}
can be recast as \hphantom{xxxxxxxxxxxxxxx}
\begin{equation}
A=\hat a^*\pi.
\label{A=a*pi/1}
\end{equation}
Similarly, we can rewrite the expression \ceqref{FA=a*pi} of 
the curvature $F_A$ of $A$ as 
\begin{equation}
F_A=d\hat a^*\pi+\tfrac{1}{2}[\hat a^*\pi,\hat a^*\pi]=d\hat a^*\pi-\hat a^*(d+Q_{\mathfrak{g}})\pi.
\label{FA=a*pi/1}
\end{equation}
We call a vector field $w$ on $T[1]M\times\mathfrak{g}[1]$ vertical, if it is everywhere 
directed along $\mathfrak{g}[1]$. An infinitesimal gauge transformation $\epsilon$ is encoded 
in a degree $-1$ vertical vector field $w$ on $T[1]M\times\mathfrak{g}[1]$.
The relation of $\epsilon$ to $w$ is 
\begin{equation}
\epsilon=w\pi. 
\label{ewpi}
\end{equation}
Taking \ceqref{A=a*pi/1} into account, the gauge variation \ceqref{deltaA} of $A$ is given by 
\begin{equation}
\delta A=-dw\pi-[\hat a^*\pi,w\pi] 
=-\hat a^*[d+Q_{\mathfrak{g}},w]\pi
\label{bksdeltaA}
\end{equation}
where in the last term $[-,-]$ denotes the Lie bracket of the vector field Lie algebra 
of $T[1]M\times\mathfrak{g}[1]$. 
Similarly, on account of \ceqref{FA=a*pi/1}, the gauge variation \ceqref{deltaFA} of $F_A$ reads as
\begin{equation}
\delta F_A=[w\pi,d\hat a^*\pi-\hat a^*(d+Q_{\mathfrak{g}})\pi] 
=-(d\hat a^*-\hat a^*(d+Q_{\mathfrak{g}}))[d+Q_{\mathfrak{g}},w]\pi.
\label{bksdeltaFA}
\end{equation}

The above reformulation of ordinary Yang--Mills theory immediately points to its 
higher generalization. One replaces the Lie algebra $\mathfrak{g}$ with an $L_\infty$ 
algebra $\mathfrak{v}$, a graded vector space equipped with a set of brackets
satisfying generalized Jacobi identities. Similarly to the ordinary Lie case, 
the algebraic structure of $\mathfrak{v}$ is encoded in the $1$--shifted space 
$\mathfrak{v}[1]$ and a Chevalley--Eilenberg differential $Q_{\mathfrak{v}}$
with an action on $\mathfrak{v}[1]$ expressed via a coordinate vector $\pi$. 
The datum of a higher connection polyform $A$
is now equivalent to that of a graded manifold morphism $a:T[1]M\rightarrow \mathfrak{v}[1]$
with the relationship of $A$ to $a$ still expressed by \ceqref{A=a*pi}. The curvature polyform
$F_A$ is then given by the last term of \ceqref{FA=a*pi} with $Q_{\mathfrak{g}}$ replaced by
$Q_{\mathfrak{v}}$. 

Higher gauge transformations are encoded by degree $-1$ vertical vector fields
on $T[1]M\times\mathfrak{v}[1]$. The gauge variations of the higher connection and 
curvature polyforms $A$ and $F_A$ are then given by the last terms of eqs. \ceqref{bksdeltaA} 
and \ceqref{bksdeltaFA}, respectively, again with $Q_{\mathfrak{g}}$ replaced by
$Q_{\mathfrak{v}}$. 

%\vfil\eject

\subsection{\textcolor{blue}{\sffamily %Review of 
Superfield approach to BRST symmetry}}\label{subsec:brsttheory}

The basic idea of BRST theory is to replace a local gauge symmetry by a degree $1$ global symmetry
acting on an extended field space containing ghost fields in addition to the original 
gauge fields. BRST symmetry transformation is essentially infinitesimal and is required to
satisfy the Wess--Zumino consistency condition. 

In BRST theory, fields are bigraded: in addition to form degree, they have ghost degree.
While form degree is definite non negative, ghost degree can have any sign.
A field $\phi^{(p,k)}$ of form degree $p$ and ghost degree $k$ is said to have bi\-degree 
$(p,k)$. The total degree, or simply degree, is the sum of the form and ghost degrees.
A field $\phi^{(n)}$ of degree $n$ is a sum of fields $\phi^{(p,n-p)}$ 
of bidegree $(p,n-p)$ with $0\leq p\leq d$, $d$ being the dimension of the space--time manifold $M$. 
We can then view $\phi^{(n)}$ as a BRST superfield and the $\phi^{(p,n-p)}$ as its components
\ccite{Baulieu:1981sb,Baulieu:1984iw,Baulieu:1984ih,Bonora:1980pt,
Bonora:1980ar,Bonora:1981rw}.

The de Rham differential $d_{dR}$ has bidegree $(1,0)$. 
BRST transformation is given by the action of a bidegree $(0,1)$ derivation $s_{BRST}$,
the BRST variation operator, 
which by virtue of the Wess--Zumino consistency condition is nilpotent, 
\begin{equation}
s_{BRST}{}^2=0.
\label{wzcons}
\end{equation}
In BRST theory, we have therefore two odd derivations. 
$d_{dR}$ and $s_{BRST}$ anticommute, $s_{BRST}d_{dR}+d_{dR}s_{BRST}=0$. 

The above can be stated in the language of NQ--manifold theory 
employed in the BGKS theory of subsect. \cref{subsec:bgkstheory}. 
While ordinary fields belong to the graded commutative function algebra $C^\infty(T[1]M)$,
BRST superfields span the algebra $C^\infty(T[1]M)\otimes G_{\mathbb{R}}$, where 
$G_{\mathbb{R}}$ is the ghost algebra \hphantom{xxxxxxxx}%raded commutative algebra
\begin{equation}
G_{\mathbb{R}}=\ddd_{k=-\infty}^\infty\mathbb{R}[k]
\label{ghostreal}
\end{equation}
with $\mathbb{R}[k]$ the $k$--shifted real line. 
$C^\infty(T[1]M)\otimes G_{\mathbb{R}}$ has indeed two distinct gradings. The first stems from the grading of 
$T[1]M$ in the usual way and can thus be identified with form degree. The second originates 
from the independent grading of $G_{\mathbb{R}}$ and corresponds to ghost degree. 
The de Rham differential $d_{dR}$ and the BRST variation operator $s_{BRST}$ are expressed as anticommuting 
nilpotent bidegree $(1,0)$ and $(0,1)$ derivations $d$ and $s$ on $C^\infty(T[1]M)\otimes G_{\mathbb{R}}$,
respectively.  

In the BRST formulation of ordinary gauge theory, the gauge field $A$ is promoted to a 
$\mathfrak{g}[1]$--valued BRST superfield $\mathcal{A}$, the BRST gauge superfield. 
The total field strength $\mathcal{F}_{\mathcal{A}}$ of $\mathcal{A}$, a $\mathfrak{g}[2]$--valued BRST superfield, 
is required to vanish, %\hphantom{xxxxxxxxxxxxxxx}
\begin{equation}
\mathcal{F}_{\mathcal{A}}=(d+s)\mathcal{A}+\tfrac{1}{2}[\mathcal{A},\mathcal{A}]=0. 
\label{curlyFA0}
\end{equation}
This relation defines the BRST variation $s\mathcal{A}$ of $\mathcal{A}$, 
\begin{equation}
s\mathcal{A}=-d\mathcal{A}-\tfrac{1}{2}[\mathcal{A},\mathcal{A}]\equiv-F_{\mathcal{A}}. 
\label{scurlyA}
\end{equation}
It is straightforward to check that $s^2\mathcal{A}=0$ as required by \ceqref{wzcons}. 

It is illuminating to write the BRST gauge superfield $\mathcal{A}$ and its 
BRST variation $s\mathcal{A}$ in components. $\mathcal{A}$ has a component expansion of the form \pagebreak 
\begin{equation}
\mathcal{A}=c+A+\gamma+\ldots
\label{curlyAcomps}
\end{equation}
Here, $c$ is the component of bidegree $(0,1)$ customarily called Faddeev--Popov ghost.
$A$ is the component of bidegree $(1,0)$ and is nothing but the familiar gauge field,
as indicated by the notation. $\gamma$ is a component of bidegree $(2,-1)$. The ellipses 
denote components of higher form and lower ghost degree components whose number depends on the space--time 
dimension $d$. Inserting \ceqref{curlyAcomps} into \ceqref{scurlyA}, the BRST variations 
$sc$, $sA$, $s\gamma$, ... of $c$, $A$, $\gamma$, ... are easily obtained. 

The BRST variation of the gauge field $A$ is 
\begin{equation}
sA=-dc-[A,c],
\label{sA}
\end{equation}
while that of the field strength $F_A$ of $A$  
\begin{equation}
sF_A=-[c,F_A].
\label{sFA}
\end{equation}
They are formally identical to \ceqref{deltaA} and \ceqref{deltaFA}. 
The Faddeev--Popov ghost field $c$ behaves therefore as a degree $1$ infinitesimal 
gauge transformation parameter. $c$ has however a non trivial BRST variation 
\begin{equation}
sc=-\tfrac{1}{2}[c,c].
\label{sc}
\end{equation}
The BRST variation of the ghost field $\gamma$,
\begin{equation}
s\gamma=-F_A-[c,\gamma],
\label{sgamma}
\end{equation}
is interesting because it yields the field strength $F_A$ of the gauge field  $A$. 

The possibility of a BRST formulation of higher gauge theory on the same lines should now be quite evident.
The higher gauge field $A$ is promoted to a $\mathfrak{v}[1]$--valued BRST higher gauge
superfield $\mathcal{A}$.
Imposing the vanishing of the total BRST field strength $\mathcal{F}_{\mathcal{A}}$ of $\mathcal{A}$,
generalizing condition \ceqref{curlyFA0}, it is then possible to obtain the 
BRST variation $s\mathcal{A}$ of $\mathcal{A}$, generalizing \ceqref{scurlyA}. 
More specifically, unfolding $A$ as a collection of fields $A_p$ with  $A_p$ 
a $p+1$--form valued in the degree $p$ subspace $\mathfrak{v}_p$
of $\mathfrak{v}$, $\mathcal{A}$ gets expressed as a collection of degree $p+1$ $\mathfrak{v}_p$--valued
BRST superfield $\mathcal{A}_p$. For the vanishing of \pagebreak the total BRST field strength 
$\mathcal{F}_{\mathcal{A}p}$ of $\mathcal{A}_p$, one obtains the BRST variations $s\mathcal{A}_p$
of $\mathcal{A}_p$. 

\vspace{-.4mm}
%\vfil\eject

\subsection{\textcolor{blue}{\sffamily BRST reformulation of BGKS theory}}\label{subsec:ztheory}

\vspace{-.2mm}
%It must be noted that t
The way gauge symmetry emerges in BGKS theory is
quite different from that it does in BRST theory, though the two
formulations are obviously related. %by the concordance of their 
BGKS theory is simple, elegant and geometrical, but in its present
form cannot be used in perturbative quantum field theory.
BRST theory treats ghost and gauge fields on an equal 'democratic'
footing. By design it provides the natural setting for gauge fixing,
but it lacks a clear geometrical underpinning. 

%\vspace{-.1mm}
Consider again the BGKS formulation of ordinary gauge theory
expounded in subsect. \cref{subsec:bgkstheory}. 
From inspecting \ceqref{A=a*pi} and \ceqref{FA=a*pi}, it appears that
both the connection $A$ and its curvature $F_A$ depend on the map 
$a:T[1]M\rightarrow \mathfrak{g}[1]$ through
its pull-back operator $a^*:C^\infty(\mathfrak{g}[1])\rightarrow C^\infty(T[1]M)$.
$a^*$ is a graded commutative algebra morphism. By \ceqref{A=a*pi}, on one hand 
$a^*$ determines $A$, on the other $a^*$ is determined by $A$, 
since $C^\infty(\mathfrak{g}[1])=S(\mathfrak{g}[1])$, the symmetric algebra of $\mathfrak{g}[1]$,
and hence $a^*$ is fixed by its action on the degree $1$ subspace $\mathfrak{g}[1]\subset S(\mathfrak{g}[1])$. 
Therefore, we can identify the connection $A$ with the morphism $a^*$. 
The operator $F_{a^*}:C^\infty(\mathfrak{g}[1])\rightarrow C^\infty(T[1]M)$ 
defined by \hphantom{xxxxxxxxxx}
\begin{equation}
F_{a^*}=da^*-a^*Q_{\mathfrak{g}}. \vphantom{\bigg]}
\label{deficit}
\end{equation}
appearing in \ceqref{FA=a*pi} is a degree $1$ derivation of $C^\infty(\mathfrak{g}[1])$ 
valued in $C^\infty(T[1]M)$ over $a^*$. By \ceqref{FA=a*pi}, on one hand 
$F_{a^*}$ determines $F_A$, on the other $F_{a^*}$ is determined by $F_A$, 
as, being $C^\infty(\mathfrak{g}[1])=S(\mathfrak{g}[1])$,  
$F_{a^*}$ is fixed by its action on the degree $1$ subspace $\mathfrak{g}[1]$
once $a^*$ is given. Therefore, we can identify the curvature $F_A$ with the derivation≈ß
$F_{a^*}$. 

\vspace{-.1mm}
The derivation $F_{a^*}$ encodes a basic property of  the algebra morphism $a^*$.
$F_{a^*}$ vanishes precisely when $a^*$ is 
a differential morphism. It thus measures the failure of $a^*$ to be so. 

\vspace{-.1mm}
In higher gauge theory, the Lie algebra $\mathfrak{g}$ is replaced by an 
$L_\infty$--algebra $\mathfrak{v}$ and the higher connection $A$ and curvature $F_A$ 
depend again on the mapping $a:T[1]M\rightarrow \mathfrak{v}[1]$ through its pull-back operator 
$a^*:C^\infty(\mathfrak{v}[1])\rightarrow C^\infty(T[1]M)$. The above considerations extend 
essentially unchanged. 

\vspace{.33mm}
The maps $a:T[1]M\rightarrow \mathfrak{v}[1]$ employed in BGKS theory
are ordinary graded manifold morphisms. 
As we shall endeavour to show in this paper, a geometrical framework 
allowing for more general {\it internal} graded manifold morphisms $a:T[1]M\rightarrow \mathfrak{v}[1]$
paves the way to a reformulation of BGKS theory that is interesting on its own and moreover 
subsumes BRST theory from the start in an efficient and elegant way. Let us see how.

\vspace{.33mm}
Internal morphisms differ from ordinary ones in that their local expression
contains smooth functions of the base coordinates $x^i$ of possibly non zero degree.
For an internal $a:T[1]M\rightarrow \mathfrak{v}[1]$, one can define an internal pull--back operator
$a^\#:C^\infty(\mathfrak{v}[1])\rightarrow C^\infty(T[1]M)\otimes G_{\mathbb{R}}$
having the {\it internal} function algebra $C^\infty(T[1]M)\otimes G_{\mathbb{R}}$
as its range, where $G_{\mathbb{R}}$ is the graded algebra \ceqref{ghostreal}. Since $C^\infty(T[1]M)\otimes G_{\mathbb{R}}$ 
is precisely the algebra of BRST superfields, as explained in subsect. \cref{subsec:brsttheory}, 
we are now in the position of merging BGKS and BRST theory.

\vspace{.33mm}
In the BRST formulation of ordinary gauge theory, the connection $A$ is promoted to a
BRST connection superfield $\mathcal{A}$, which we describe as a degree $1$ 
element of the function space $C^\infty(T[1]M)\otimes G_{\mathbb{R}}\otimes\mathfrak{g}$ containing 
components of all possible form and ghost degree compatible with it having total degree $1$
(cf. subsect. \cref{subsec:brsttheory}). Setting \hphantom{xxxxxxxxxxxxxxxxxxxxx}
\begin{equation}
\mathcal{A}=a^\#\pi
\label{curlyAapi}
\end{equation}
defines an internal graded manifold morphism $a:T[1]M\rightarrow \mathfrak{v}[1]$ 
whose datum is equivalent to that of $\mathcal{A}$. In this way, thinking {\it ab initio} 
in terms of internal morphisms allows us to reformulate the BGKS theory in a way 
that fully incorporates BRST theory, as we anticipated. Upon replacing the Lie algebra
$\mathfrak{g}$ with an $L_\infty$--algebra $\mathfrak{v}$, the same conclusion is reached 
with regard to higher gauge theory.

This reformulation is not as straightforward as it may seem at first glance.
As the reader will see, a thorough reconsideration of the part relative 
to gauge transformation is involved. Further, 
the resulting theory will allow us also to make contact with BV approaches 
to higher gauge theory \ccite{BV1,BV2,Gomis:1994he} and in particular in the 
AKSZ formulation \ccite{Alexandrov:1995kv,Ikeda:2012pv}. %Alexandrov:1995kv}.

%\vfil\eject

\vspace{-1mm}
 
\subsection{\textcolor{blue}{\sffamily Plan of the paper
}}\label{subsec:plan}

%\pagebreak 

In sect. \cref{sec:higau}, we show how the intuitions about higher gauge theory
presented in the previous subsections can be systematically organized into a 
purely algebraic formulation based on the abstract theory of graded commutative algebras 
and their morphisms. Differential graded commutative algebras algebras, that is algebras
endowed with a nilpotent differential, play a basic role. The formulation incorporates naturally 
BRST symmetry and is also suitable for an AKSZ type construction of higher gauge theoretic 
models. 

In sect. \cref{sec:hisigma}, 
we work out an explicit formulation of higher gauge theory and gauged sigma models  
relying on the formal framework developed in  sect. \cref{sec:higau}. 
The characterizing point of our construction is the realization that for a full--fledged 
BV formulation incorporating ghost degrees of freedom higher fields must be viewed as
internal smooth functions from the shifted tangent bundle 
of a space time manifold to a shifted $L_\infty$--algebroid encoding the higher gauge symmetry. 
We analyze also the case where the $L_\infty$--algebroid arises from a higher Lie groupoid 
by Lie differentiation and elucidate the relation between higher gauge symmetry 
expressed through simplicial homotopy and BRST symmetry. 

Throughout the paper, we rely heavily on graded differential geometry, which 
through its multiple ramifications is the most
natural framework within which carrying out our program is possible.

\vspace{-1.5mm}

\subsection{\textcolor{blue}{\sffamily Outlook
}}\label{subsec:outlook}

Our analysis, albeit inspired by geometry, indicates that the most basic features 
of higher gauge theory are ultimately algebraic.
The formulation of sect. \cref{sec:higau} captures many of the most basic properties 
of standard higher gauge theoretic models without committing
to the assumption that the relevant differential graded commutative algebras 
are algebras of smooth functions on appropriate graded manifolds as in BGKS and BRST 
theory reviewed above. Sometimes in the future, it is our hope, this may allow to work out models with a 
definite higher gauge theoretic outlook in contexts quite different from the
supergravity, string and brane theoretic ones where higher gauge theory 
was elaborated in the first place, with innovative applications to
the study of confining and gapped phases of gauge theory \ccite{Kapustin:2013qsa,Kapustin:2013uxa}, 
statistical mechanics and topological phases of matter \ccite{Bullivant:2016clk,Bullivant:2017sjz}
and more.

\vfil\eject

\section{\textcolor{blue}{\sffamily Algebraic higher gauge theory}}\label{sec:higau}

\vspace{.15mm}

In this section, we present an abstract purely algebraic formulation of higher gauge theory
based on the theory of graded commutative algebras and their dif\-ferential enhancements.
Our framework lends itself to multiple readings, each of which is interesting on its own.  
%See ref. for background material. 
%We assume the reader is familiar with the basics of graded commutative algebra. 
%
On the geometrical side,
graded commutative algebras are models for the algebras of smooth functions on N--manifolds.
Similarly, differential graded commutative algebras are models for the algebras of smooth functions 
on NQ--manifolds.
On the physical mathematical side, non differential morphisms between differential 
graded commutative algebras formalize higher gauge fields, their defects encode their higher curvatures and 
the defect identities they obey are the analog of the basic Bianchi identities.
The formulation incorporates naturally BRST symmetry and is also suitable 
for an AKSZ type construction of higher gauge theory. 
%\ccite{%BV1,BV2,Gomis:1994he,Alexandrov:1995kv,Ikeda:2012pv}.

%\vfil\eject

\subsection{\textcolor{blue}{\sffamily Differential algebras, 
graded algebra morphisms and  defect %and defect identity
}}\label{subsec:algdef}

\vspace{.25mm}

A graded vector space $V$ is a vector space endowed with a direct sum decomposition $V=\bigoplus_{p\in\mathbb{Z}}V_p$.
For $p\in\mathbb{Z}$, a degree $p$ element $f$ of $V$ is just an element of $V_p$. An element $f$ of $V$ 
is called homogeneous if it is of degree $p$ for some $p$. In that case, we write $|f|=p$.  

\vspace{.15mm}

For $p\in\mathbb{Z}$, a degree $p$ linear map $T:V'\rightarrow V$ of graded vector spaces is a linear map 
such that $TV'{}_q\subseteq V_{p+q}$ for all $q\in\mathbb{Z}$. A linear map $T:V'\rightarrow V$ is said homogeneous
if it is of degree $p$ for some $p$. In that case, we write $|T|=p$.  

%If $f$ is an element of $C_p$, we say that $f$ is homogeneous of degree $p$ and write $|f|=p$.  
%If $T:C'\rightarrow C$ is a linear map of graded commutative algebras % $C$, $C'$ 
%such that $TC'{}_q\subseteq C_{p+q}$ for all $q$, we say that $T$ is homogeneous of degree $p$ 
%and write $|T|=p$. 

%In this paper, we shall work mainly with homogeneous objects. %, though not all of them are.

%if for any homogeneous $f\in C_{2q}$ of degree $q$, $Tf\in C_{1$ 
%is homogeneous of degree $p+q$. In that case, we write $|T|=p$. 

\vspace{.15mm}

A graded algebra $A$ is a graded vector space equipped with an associative and distributive
product such that $A_pA_q\subseteq A_{p+q}$ for all $p,q\in\mathbb{Z}$ and a %multiplicative 
unit $1$ with $|1|=0$. A {\it graded commutative algebra} $C$ is a graded algebra such that 
\begin{equation}
fg=(-1)^{|f||g|}gf
\label{grcom}
\end{equation}
for all homogeneous $f,g\in C$. %Below, we consider only graded commutative algebras. 
In the following, we shall consider exclusively graded commutative algebras. 
%
%A linear map $\varUpsilon:C_2\rightarrow C_1$ of graded commutative algebra $C_1$,
%$C_2$ is said to have degree $p$ if for any homogeneous $f\in C_2$ one has 
%$|\varUpsilon f|=p+|f|$. We then write $\varUpsilon|=p$. 
A morphism $\varUpsilon$ of a graded commutative algebra $C_2$ into another 
$C_1$ is a degree $0$ linear map $\varUpsilon:C_2\rightarrow C_1$ 
such that for $f,g\in C_2$ \hphantom{xxxxxxxxxx}
\begin{equation}
\varUpsilon(fg)=\varUpsilon f \varUpsilon g.
\label{}
\end{equation}
Graded commutative algebras and their morphisms from a category $\mathsans{grcAlg}$.

\vspace{.15mm}

For $p\in\mathbb{Z}$, a degree $p$ {\it derivation} $D$ of a graded commutative algebra $C$ is a degree 
$p$ linear map $D:C\rightarrow C$ such that for homogeneous $f,g\in C$
\begin{equation}
D(fg)=Df g+(-1)^{p|f|}f Dg.
\label{grder}
\end{equation}
The degree $p$ derivations $D$ of $C$ form a vector space $\Der_p(C)$.
The derivations $D$ of all possible degrees $p$ span 
the graded vector space $\Der(C)=\bigoplus_{p\in\mathbb{Z}}\Der_p(C)$
of all derivations of $C$. A derivation $D$ of $\Der(C)$ is called homogeneous if it is of
degree $p$ for some $p$. In that case, we write $|D|=p$.

$\Der(C)$ is a graded Lie algebra with the Lie brackets
\begin{equation}
[X,Y]=XY-(-1)^{|X||Y|}YX
\label{}
\end{equation}
for homogeneous derivations $X, Y\in\Der(C)$. These brackets are graded antisymmetric 
and satisfy the graded Jacobi identity, 
\begin{align}
&[X,Y]+(-1)^{|X||Y|}[Y,X]=0, \hspace{3cm}
\vphantom{\Big]}
\label{gransym}
\\
&(-1)^{|Z||X|}[X,[Y,Z]]+(-1)^{|X||Y|}[Y,[Z,X]]
\vphantom{\Big]}
\label{grjac}
\\
&\hspace{5.9cm}+(-1)^{|Y||Z|}[Z,[X,Y]]=0,
\vphantom{\Big]}
\nonumber
\end{align}
for homogeneous derivations $X, Y,Z\in\Der(C)$, as it is straightforward to verify. 
%If $X\in \Der_p(C)$, we say that $X$ is a homogeneous derivation of degree $p$ and write $|X|=p$. 
%In this paper, we normally consider fixed degree homogeneous derivations only. 

A graded algebra possesses a distinguished degree $0$ derivation $E$, the {\it Euler derivation},
characterized by the property that
\begin{equation}
Ef=|f|f
\label{}
\end{equation}
for homogeneous $f\in C$. If $D\in \Der(C)$ is a homogeneous derivation of $C$, then
\begin{equation}
[E,D]=|D|D.
\label{}
\end{equation}

The notion of derivation has the following generalization that will turn out to be  
important in our analysis. 
A degree $p$ derivation $D$ of a graded commutative algebra $C_2$ valued in another %graded commutative algebra 
$C_1$ over a morphism $\varUpsilon:C_2\rightarrow C_1$ is a degree $p$ linear map $D:C_2\rightarrow C_1$ such that
\begin{equation}
D(fg)=Df \varUpsilon g+(-1)^{p|f|}\varUpsilon f Dg
\label{}
\end{equation}
for homogeneous $f,g\in C_2$. 
The degree $p$ derivations $D$ of $C_2$ valued in $C_1$ over $\varUpsilon$ 
form a vector space $\Der_{p\varUpsilon}(C_2,C_1)$.
The derivations $D$ of all possible degrees $p$ span 
the graded vector space $\Der_\varUpsilon(C_2,C_1)=\bigoplus_{p\in\mathbb{Z}}\Der_{p\varUpsilon}(C_2,C_1)$
of all derivations of $C_2$ valued in $C_1$ over $\varUpsilon$. 
A derivation $D$ of $\Der_\varUpsilon(C_2,C_1)$ is called homogeneous if it is of 
degree $p$ for some $p$. In that case, we write $|D|=p$.
A degree $p$ derivation $D$ of a graded commutative algebra $C$ 
is just a degree $p$ derivation of $C$ valued in $C$ over the identity $\id_C$. %:C\rightarrow C$.

A {\it differential graded commutative algebra} $C$ is a graded commutative algebra equipped with
a {\it differential}, a degree $1$ derivation $Q$ of $C$ that is nilpotent,
\begin{equation}
Q^2=0.
\label{q20}
\end{equation}
With $C$, there are associated two cochain complexes and their cohomologies. The first is 
$(C,Q)$. The second is $(\Der(C),\ad Q)$, where $\ad Q=[Q,-]$. 
A morphism $\varUpsilon:C_2\rightarrow C_1$ of a differential graded commutative algebra $C_2$ into another 
$C_1$ is a graded commutative algebra morphism such that 
\begin{equation}
Q_1 \varUpsilon-\varUpsilon  Q_2=0.
\label{}
\end{equation}
Differential graded commutative algebras and their morphisms constitute a category 
$\mathsans{dgrcAlg}$ that is a non full subcategory of $\mathsans{grcAlg}$.

Let $C_1$, $C_2$ be differential graded commutative algebras with differentials $Q_1$, $Q_2$, 
respectively. For any morphism $\varUpsilon:C_2\rightarrow C_1$ of graded commutative algebras,
we define its {\it defect} $F_\varUpsilon:C_2\rightarrow C_1$ to be the degree $1$ linear map
\begin{equation}
F_\varUpsilon=Q_1\varUpsilon-\varUpsilon  Q_2. 
\label{defect}
\end{equation}
%Since as it readly verified 
%\begin{equation}
%F_\varUpsilon(fg)=F_\varUpsilon f \varUpsilon g+(-1)^{|f|}\varUpsilon f F_\varUpsilon g
%\label{}
%\end{equation}
%for $f,g\in C_2$, 
$F_\varUpsilon$ is in fact a degree $1$ derivation of $C_2$ valued in $C_1$ 
over the $\varUpsilon$. Intuitively, $F_\varUpsilon$ measures the failure of $\varUpsilon$ to be a morphism of  
differential graded commutative algebras: 
$\varUpsilon$ is one precisely when its defect $F_\varUpsilon=0$. 

The defect $F_\varUpsilon$ of $\varUpsilon$ satisfies the {\it defect identity}
\begin{equation}
Q_1F_\varUpsilon+F_\varUpsilon Q_2=0, \vphantom{\bigg]}
\label{bianchi}
\end{equation}
which follows from the nilpotence of $Q_1$, $Q_2$. 

%The fact that a morphism $\varUpsilon$ and its defect $F_\varUpsilon$ have respectively degree $0$ and $1$
%imply that $\varUpsilon$ and $F_\varUpsilon$ intertwine between the Euler derivations $E_1$, $E_2$ of $C_1$, $C_2$
%in a precise manner, namely \hphantom{xxxxxxxxxxxx}
%\begin{align}
%&E_1\varUpsilon-\varUpsilon E_2=0,
%\vphantom{\Big]}
%\label{euphi}
%\\
%&E_1F_\varUpsilon-F_\varUpsilon E_2=F_\varUpsilon.
%\vphantom{\Big]}
%\label{eufphi}
%\end{align}
%These relations will turn out to be useful to us. 

As we shall see, any higher gauge theory belongs to the realm of differential graded commutative algebras. 
The theory's field space is a graded commutative algebra $C_1$ and 
the theory's field type space is another $C_2$.
In the Lagrangian, the kinetic term is determined by a differential $Q_1$ on $C_1$, while the gauge 
field self--interaction is by another $Q_2$ on $C_2$. 
The theory's higher gauge field content is encoded in a graded commutative algebra 
morphism $\varUpsilon:C_2\rightarrow C_1$ in the sense that, for any type $f\in C_2$, $\varUpsilon f\in C_1$ 
is the gauge field of type $f$. %Furthermore, $C_1$, $C_2$ are differential with differentials $Q_1$, $Q_2$. 
The defect $F_\varUpsilon$ of $\varUpsilon$ encodes the higher gauge field curvatures 
and the defect identity the higher Bianchi identities these obey meaning that, for  $f\in C_2$, $F_\varUpsilon f\in C_1$ 
is the gauge curvature of the gauge field of type $f$ and the corresponding defect identity is the Bianchi identity 
this obeys. It is therefore important to study these objects in greater detail.

%\vfil\eject

\subsection{\textcolor{blue}{\sffamily The graded %algebra 
morphism manifold and the defect vector field}}\label{subsec:geodef}

We have found an elegant geometric interpretation of the defect of a graded com\-mutative algebra morphism
and defect identity, which now we illustrate. 

Let $C_1$, $C_2$ be graded commutative algebras. We assume that the set of (non differential) graded commutative 
algebra morphisms 
\begin{equation}
M(C_2,C_1)=\Hom_{\mathsans{grcAlg}}(C_2,C_1) \vphantom{\bigg]}
\label{}
\end{equation}
from $C_2$ to $C_1$ is an ordinary manifold (at least at a formal level). We aim to study its geometry. 

$M(C_2,C_1)$ is characterized by certain distinguished vector bundles.
For $p\in\mathbb{Z}$, the vector bundle $\Der_p(C_2,C_1)$ of degree $p$ derivations $D$ of $C_2$ valued in $C_1$ 
is the vector bundle over $M(C_2,C_1)$ whose fiber at a point $\varUpsilon$ is the vector space 
$\Der_{p\varUpsilon}(C_2,C_1)$ of degree $p$ derivations $D$ of $C_2$ valued in $C_1$ over $\varUpsilon$. 
Similarly, the vector bundle $\Der(C_2,C_1)$ of derivations $D$ of $C_2$ valued in $C_1$ 
is the vector bundle over $M(C_2,C_1)$ whose fiber at a point $\varUpsilon$ is the vector space 
$\Der_\varUpsilon(C_2,C_1)$ of all derivations $D$ of $C_2$ valued in $C_1$ over $\varUpsilon$. Clearly, one has 
$\Der(C_2,C_1)=\bigoplus_{p\in\mathbb{Z}}\Der_p(C_2,C_1)$. 

To study the geometry of the manifold $M(C_2,C_1)$, it is necessary to describe its tangent bundle $TM(C_2,C_1)$.  
The tangent space $T_\varUpsilon M(C_2,C_1)$ to $M(C_2,C_1)$ at a point $\varUpsilon$ can be characterized as follows. Since
\begin{equation}
\varUpsilon(fg)=\varUpsilon f\varUpsilon g \vphantom{\bigg]}
\label{}
\end{equation}
for $f,g\in C_2$, a tangent vector $\dot\varUpsilon\in T_\varUpsilon M(C_2,C_1)$ is just a degree $0$ linear map 
$\dot\varUpsilon:C_2\rightarrow C_1$ obeying the condition
\begin{equation}
\dot\varUpsilon(fg)=\dot\varUpsilon f\varUpsilon g+\varUpsilon f\dot\varUpsilon g \vphantom{\bigg]}
\label{phidot}
\end{equation}
that is a degree $0$ derivation of $C_2$ valued in $C_1$ over $\varUpsilon$ (cf. subsect. \cref{subsec:algdef}). The tangent bundle
$TM(C_2,C_1)$ is therefore identified with the vector bundle $\Der_0(C_2,C_1)$ of degree $0$ derivations of $C_2$
valued in $C_1$.  

%the set of the pairs $(\varUpsilon,\dot\varUpsilon)$ with $\varUpsilon$ a morphism of $C_2$ into $C_1$ 
%and $\dot \varUpsilon$ a degree $0$ derivation of $C_2$ valued in $C_1$ over $\varUpsilon$. 

In a graded geometric description of $M(C_2,C_1)$, one needs also 
to consider the degree shifted forms of the tangent bundle $TM(C_2,C_1)$.
Let $p\in\mathbb{Z}$. The $p$--shifted tangent space $T_\varUpsilon[p]M(C_2,C_1)$ to $M(C_2,C_1)$ 
at a point $\varUpsilon$ can be characterized by the natural graded generalization of condition \ceqref{phidot}. 
A tangent vector of $T_\varUpsilon[p]M(C_2,C_1)$ is a degree $p$ linear map 
$\dot\varUpsilon:C_2\rightarrow C_1$ obeying 
\begin{equation}
\dot\varUpsilon(fg)=\dot\varUpsilon f\varUpsilon g+(-1)^{p|f|}\varUpsilon f\dot\varUpsilon g \vphantom{\bigg]}
\label{}
\end{equation}
that is a degree $p$ derivation of $C_2$ valued in $C_1$ over $\varUpsilon$. The 
$p$--shifted tangent bundle $T[p]M(C_2,C_1)$ is therefore identified with the vector bundle $\Der_p(C_2,C_1)$ 
of degree $p$ derivations of $C_2$ valued in $C_1$.

%the set of the pairs $(\varUpsilon,\dot\varUpsilon)$ with $\varUpsilon$ a morphism of $C_2$ into $C_1$ 
%and $\dot\varUpsilon$ a degree $p$ derivation of $C_2$ valued in $C_1$ over $\varUpsilon$. 

Suppose now that $C_1$, $C_2$ are differential graded commutative algebras with differentials $Q_1$, $Q_2$, respectively. 
As we saw in subsect. \cref{subsec:algdef}, the defect $F_\varUpsilon$ of a morphism $\varUpsilon\in M(C_1,C_2)$ 
is a degree $1$ derivation of $C_2$ valued in $C_1$ over $\varUpsilon$, hence 
a tangent vector of $T_\varUpsilon[1]M(C_2,C_1)$. 
Therefore, the defect map $\varPhi\rightarrow F_\varPhi$ 
\footnote{$\vphantom{\dot{\dot{f}}}$ Here and in the following, we distinguish notationally between 
an arbitrary chosen but fixed morphism $\varUpsilon\in M(C_2,C_1)$ and the variable $\varPhi\in M(C_2,C_1)$.\vspace{-3mm}}     
encodes geometrically a degree $1$ vector field $F$ over $M(C_2,C_1)$, that is a section of the $1$--shifted tangent bundle 
$T[1]M(C_2,C_1)$. We shall call $F$ the {\it defect vector field} for obvious reasons. $F$ can be written as 
\begin{equation}
F=\langle F_\varPhi,\partial_\varPhi\rangle,
\label{fvect}
\end{equation}
where $\partial_\varPhi$ denotes derivation with respect to $\varPhi$ and 
$\langle -,-\rangle$ denotes the tangent--cotangent duality pairing (index contraction).
$F$ is in fact nilpotent
\begin{equation}
F^2=0
\label{f20}
\end{equation}
as a consequence of the defect identity \ceqref{bianchi}.

\vspace{-.5mm}

\begin{proof}
From \ceqref{defect}, \ceqref{fvect} and \ceqref{bianchi}, we have indeed 
\begin{equation}
F^2=-\langle Q_1F_\varPhi+F_\varPhi Q_2,\partial_\varPhi\rangle=0
\label{}
\end{equation}
showing \ceqref{f20}. 
\end{proof}

\vspace{-.5mm}

\noindent
$M(C_2,C_1)$ equipped with the homological defect vector field $F$ is in this way an NQ--manifold 
(cf. subsect. \cref{subsec:lgraman}).
%\vspace{-1mm}
%\noindent
%Equipped with $F$, $C^\infty(M(C_2,C_1))$ is so a differential graded commutative algebra.

The shifted tangent bundle $T[1]M(C_2,C_1)$ of $M(C_2,C_1)$ has itself a rich geometry 
stemming from the defect map and identity. As a manifold, 
$T[1]M(C_2,C_1)$ is described by the base coordinate $\varPhi$ and degree $1$ fiber coordinate $\delta\varPhi$ 
viewed as a formal vector. $T[1]M(C_2,C_1)$ is characterized by two distinguished degree $1$ vector fields,
that is sections of the twice iterated $1$--shifted tangent bundle $T[1]^2M(C_2,C_1)$.  
The first one is the canonical vector field 
\begin{equation}
\delta=\langle \delta\varPhi,\partial_\varPhi\rangle. \vphantom{\bigg]}
\label{}
\end{equation}
The second one is the derivative vector field $l_F$ of the defect vector field $F$. 
Ex\-plicitly, it reads as   \hphantom{xxxxxxxxxxxxxxxx}
\begin{equation}
l_F=\langle F_\varPhi,\partial_\varPhi\rangle-\langle \delta F_\varPhi,\partial_{\delta\varPhi}\rangle,
\label{lfvect}
\end{equation}
where $\delta F_\varPhi$ is given by the expression %\hphantom{xxxxxxxxxxxx}
\begin{equation}
\delta F_\varPhi=-Q_1\delta\varPhi-\delta\varPhi Q_2. 
\label{ddefect}
\end{equation}
and $\partial_{\delta\varPhi}$ denotes derivation with respect to $\delta\varPhi$ of degree $-1$. 
%\footnote{$\vphantom{\dot{\dot{\dot{x}}}}$ 
%It is a standard result of graded geometry that any section $V$ of the shifted tangent bundle $T[1]X$ 
%of a mainfold $X$ can be lifted to one $l_V$ of $T[1]^2X$, the Lie derivative vector field of $V$.}. 
$\delta$ and $l_F$ have both degree $1$, are both nilpotent and mutually anticommute, 
\begin{align}
&\delta^2=0,
\vphantom{\Big]}
\label{delta2=0}
\\
&l_F{}^2=0,
\vphantom{\Big]}
\label{lF20}
\\
&l_F\delta+\delta l_F=0.
\vphantom{\Big]}
\label{deltalFcom}
\end{align}

\begin{proof} \ceqref{delta2=0} is obvious. 
From \ceqref{defect}, \ceqref{lfvect}, \ceqref{ddefect} and \ceqref{bianchi}, we have 
\begin{equation}
l_F{}^2=-\langle Q_1F_\varPhi+F_\varPhi Q_2,\partial_\varPhi\rangle
-\langle \delta(Q_1F_\varPhi+F_\varPhi Q_2),\partial_{\delta\varPhi}\rangle=0,
\label{}
\end{equation}
showing \ceqref{lF20}. Finally, one has
\begin{equation}
l_F \delta+\delta l_F
=\langle \delta F_\varPhi,\partial_\varPhi\rangle-\langle \delta F_\varPhi,\partial_\varPhi\rangle=0
\label{}
\end{equation}
leading to \ceqref{deltalFcom}. 
\end{proof}

\noindent
So, $T[1]M(C_2,C_1)$ has two compatible NQ--manifold structures
associated with the homological canonical and defect Lie derivative vector fields 
$\delta$ and $l_F$.
%\footnote{$\vphantom{\dot{\dot{\dot{x}}}}$ The reader that is not familiar with this type
%graded geometric formulation of differential forms may find it useful to read subsect. 
%\cref{subsec:graman}\vspace{-3mm}}, 

%On general grounds, t

In general, the algebra $\Omega^*(X)$ of exterior differential forms of a manifold $X$ can be identified 
with the graded commutative algebra $C^\infty(T[1]X)$ of functions of the shifted tangent bundle 
$T[1]X$, the de Rham differential $d_{X}$ of $X$ with the 
canonical homological vector field $\delta$ of $T[1]X$
and the de Rham cohomology of $X$ with the $\delta$ cohomology of $T[1]X$. 
In the case of $M(C_2,C_1)$, the algebra $\Omega^*(M(C_2,C_1))$ is acted upon 
by two compatible differentials, the de Rham and the defect Lie derivative ones, 
and features the associated cohomologies. 

%The elements of $C^\infty(T[1]M(C_2,C_1))$ can be expressed as functions of $\varPhi$, $\delta\varPhi$ 
%polynomials in $\delta\varPhi$ and \noindent
%$(C^\infty(T[1]M(C_2,C_1)),\delta,l_F)$ By \ceqref{delta2=0}, \ceqref{lF20}, \ceqref{deltalFcom}, 
%$C^\infty(T[1]M(C_2,C_1))$ equipped with $\delta$ and $l_F$ 
%is thus a bidifferential graded commutative algebra.

As explained in subsect. \cref{subsec:algdef}, in a higher gauge theory characterized 
by a differential graded commutative algebra $C_1$ of gauge fields and another $C_2$ of field types, 
a graded commutative algebra morphism $\varUpsilon:C_2\rightarrow C_1$ represents
a collection of higher gauge fields, its defect $F_\varUpsilon$ the associated collection of higher gauge 
curvatures and the defect identity of $F_\varUpsilon$ the Bianchi identities these obey. 
The morphism manifold $M(C_2,C_1)$ can so be regarded as the space of all higher gauge field configurations,
the vector field $F$ of $M(C_2,C_1)$ as the higher gauge curvature map and its nilpotence as a condition
encoding the Bianchi identities.

%\vfil\eject

\subsection{\textcolor{blue}{\sffamily BRST theory of graded %algebra 
morphisms}}\label{subsec:brst}

We have just seen in subsect. \cref{subsec:geodef} that in higher gauge theory 
the higher gauge field configuration space can be identified with the manifold $M(C_2,C_1)$ 
of non differential morphisms 
of two differential graded commutative algebras $C_2$, $C_1$. The BRST analysis
of higher gauge symmetry thus reduces to that of the symmetry of $M(C_2,C_1)$.
The resulting BRST theory turns out to be very rich.
%We have found BRST theory a very powerful approach to the study of higher gauge symmetry. 

Let $C_1$, $C_2$ be graded commutative algebras and let $\varPhi\in M(C_2,C_1)$
be a graded commutative algebra morphism variable. Then, 
\begin{equation}
\varPhi(fg)=\varPhi f\varPhi g
\label{phimorp}
\end{equation}
for $f,g\in C_2$. 

On general grounds, the BRST symmetry of $M(C_2,C_1)$ is encoded in degree $1$ derivation $s$ of 
$M(C_2,C_1)$. By consistency with \ceqref{phimorp}, $s\varPhi$ must satisfy
\begin{equation}
s\varPhi(fg)=s\varPhi f\,\varPhi g+(-1)^{|f|}\varPhi f\,s\varPhi g \vphantom{\bigg]}
\label{sphider}
\end{equation}
for $f,g\in C_2$. $s\varPhi$ is therefore a degree $1$ derivation of $C_2$ valued in $C_1$ 
over $\varPhi$. 

The twice iterated variation $s^2\varPhi$ is a degree $2$ derivation of $C_2$ valued in $C_1$ over $\varPhi$ 
as it is readily checked from \ceqref{sphider}. Complying with general principles of gauge theory, we impose the 
Wess--Zumino consistency condition
\begin{equation}
s^2\varPhi=0.
\label{wzcc}
\end{equation}

Following Kotov and Strobl \ccite{Kotov:2007nr}, we consider the case when $s\varPhi$ has the form
\begin{equation}
s\varPhi=-U_1\varPhi+\varPhi U_2,
\label{sphi}
\end{equation}
where $U_1\in \Der(C_1)$, $U_2\in \Der(C_2)$ are degree $1$ derivation variables. 
It is immediately checked that $s\varPhi$ is a degree $1$ derivation of $C_2$ valued in $C_1$ 
over $\varPhi$, as required. 

To impose the nilpotence condition \ceqref{wzcc}, we have to extend the degree $1$ derivative action of $s$ to 
$U_1$, $U_2$. %we have to allow $s$ to act on $U_1$, $U_2$ as a degree $1$ derivation. 
A sufficient condition for \ceqref{wzcc} to hold is that 
\begin{equation}
sU_1=-\tfrac{1}{2}[U_1,U_1], \qquad sU_2=-\tfrac{1}{2}[U_2,U_2]. 
%, \qquad sU_2=-\tfrac{1}{2}[U_2,U_2].
%sU_2=-U_2{}^2.
%sU_i+U_i{}^2=0.
\label{sui}
\end{equation}

\begin{proof} A simple calculation furnishes
\begin{equation}
s^2\varPhi=-(sU_1+U_1{}^2)\varPhi+\varPhi(sU_2+U_2{}^2).
\label{}
\end{equation}
Thus, if the \ceqref{sui} hold, $s^2\varPhi=0$ as required. 
\end{proof}

\noindent
The nilpotence condition \ceqref{wzcc} extends to $U_1$, $U_2$, 
\begin{equation}
s^2U_1=0, \qquad s^2U_2=0, %s^2U_i=0
\label{wzccgh}
\end{equation}

\begin{proof}
As the proof of the statement is identical for $U_1$, $U_2$, we suppress indexes. 
Using the \ceqref{sui} and the graded Jacobi identity \ceqref{grjac}, we obtain 
\begin{equation}
s^2U=-\tfrac{1}{2}[U,[U,U]]=0,
\label{}
\end{equation}
showing the \ceqref{wzccgh}. %as it is readily verified from \ceqref{sui}. 
\end{proof}

\vspace{-1.3mm}

Next, we assume that a differential structure is added to our graded commutative algebras.
We thus consider two differential graded commutative algebras $C_1$, $C_2$ with differentials 
$Q_1$, $Q_2$. We want to compute the variation $sF_\varPhi$ of the defect $F_\varPhi$ of $\varPhi$. 
As it is reasonable, we shall assume that
\begin{equation}
sQ_1=0,\qquad sQ_2=0,
%sQ_i=0
\label{sqi0}
\end{equation}
since $Q_1$, $Q_2$ are fixed data of our construction. From \ceqref{defect}, we obtain, 
\begin{equation}
sF_\varPhi=-U_1F_\varPhi-F_\varPhi U_2+[Q_1,U_1]\varPhi-\varPhi[Q_2,U_2].
\label{sfphiex}
\end{equation}

\begin{proof} By \ceqref{defect} and \ceqref{sqi0}, 
\begin{equation}
sF_\varPhi=-Q_1s\varPhi-s\varPhi Q_2.
\label{sFgen}
\end{equation}
Inserting \ceqref{sphi} into \ceqref{sFgen} and using %rearranging according to 
\ceqref{defect}, we obtain \ceqref{sfphiex} through a simple rearrangement. 
\end{proof}

\noindent
Inspired by gauge theory, it is natural to require that the variation $sF_\varPhi$ of $F_\varPhi$
should depend on $\varPhi$ only through $F_\varPhi$ itself \ccite{Kotov:2007nr}. For this reason,  
the last two terms of the right hand side of \ceqref{sfphiex} should be absent. \pagebreak A sufficient condition
for this to be the case is that  
\begin{equation}
[Q_1,U_1]=0, \qquad [Q_2,U_2]=0.
%[Q_i,U_i]=0, 
\label{uicycle}
\end{equation}
%i. e that $U_1$, $U_2$ are degree $1$--cocycles in the $\ad Q_1$, $\ad Q_1$ cohomologies, respectively. 
When these hold, \ceqref{sfphiex} takes indeed the required form
\begin{equation}
sF_\varPhi=-U_1F_\varPhi-F_\varPhi U_2. 
\label{sfphi}
\end{equation}

The restrictions \ceqref{uicycle} are compatible with the variations \ceqref{sui}, as required by consistency.

\vspace{-.5mm}

\begin{proof} 
Again, since the proof of the statement is identical for $U_1$, $U_2$, we suppress indexes. 
A simple calculation using only the \ceqref{sui} furnishes
\begin{equation}
s[Q,U]=[[Q,U],U].  %_1
\label{}
\end{equation}
The relations \ceqref{sui} and the conditions \ceqref{uicycle} are therefore compatible.
%The proof of the analogous statement for $U_2$ is formally identical. 
\end{proof}

\vspace{-.5mm}

The expressions \ceqref{sphi} and \ceqref{sui} and the conditions \ceqref{uicycle} together define the appropriate
BRST variation $s$ of $M(C_2,C_1)$ when the graded commutative algebras $C_1$, $C_2$
are differential. 

\ceqref{uicycle} entails that $U_1$, $U_2$ are $1$--cocycles of 
the derivation cochain complexes $(\Der(C_1),\ad Q_1)$, $(\Der(C_2),\ad Q_2)$, respectively
(cf. subsect. \cref{subsec:algdef}). Thus suggests that the cohomology of these %complexes 
should play a basic role in the analysis of the BRST symmetry of $M(C_2,C_1)$. 

Requiring $U_1$, $U_2$ to be cohomologically trivial yields a restricted variant of the BRST symmetry 
with special properties. If $U_1$, $U_2$ are $1$--coboundaries, then 
\begin{equation}
U_{\mathrm{res}1}=[Q_1,X_1], \qquad U_{\mathrm{res}2}=[Q_2,X_2]. 
%U_i[Q_i,X_i]
\label{uibndry}
\end{equation}
where $X_1\in\Der(C_1)$, $X_2\in\Der(C_2)$ are degree $0$ derivation variables. $X_1$, $X_2$
are however defined only modulo $0$--cocycles of $\Der(C_1)$, $\Der(C_2)$, respectively.
In order \ceqref{sui} to be satisfied, it is sufficient that
\begin{equation}
sX_1=\tfrac{1}{2}[X_1,[Q_1,X_1]],
\qquad sX_2=\tfrac{1}{2}[X_2,[Q_2,X_2]]
% \quad \text{mod $\ad Q_2$}. %sX_i=-\tfrac{1}{2}[[Q_i,X_i],X_i].
\label{sxi}
\end{equation}
modulo $1$--cocycles of $\Der(C_1)$, $\Der(C_2)$, respectively.

\vspace{-.5mm}

\begin{proof} Again, we suppress all indexes as they are 
inessential to the argument. Recalling \ceqref{sqi0} and using \ceqref{sxi}, we find 
\begin{align}
&s[Q,X]=-[Q,sX]
\vphantom{\Big]}
\label{sqx}
\\
&\hspace{2.5cm}=-\tfrac{1}{2}[Q,[X,[Q,X]]]=-\tfrac{1}{2}[[Q,X],[Q,X]], %=-[Q_1,X_1]^2,  
\vphantom{\Big]}
\nonumber
\end{align}
where we used the graded Jacobi identity \ceqref{grjac} and the nilpotence relation 
$[Q,[Q,-]]=0$. Substituting the relation \ceqref{uibndry} into \ceqref{sqx}, we recover the 
the relations \ceqref{sui}. 
%The analogous statement for $U_2$ and $X_2$ is proven similarly. %in the same way. 
\end{proof}

%\vspace{-.7mm}

%The fact that $X_1$ $X_2$ are defined only modulo $0$--cocycles of $\Der(C_1)$, $\Der(C_2)$
%entails that $sX_1$, $sX_2$ as given by \ceqref{sxi} are defined modulo $1$--coboundaries of 
%\Der(C_1)$, $\Der(C_2)$ and that the nilpotence conditions 

\noindent
As observed in \ccite{Kotov:2007nr}, in the left hand sides of the first relation \ceqref{sxi} there appears 
the degree $1$ derived brackets $[[Q_1,-],-]$ on the derivation space $\Der(C_1)$ associated 
with the differentials $Q_1$ of $C_1$ and similarly for the other relation. 

The BRST variations \ceqref{sui} are nilpotent in the appropriate sense, namely 
\begin{equation}
s^2X_1=0, \qquad s^2X_2=0
%sX_i=-\tfrac{1}{2}[[Q_i,X_i],X_i]
\label{sxi2eq0}
\end{equation}
modulo $2$--coboundaries of $\Der(C_1)$, $\Der(C_2)$, respectively.

%\vspace{-.7mm}

\begin{proof} Again, we suppress all indexes as they are inessential to the argument. 
From relations \ceqref{sui} and \ceqref{sxi}, written as $sX=[X,U_{\mathrm{res}}]/2$ using 
\ceqref{uibndry} for brevity, we have 
\begin{align}
&s^2X=\tfrac{1}{2}\big\{[sX,U_{\mathrm{res}}]+[X,sU_{\mathrm{res}}]\big\} \hspace{4cm}
\vphantom{\Big]}
\label{sxi60p2}
\\
&\hspace{2.5cm}=\tfrac{1}{4}\big\{[[X,U_{\mathrm{res}}],U_{\mathrm{res}}]-[X,[U_{\mathrm{res}},U_{\mathrm{res}}]]\big\}
=-\tfrac{1}{4}[[X,U_{\mathrm{res}}],U_{\mathrm{res}}],
\vphantom{\Big]}
\nonumber
\end{align}
where the graded Jacobi identity \ceqref{grjac} was used. From  \ceqref{uicycle} and \ceqref{uibndry}, we
also have 
\begin{align}
[Q,[[X,U_{\mathrm{res}}],X]]&=[[[Q,X],U_{\mathrm{res}}],X]-[[X,U_{\mathrm{res}}],[Q,X]]
\vphantom{\Big]}
\label{sxi60p3}
\\
&=[[U_{\mathrm{res}},U_{\mathrm{res}}],X]-[[X,U_{\mathrm{res}}],U_{\mathrm{res}}]]=-3[[X,U_{\mathrm{res}}],U_{\mathrm{res}}]. 
\vphantom{\Big]}
\nonumber
\end{align} %W_
From \ceqref{sxi60p2}, \ceqref{sxi60p3}, it follows that  
\begin{equation}
s^2X=\tfrac{1}{12}[Q,[[X,U_{\mathrm{res}}],X]].
\label{sxi60p4}
\end{equation}
In the above calculation, we did not take into account 
the indeterminacy of $X$ and $sX$ modulo $0$-- and $1$--cocycles of $\Der(C)$,
which may affect the expression found. Suppose that we shift $X$ and $sX$ 
by a $0$-- and a $1$--cocycles $\varDelta X$ and $\varDelta sX$, respectively. Then, by \ceqref{sui}, \ceqref{uicycle}
\ceqref{uibndry} and \ceqref{sxi60p2}, $s^2X$ is shifted by the amount 
\begin{align}
&\varDelta s^2X=\tfrac{1}{2}[\varDelta sX,[Q, X]]-\tfrac{1}{4}[\varDelta X,[Q,[X,U_{\mathrm{res}}]]]
\vphantom{\Big]}
\label{sxi60p1}
\\
&\hspace{2.5cm}=-\tfrac{1}{2}[Q,[\varDelta sX,X]]-\tfrac{1}{4}[Q,[\varDelta X,[X,U_{\mathrm{res}}]]],
\vphantom{\Big]}
\nonumber
\end{align}
where we used the cocycle relations $[Q,\varDelta X]=0$ and $[Q,\varDelta sX]=0$, 
the nilpotence relation $[Q,[Q,-]]=0$ and the graded Jacobi identity \ceqref{grjac}. 
By \ceqref{sxi60p4} and \ceqref{sxi60p1}, it follows that $s^2X$ 
vanishes modulo a $2$--coboundary of $\Der(C)$.
\end{proof}

%\vspace{-.75mm}

A restricted variation operation $s_{\mathrm{res}}$ is yielded from $s$ 
when $U_1$, $U_2$ are given by \ceqref{uibndry}.  From \ceqref{sphi}, the restricted variation  
$s_{\mathrm{res}}\varPhi$ takes the form 
\begin{equation}
s_{\mathrm{res}}\varPhi=
X_1F_\varPhi-F_\varPhi X_2-Q_1(X_1\varPhi-\varPhi X_2)+(X_1\varPhi-\varPhi X_2)Q_2,
\label{sphires}
\end{equation}
as follows form a straightforward calculation. Similarly, from \ceqref{sfphi},
the restricted variation $s_{\mathrm{res}}F_\varPhi$ reads as 
\begin{equation}
s_{\mathrm{res}}F_\varPhi=-Q_1(X_1F_\varPhi-F_\varPhi X_2)-(X_1F_\varPhi-F_\varPhi X_2)Q_2. 
\label{sfphires}
\end{equation}
Note that these variations are not affected by the indetermination of $X_1$, $X_2$ by $0$--cocycles of $\Der(C_1)$, $\Der(C_2)$,
since they are ultimately expressible in terms of $U_{\mathrm{res}1}$, $U_{\mathrm{res}2}$, which are not. 

We have thus two forms of BRST symmetry in $M(C_2,C_1)$, the {\it primary} and the
{\it restricted}. The primary BRST complex is the graded commutative algebra 
\begin{equation}
C^\infty(M(C_2,C_1))\otimes \Poly(\Der_1(C_1))\otimes \Poly(\Der_1(C_2))
\label{brstcmplx}
\end{equation}
with a degree $1$ derivative action of the primary BRST variation operator $s$.
The restricted  BRST complex is the graded commutative algebra 
\begin{equation}
C^\infty(M(C_2,C_1))\otimes \Poly\Big(\frac{\Der_0(C_1)}{Z\Der_0(C_1)}\Big)\otimes \Poly\Big(\frac{\Der_0(C_2)}{Z\Der_0(C_2)}\Big)
\label{resbrstcmplx}
\end{equation}
with a degree $1$ derivative action of the restricted BRST variation operator $s_{\mathrm{res}}$, where 
$Z\Der_0(C_1)$, $Z\Der_0(C_2)$ denote the vector spaces of $0$--cocycles of the derivation cochain complexes 
$(\Der(C_1),\ad Q_1)$, $(\Der(C_2),\ad Q_2)$, respectively.

A determination $s_D$ of $s$ is the degree $1$ variation operation resulting from 
an assignment of particular values $U_{1D}$, $U_{2D}$ to the degree $1$ derivation variables 
$U_1$, $U_2$. While $s$ is nilpotent, a generic determination $s_D$ of it is not in general,
because $U_{1D}$, $U_{2D}$ do not satisfy \ceqref{sui}. 

A restricted determination $s_{\mathrm{res}D}$ of $s_{\mathrm{res}}$ is similarly the degree $1$ 
restricted variation operation resulting from 
an assignment of particular values $X_{1D}$, $X_{2D}$ to the degree $0$ derivation variables 
$X_1$, $X_2$. Again, albeit $s_{\mathrm{res}}$ is nilpotent, a generic determination $s_{\mathrm{res}D}$ 
of it is not in general. 

An important case considered in the following is the determination $s_Q$ of $s$, which we shall call {\it canonical},  
specified by the value assignment 
\begin{equation}
U_{1Q}=Q_1,\qquad U_{2Q}=Q_2.
\label{uiq}
\end{equation}
By \ceqref{defect} and \ceqref{sphi}, $s_Q$ acts on $\varPhi$ as 
\begin{equation}
s_Q\varPhi=-F_\varPhi
\label{sphiq}
\end{equation} 
%to be compared with \ceqref{scurlyA}, 
while, by the defect identity \ceqref{bianchi} and \ceqref{sfphi}, $s_Q$ is inert on $F_\varPhi$, 
\begin{equation}
s_QF_\varPhi=0.
\label{sfphiq}
\end{equation}
By \ceqref{sfphiq} the determination $s_Q$ is actually nilpotent
\begin{equation}
s_Q{}^2\varPhi=-s_QF_\varPhi=0.
\label{sq20}
\end{equation}
The determination $s_Q$ equals secretly the restricted determination $s_{\mathrm{res}Q}$ 
specified by the value assignment 
\begin{equation}
X_{1Q}=-E_1,\qquad X_{2Q}=-E_2,
\label{}
\end{equation}
where $E_1$, $E_2$ are  the Euler derivations of $C_1$, $C_2$, respectively,
as follows readily from \ceqref{uibndry} recalling that $Q_1=[E_1,Q_1]$, $Q_2=[E_2,Q_2]$.

The primary and restricted BRST symmetries are distinct if the degree $1$ cohomology 
of at least one of the derivation cochain complexes $(\Der(C_1),\ad Q_1)$, $(\Der(C_2),\ad Q_2)$
is non trivial. In such a case, the natural question arises about their different interpretation 
and role in higher gauge theory. At the present level of abstractness, % of the analysis carried out here, 
it seems unlikely that such a question can be answered. 

The determination $s_Q$ of $s$ %and its action on the morphism variable $\varPhi$ 
given by \ceqref{sphiq} is to be compared with the familiar BRST transformations
of ordinary gauge theory described in subsect. \cref{subsec:brsttheory} 
and expressed by \ceqref{scurlyA} in BRST superfield form. The formal correspondence
is manifest. We have given in this way a graded geometrical foundation to the BRST 
formulation of higher gauge theory. 

%\vfil\eject

\subsection{\textcolor{blue}{\sffamily BRST symmetry and the differential cone}}\label{subsec:brstcone}

Let $C$ be a graded commutative algebra. A {\it differential structure} on $C$ is a choice of a 
differential $Q$ on $C$ making $C$ a differential graded commutative algebra (cf. subsect. 
\cref{subsec:algdef}). The differential structures of $C$ form a manifold $\mathsans{Q}_C$. 

$\mathsans{Q}_C$ is contained in the linear space $\Der_1(C)$ of degree $1$ derivation of $C$,
but it is not itself a linear space because of the quadratic nature of the nilpotence condition 
$Q^2=0$ (cf. eq. \ceqref{q20}). While nilpotence is broken by addition (if $Q,Q'$ are differentials on $C$ 
their sum $Q+Q'$ in general is not), it is preserved by rescaling (if $Q$ is a differential on $C$, so is $tQ$ 
for $t\in \mathbb{R}$). This shows that $\mathsans{Q}_C$ is \linebreak a 
cone contained in $\Der_1(C)$. We thus call $\mathsans{Q}_C$ the {\it differential cone} of $C$. 

Let $Q$ be a fixed differential in $\mathsans{Q}_C$. We want to deform $Q$ within $\mathsans{Q}_C$.
This involves adding to $Q$ a deformation $W$ such that 
\begin{equation}
Q_W=Q+W
\label{qdefrm}
\end{equation}
is another differential in $\mathsans{Q}_C$. In order this to be the case, $W$ must be a degree $1$ derivation 
of $C$, so that $Q_W$ also is, and must satisfy the {\it deformation equation}
\begin{equation}
[Q,W]+\tfrac{1}{2}[W,W]=0,
\label{defrm}
\end{equation}
so that $Q_W$ is nilpotent as required. 
For infinitesimal deformations, we can drop the term $\frac{1}{2}[W,W]$ and \ceqref{defrm} reduces to 
the infinitesimal deformation equation \pagebreak 
\begin{equation}
[Q,W]=0
\label{defrminf}
\end{equation}
stating that $W$ is a $1$--cocycle in the $\ad Q$ cohomology. 
\ceqref{defrminf} admits in particular the cohomologically trivial restricted solution 
\begin{equation}
W_{\mathrm{res}}=[Q,N], 
\label{wtriv}
\end{equation}
with $N$ a degree $0$ derivation of $C$. 

A special solution of the deformation equation \ceqref{defrm} as well as its 
infinitesimal form \ceqref{defrminf} is \hphantom{xxxxxxxxxxxxxxxxx}
\begin{equation}
W_Q=u Q, 
\label{wq}
\end{equation}
with $u\in\mathbb{R}$. This corresponds to a rescaling $Q\rightarrow (1+u)Q$ of $Q$. 
$W_Q$ is actually of the restricted form \ceqref{wtriv} with 
\begin{equation}
N_Q=-u E, 
\label{nq}
\end{equation}
$E$ being the Euler derivation of $C$. 

Next, suppose that $C_1$, $C_2$ are graded commutative algebras and that 
$\mathsans{Q}_{C_1}$, $\mathsans{Q}_{C_2}$ are their respective differential cones.
Pick reference differentials $Q_1$, $Q_2$ for $C_1$, $C_2$ and consider infinitesimal deformations
$W_1$, $W_2$ of $Q_1$, $Q_2$, so that $W_1$, $W_2$  satisfy eq. \ceqref{defrminf}. 
If $\varPhi:C_2\rightarrow C_1$ is a graded algebra morphism, $\varPhi$ is not affected by 
the deformations, but its defect $F_\varPhi$ is: \ceqref{defect} implies indeed that 
\begin{equation}
F_{\varPhi W}=F_\varPhi+W_1\varPhi-\varPhi W_2. 
\label{fphiw}
\end{equation}
Now, by \ceqref{sphi}, the deformation terms in the right hand side are just the variation $-s_W\varPhi$ 
of $\varPhi$ yielded by the determination $s_W$ of the BRST variation operator $s$ specified by
the value assignment 
\begin{equation}
U_{1W}=W_1,\qquad U_{2W}=W_2
\label{uiw}
\end{equation}
(cf. subsect. \cref{subsec:brst}). We can thus write \ceqref{fphiw} suggestively as 
\begin{equation}
F_{\varPhi W}=F_\varPhi-s_W\varPhi. 
\label{fphidefrm}
\end{equation}
In this way, there exists a one--to--one correspondence between infinitesimal deformations $W_1$, $W_2$ 
of the differential structures $Q_1$, $Q_2$ of $C_1$, $C_2$ and the determinations $s_W$ of the BRST variation operator
$s$ such that \ceqref{fphidefrm} holds. Further, by \ceqref{uibndry} and \ceqref{wtriv}, the correspondence 
maps restricted deformations $W_{\mathrm{res}1}$, $W_{\mathrm{res}2}$ to restricted determinations $s_{\mathrm{res}W}$. %%%@@@

When $W_1$, $W_2$ are of the special form \ceqref{wq},
\begin{equation}
W_{1Q}=u Q_1,\qquad  W_{2Q}=u Q_2,
\label{wqi}
\end{equation}
the variation $s_W\varPhi$ takes the form 
\begin{equation}
s_{W_Q}\varPhi=u s_Q\varPhi,
\label{}
\end{equation}
where $s_Q$ is the canonical determination of $s$ (cf. eq. \ceqref{sphiq}).  
A concurrent rescaling $Q_1\rightarrow (1+u)Q_1$, $Q_2\rightarrow (1+u)Q_2$ of the %two 
differentials $Q_1$, $Q_2$ therefore corresponds essentially to $s_Q$
under the aforementioned one--to--one correspondence. 

%This suggests an interpretation of the variation $s_Q\varPhi$: $-s_Q\varPhi=F_\varPhi$ is the rate of change
%of $F_\varPhi$ under a concurrent rescaling $Q_1\rightarrow (1+u)Q_1$, $Q_2\rightarrow (1+u)Q_2$ of the %two 
%differentials $Q_1$, $Q_2$. 

In higher gauge theory, the differentials $Q_1$, $Q_2$ of the differential graded commutative 
algebras $C_1$ $C_2$ yield the kinetic and the self-interaction terms of the higher gauge fields,
respectively. They constitute the basic data upon which the theory rests. 
This suggests that the theory space consists of the Cartesian product $\mathsans{Q}_{C_1}\times\mathsans{Q}_{C_2}$
of the differential cones of $C_1$, $C_2$ or, equivalently, the differential cone $\mathsans{Q}_{C_1\,\oplus \,C_2}$ 
of the differential graded commutative algebra direct sum $C_1\oplus C_2$ of $C_1$, $C_2$. This claim however is not
completely correct as %for reasons which 
we explain next. 

The relative normalization of 
$Q_1$, $Q_2$ determines the value of the higher gauge fields' self-interaction coupling strength and is 
thus physically relevant. The overall normalization of $Q_1$, $Q_2$ is instead 
physically irrelevant, as a change of its value amounts to a change of the overall
normalization of the Lagrangian. For this reason, the theory space %can be identified 
is not quite the differential cone $\mathsans{Q}_{C_1\,\oplus \,C_2}$ but the projective cone 
$\mathbb{P}\mathsans{Q}_{C_1\,\oplus \,C_2}=\mathsans{Q}_{C_1}\times\mathsans{Q}_{C_2}/\mathbb{R}^\times$, 
where $\mathbb{R}^\times$ is the multiplicative 
group of non zero reals acting on $\mathsans{Q}_{C_1}\times\mathsans{Q}_{C_2}$ by simultaneous rescaling. 

Therefore, deformations $W_1$, $W_2$ of the differential structures $Q_1$, $Q_2$ of $C_1$, $C_2$ 
modulo those of the form \ceqref{wqi} yield infinitesimal translations in the theory space 
$\mathbb{P}\mathsans{Q}_{C_1\,\oplus \,C_2}$. The ineffective deformations \ceqref{wqi} are precisely 
those which correspond to the canonical determination $s_Q$ of $s$.

There are two archetypal higher gauge theories whose gauge symmetry is described by $s_Q$:
higher BF gauge theory and higher Chern--Simons gauge theory. These are studied next after a 
brief review of classical BV theory.

%\vfil\eject 

\subsection{\textcolor{blue}{\sffamily Classical BV theory}}\label{subsec:bvalg}

In this subsection, we recall briefly the basic definitions of formal BV theory used 
in the following. See ref. \ccite{Gomis:1994he} for a physically motivated introduction. 

Let $A$ be a graded commutative algebra and $n\in\mathbb{Z}$.
A degree $n$ {\it Poisson--Gerstenhaber structure} on $A$ 
is a bilinear bracket $(-,-):A\times A\rightarrow A$ with the following properties. 
$|(u,v)|=|u|+|v|+n$ for homogeneous $u,v\in A$. Further, 
\begin{align}
&(u,v)+(-1)^{(|u|+n)(|v|+n)}(v,u)=0, \hspace{4cm}
\vphantom{\Big]}
\label{pgh1}
\\
&(-1)^{(|w|+n)(|u|+n)}(u,(v,w))+(-1)^{(|u|+n)(|v|+n)}(v,(w,u))
\vphantom{\Big)}
\label{pgh2}
\\
&\hspace{5.9cm}+(-1)^{(|v|+n)(|w|+n)}(w,(u,v))=0,
\vphantom{\Big]}
\nonumber
\\
&(u,vw)=(u,v)w+(-1)^{(|u|+n)|v|}v(u,w)
\vphantom{\Big]}
\label{pgh3}
\end{align}  
for homogeneous $u,v,w\in A$. A graded commutative algebra $A$ equipped with a degree $n$  
Poisson--Gerstenhaber structure $(-,-)$ is called a degree $n$ Poisson--Gerstenhaber algebra. 
$(-,-)$ is also called {\it Poisson--Gerstenhaber brackets}. 
One usually employs the term Poisson for $n$ even and Gerstenhaber for $n$ odd. 

Let $A$ be a graded commutative algebra and $n\in\mathbb{Z}$ as before.  
A degree $n$ {\it classical Batalin-Vilkovisky or BV structure} on $A$ consists
of a degree $n$ Poisson--Gerstenhaber structure $(-,-)$ on $A$ and
a distinguished degree $1-n$ element $S\in A$ satisfying the condition \hphantom{xxxxxxxx}
\begin{equation}
(S,S)=0. \vphantom{\bigg]}
\label{clmaster}
\end{equation}
A graded commutative algebra $A$ equipped with a degree $n$ classical
BV structure $((-,-),S)$ is called a degree $n$ {\it classical BV algebra}. 
$S$ is called {\it classical BV master action} and \ceqref{clmaster} 
{\it classical BV master equation}. \vspace{-3mm} \eject

If $A$ is a degree $n$ classical BV algebra, the operator $\delta:A\rightarrow A$ defined by 
\begin{equation}
\delta u=(S,u)
\label{deltabv}
\end{equation}
has degree $1$ and is nilpotent, so that  \hphantom{xxxxxxxxx}
\begin{equation}
\delta^2=0.
\label{delta20}
\end{equation}
$\delta$ is called {\it classical BV differential}. 
A classical BV algebra is in this way a differential graded commutative algebra
whose differential $\delta$ is Hamiltonian with respect the underlying Poisson--Gerstenhaber 
structure with Hamiltonian element $S$. As a cochain complex, $(A,\delta)$ is called the {\it classical BV complex} of $A$. 
Its cohomology is the {\it classical BV cohomology} of $A$. 

\vspace{.44mm}

In applications of BV theory to field theoretic models, $A$ is as a rule a graded commutative algebra
$A_{\mathscr{F}}$ of field functionals on a field space $\mathscr{F}$ endowed with a degree $-1$ 
symplectic form $\omega$, called {\it BV form}. Just as in ordinary Hamiltonian mechanics 
the canonical symplectic form of phase space yields the canonical Poisson brackets, 
$\omega$ yields a degree $1$ Gerstenhaber structure $(-,-)_\omega$ on $A_{\mathscr{F}}$, 
called {\it BV antibrackets}, rendering in this way $A_{\mathscr{F}}$ a degree $1$ Gerstenhaber algebra. 
Further, the data which define $\mathscr{F}$ and $\omega$ contain also elements allowing for the 
construction of a natural BV master action $S_{\mathscr{F}\omega}$ and the associated 
BV differential $\delta_{\mathscr{F}\omega}$, turning $A_{\mathscr{F}}$ into a degree $1$ 
classical BV algebra. However, the datum of the symplectic form 
$\omega$ is not always strictly necessary. In certain cases it is possible to define
a degree $1$ Gerstenhaber structure on a algebra $A_{\mathscr{F}}$ of field functionals 
by directly and consistently assigning the value of the brackets 
$(z_i,z_j)$ for a set of basic field functionals $z_i$ of $A_{\mathscr{F}}$ in terms of which all
other field functionals can be expressed compatibly with the properties \ceqref{pgh1}, \ceqref{pgh2}. 
Further, it is possible to 
construct a master action $S_{\mathscr{F}}$ as a distinguished function of the $z_i$ and obtain 
the associated differential $\delta_{\mathscr{F}}$. A BV algebra of field functionals is so obtained again. 
The AKSZ formulation of the classical BV theory of ref. \ccite{Alexandrov:1995kv,Ikeda:2012pv}
provides an elegant geometrical implementation of this program. 

\vspace{.44mm}

The field functionals $f$ composing the graded algebra $A_{\mathscr{F}}$ are characterized by ghost degree and so 
can be viewed as maps $f:\mathscr{F}\rightarrow G_{\mathbb{R}}$, where $G_{\mathbb{R}}$
is the graded commutative algebra 
\begin{equation}
G_{\mathbb{R}}=\ddd_{k=-\infty}^\infty\mathbb{R}[k]
\label{gradreal}
\end{equation}
(cf. eq. \ceqref{ghostreal}), called {\it ghost algebra}. In general, however, not all such maps $f$ belong to 
$A_{\mathscr{F}}$. Restrictions on the content of $A_{\mathscr{F}}$ may be required. 

$L_\infty$ algebras are extensions of ordinary Lie algebras. An $L_\infty$--algebra is a graded vector space $L$ 
equipped with a collection of multiple argument brackets satisfying 
generalized Jacobi identities. The $L_\infty$--algebra structure is encoded in a degree $1$ nilpotent 
Chevalley--Eilenberg differential $Q_L$ acting on the $1$--shift $L[1]$ of $L$.  
A classical BV algebra $A_{\mathscr{F}}$ of functionals on a field space $\mathscr{F}$
of a field theory with BV differential $\delta_{\mathscr{F}}$ therefore secretly supports 
an infinite dimmensional $L_\infty$--algebra structure on the $-1$--shift $A_{\mathscr{F}}[-1]$ of $A_{\mathscr{F}}$.
This describes the full symmetry of the field theory at the most basic level. 
%We shall consider again $L_\infty$ algebras and algebroids in subsect. \cref{subsec:linf} below
%from another perspective. 

%\vfil\eject +++++

\subsection{\textcolor{blue}{\sffamily Higher BF gauge theory}}\label{subsec:bv}

The first basic higher gauge theoretic model whose symmetry is described by 
the canonical determination $s_Q$ of the BRST variation operation $s$ is higher BF theo\-ry.
Below, we present an abstract BV formulation of the model based on the %formal 
BRST framework worked out in subsects. \cref{subsec:brst}, \cref{subsec:brstcone}.

Consider the $-1$--shifted cotangent bundle $T^*[-1]M(C_2,C_1)$ of the non differential morphism manifold 
$M(C_2,C_1)$ of two differential graded commutative algebras $C_1$, $C_2$. Then, $T^*[-1]M(C_2,C_1)$ is equipped with the
canonical degree $-1$ BV symplectic form $\varOmega_{BV}$
\begin{equation}
\varOmega_{BV}=\langle\delta\varPhi^*,\delta\varPhi\rangle, \vphantom{\bigg]}
\label{omegabvbf}
\end{equation}
where $\varPhi\in M(C_2,C_1)$  and $\varPhi^*\in T^*{}_\varPhi[-1]M(C_2,C_1)$ are %denote 
the base and fiber variables of $T^*[-1]M(C_2,C_1)$ and 
$\langle-,-\rangle$ stands for the natural cotangent--tangent pairing. 
With $\varOmega_{BV}$, in turn, there are associated the canonical BV antibrackets. These can be written compactly as 
\begin{equation}
(\langle A^*,\varPhi\rangle, \langle\varPhi^*,B\rangle)_{BV}=\langle A^*,B\rangle, 
\label{bvbra}
\end{equation}
where $A^*\in T^*{}_\varPhi M(C_2,C_1)$, $B\in T_\varPhi M(C_2,C_1)$ 
and the pairing with $A^*$ and $B$ is tacitly assumed to follow the BV antibracketing.

We can extend the canonical determination $s_Q$ of the BRST variation operation $s$ from $M(C_2,C_1)$
to $T^*[-1]M(C_2,C_1)$ by setting 
\begin{align}
&s_Q\varPhi=-F_\varPhi,
\vphantom{\Big]}
\label{sphiqcot}
\\
&s_Q\varPhi^*=\dot F^\vee{}_{\varPhi^*},
\vphantom{\Big]}
\label{sphi*qcot}
\end{align}
where $\dot F^\vee{}_{\varPhi^*}$ is defined by the relation
\begin{equation}
\langle \dot F^\vee{}_{\varPhi^*}, B\rangle=-\langle \varPhi^*,\dot F_B\rangle
\label{veedef}
\end{equation}
with $B\in T_\varPhi M(C_2,C_1)$ and $\dot F_B=Q_1B-BQ_2$. Since \ceqref{sphiqcot} is just 
\ceqref{sphiq}, \ceqref{sphiqcot}, \ceqref{sphi*qcot} do indeed 
define an augmentation of $s_Q$. The extension keeps the property of nilpotence. 

\begin{proof} We have $s_Q{}^2\varPhi=0$ by \ceqref{sq20}. By iterating \ceqref{sphi*qcot} twice, we get 
\begin{equation}
s_Q{}^2\langle\varPhi^*,B\rangle=s_Q\langle\dot F^\vee{}_{\varPhi^*},B\rangle=-s_Q\langle\varPhi^*,\dot F_B\rangle
=-\langle\dot F^\vee{}_{\varPhi^*},\dot F_B\rangle=\langle\varPhi^*,\dot F_{\dot F_B}\rangle.
\label{}
\end{equation}
It is now immediately verified that $\dot F_{\dot F_B}$ vanishes,
\begin{equation}
\dot F_{\dot F_B}=Q_1(Q_1B-BQ_2)+(Q_1B-BQ_2)Q_2=0.
\label{}
\end{equation}
by the nilpotence of $Q_1$, $Q_2$. So, we have $s_Q{}^2\varPhi^*=0$ as well.
\end{proof}

\noindent
The action of the extended canonical determination  $s_Q$ on $M(C_2,C_1)$ is Hamiltonian 
with degree $0$ BV master action 
\begin{equation}
S_{BV}=-\langle\varPhi^*,F_\varPhi\rangle.
\label{sbfbv}
\end{equation}
Indeed, $S_{BV}$ satisfies the classical master equation  %\hphantom{xxxxxxxxxxxxxxx}
\begin{equation}
(S_{BV},S_{BV})_{BV}=0.
\label{masteq}
\end{equation}
Further, a straightforward cal\-culation shows that 
\begin{align}
&(S_{BV},\langle A^*,\varPhi\rangle)_{BV}=-\langle A^*, F_\varPhi\rangle=\langle A^*, s_Q\varPhi\rangle,
\vphantom{\Big]}
\label{sphiqbv}
%\\
\end{align}
\vspace{-1cm}\eject\noindent
\begin{align}
&(S_{BV},\langle\varPhi^*,B\rangle)_{BV}=\langle \dot F^\vee{}_{\varPhi^*}, B\rangle%=-\langle\varPhi^*,\dot F_B\rangle
=\langle s_Q\varPhi^*, B\rangle,
\vphantom{\Big]}
\label{sphi*qbv}
\end{align}
where $A^*\in T^*{}_\varPhi[n]M(C_2,C_1)$, $B\in T_\varPhi[0]M(C_2,C_1)$ 

\begin{proof} We show only \ceqref{masteq}. Using \ceqref{bvbra}, we have 
\begin{align}
&(S_{BV},S_{BV})_{BV}
\vphantom{\Big]}
\label{}
\\
&\qquad\quad=(\langle\varPhi^*,F_\varPhi\rangle,\langle\varPhi^*,B\rangle)_{BV}|_{B=F_\varPhi}
+(\langle A^*,F_\varPhi\rangle,\langle\varPhi^*,F_\varPhi\rangle)_{BV}|_{A^*=\varPhi^*}
\vphantom{\Big]}
\nonumber
\\
&\qquad\quad=\langle\varPhi^*,\dot F_B\rangle|_{B=F_\varPhi}
+\langle A^*,\dot F_{F_\varPhi}\rangle|_{A^*=\varPhi^*}
\vphantom{\Big]}
\nonumber
\\
&\qquad\quad=2\langle\varPhi^*,\dot F_{F_\varPhi}\rangle
\vphantom{\Big]}
\nonumber
\end{align}
Here, $\dot F_{F_\varPhi}$ vanishes by the defect identity \ceqref{bianchi}, 
\begin{equation}
\dot F_{F_\varPhi}=Q_1F_\varPhi+F_\varPhi Q_2=0. 
\vphantom{\Big]}
\label{}
\end{equation}
\ceqref{masteq} is thus shown. 
\end{proof}

\vspace{-.2mm}\noindent
On account of  \ceqref{sphiqbv}, \ceqref{sphi*qbv}, 
we shall also denote the variation operation $s_Q$ by $\delta_{BV}$. 

It should be now apparent that our analysis does indeed provide an abstract higher gauge 
theoretic generalization of BF gauge theory, as we anticipated above. The BV master action 
$S_{BV}$ given by \ceqref{sbfbv} has the standard form of that of BF gauge theory.
Furthermore, the Euler--Lagrange equation ensuing from $S_{BV}$, which in the BV framework 
can be formally expressed as the vanishing of the BV variations $\delta_{BV}\varPhi$, $\delta_{BV}\varPhi^*$ 
of $\varPhi$, $\varPhi^*$, have too by \ceqref{sphiqbv}, \ceqref{sphi*qbv} the standard form of 
those of BF gauge theory. We have worked out in this way a general formal framework suitable 
for the construction of BF type higher gauge theoretic models. %n-1

%\vfil\eject

\subsection{\textcolor{blue}{\sffamily %AKSZ formulation of h
Higher Chern--Simons gauge theory}}\label{subsec:sigmamod}

The second basic higher gauge theoretic model whose symmetry is described by 
the canonical determination $s_Q$ of the BRST 
variation operation $s$ is higher Chern--Simons theory. \pagebreak 
We present now an abstract formulation of this model 
based again on the BRST framework of subsects. \cref{subsec:brst}, \cref{subsec:brstcone}.
Our approach follows cloesely the AKSZ paradigm of ref. \ccite{Alexandrov:1995kv}

Consider once more two differential graded commutative algebras $C_1$, $C_2$ together with the manifold 
$M(C_2,C_1)$ of their non differential morphisms. The formulation rests on a 
set of assumptions which we state next. 
\begin{enumerate}

\item A degree $-n$ $G_{\mathbb{R}}$--linear map $\mu:C_1\rightarrow G_{\mathbb{R}}$
\footnote{$\vphantom{\dot{\dot{\dot{x}}}}$ 
$G_{\mathbb{R}}$--linearity is conventionally defined as the property that 
$\mu(\theta u)=(-1)^{n|\theta|}\theta\mu(u)$ and $\mu(u\theta)=\mu(u)\theta$ 
with $u\in C_1$ and $\theta\in G_{\mathbb{R}}$.}, 
where $n\in \mathbb{N}$, $n>0$ and $G_{\mathbb{R}}$ is the ghost algebra defined
in \ceqref{ghostreal}, with the following properties is given. 
$\mu$ is non singular, that is 
\begin{equation}
\mu(uv)=0 \quad\text{for all $v\in C_1$}\Rightarrow u=0 \vphantom{\bigg]}
\label{nat}
\end{equation}
for any $u\in C_1$. Furthermore, the kernel of $\mu$ contains the range of $Q_1$, 
\begin{equation}
\mu\circ Q_1=0.  \vphantom{\bigg]}
\label{stokes}
\end{equation}
\end{enumerate}
To get an intuitive understanding of these conditions, think $C_1$ as the algebra of 
differential forms on a $n$--dimensional manifold without boundary and $Q_1$ as the de 
Rham differential. Then, $\mu$ can be regarded as the integration operation and \ceqref{stokes} 
as the statement of Stokes' theorem. 
\begin{enumerate}[resume]

\item A degree $-n+1$ classical BV structure on $C_2$ with Poisson--Gerstenhaber 
brackets $(-,-)_2$ and  classical BV master action $S_2$ 
(cf. subsect. \cref{subsec:bvalg}, eqs. \ceqref{pgh1}--\ceqref{pgh3} and \ceqref{clmaster})
such that 
\begin{equation}
Q_2f=(S_2,f)_2 \vphantom{\bigg]}
\label{hams2bv}
\end{equation}
for $f\in C_2$ is given (cf. eq. \ceqref{deltabv}). 

\end{enumerate}
Again, for the sake of intuition, one can think of $C_2$ as a graded commutative algebra of functions 
on a graded phase space with graded Poisson brackets $(-,-)_2$ and a distinguished degree $n$ 
function $S_2$ satisfying $(S_2,S_2)_2=0$ which is 
the Hamiltonian for a homological vector field $Q_2=(S_2,-)$. 

The AKSZ formulation of higher Chern--Simons gauge theory involves further data, which
are required for the construction of the classical BV master action. 
\vspace{-.25mm}
\begin{enumerate}[resume]

\item There are a degree $n$ map $K_1:M(C_2,C_1)\rightarrow C_1$ 
and a degree $n-1$ fiberwise linear map $B_1:TM(C_2,C_1)\rightarrow C_1$ 
enjoying the following properties. %For every $\varPhi\in M(C_2,C_1)$, one has $|K_1(\varPhi)|=n$. 
The left tangent map $TK_1:TM(C_2,C_1)\rightarrow C_1$ of $K_1$ is compatible with left multiplication in $C_1$ 
in the sense that 
\begin{equation}
TK_1(\varPhi)(u\dot\varPhi)-Q_1B_1(\varPhi)(u\dot\varPhi)=u(TK_1(\varPhi)(\dot\varPhi)-Q_1B_1(\varPhi)(\dot\varPhi))
\label{s1local}
\end{equation}
for $\varPhi\in M(C_2,C_1)$, $\dot\varPhi\in \ddd_{k=-\infty}^\infty T_\varPhi[k]M(C_2,C_1)$ and $u\in C_1$. Furthermore, $K_1$ obeys 
\begin{equation}
TK_1(\varPhi)(\varPhi(f,-)_2)-Q_1B_1(\varPhi)(\varPhi(f,-)_2)=(-1)^{|f|-n+1}Q_1\varPhi f 
\label{s1varpois}
\end{equation}
for $\varPhi\in M(C_2,C_1)$ and homogeneous $f\in C_2$. 
\end{enumerate} \vspace{-.25mm}
%In our AKSZ formulation of higher Chern--Simons gauge theory, 
As we shall see, $\mu(K_1(\varPhi))$ constitutes the `kinetic' 
term of the higher gauge fields in the BV action. 
$Q_1B_1(\varPhi)(\dot\varPhi)$ is an exact term that is produced when $K_1(\varPhi)$ is varied 
and that vanishes upon integration. \ceqref{s1local} is a minimal condition 
that $K_1$ must satisfy in a local Lagrangian field theory. 
\ceqref{s1varpois} is a restriction on the general form of $K_1$ that ensures that
the action has the required properties. %will be shown momentarily. 
%In our AKSZ formulation of higher Chern--Simons gauge theory, 
The BV action's `self--interaction' term of the higher gauge fields turns out to be $\mu(\varPhi S_2)$
and so does not involve new data. 

With the above data, it is possible to define degree $1$ Poisson--Gerstenhaber brackets
on the graded commutative algebra $C^\infty(M(C_2,C_1))\otimes G_{\mathbb{R}}$ of 
$G_{\mathbb{R}}$--valued functions on the morphism manifold $M(C_2,C_1)$ by setting
\begin{equation}
(\mu(u\,\varPhi f),\mu(v\,\varPhi g))_{BV}=(-1)^{(|f|-n+1)|v|}\mu(uv\,\varPhi(f,g)_2) %\vphantom{\bigg]}
\label{bvmc2c1}
\end{equation}
%\vspace{-9mm}\eject \noindent
with homogeneous $u,v\in C_1$, $f,g\in C_2$. 

\vspace{-.25mm}
\begin{proof}
To begin with, we notice that by the non singularity of $\mu$, eq. \ceqref{nat}, 
any function in $C^\infty(M(C_2,C_1))\otimes G_{\mathbb{R}}$ can be expressed in terms of 
the basic functions of $\mu(u\varPhi f)$ with $u\in C_1$, $f\in C_2$. 
To show that the brackets \ceqref{bvmc2c1} define a degree $1$ Gerstenhaber
structure on $C^\infty(M(C_2,C_1))\otimes G_{\mathbb{R}}$ it is sufficient to show that 
they have degree $1$ and enjoy the properties \ceqref{pgh1}, \ceqref{pgh2}. 

Since $|(f,g)_2|=-n+1+|f|+|g|$ for $f,g\in C_2$ and $|\mu(u\,\varPhi f)|=-n+|u|+|f|$ 
for $u\in C_1$, $f\in C_2$, one has %\pagebreak
\begin{equation}
|\mu(uv\,\varPhi(f,g)_2)|=1+|\mu(u\,\varPhi f)|+|\mu(v\,\varPhi g)|
\label{bvmc2c1a1}
\end{equation}
for $u,v\in C_1$, $f,g\in C_2$. The brackets \ceqref{bvmc2c1} have therefore degree $1$.
By straightforward calculations, one verifies that 
\begin{align}
&(\mu(u\,\varPhi f),\mu(v\,\varPhi g))_{BV}
\vphantom{\Big]}
\label{bvmc2c1a2}
\\
&\hspace{3cm}
+(-1)^{(-n+|u|+|f|+1)(-n+|v|+|g|+1)}(\mu(v\,\varPhi g),\mu(u\,\varPhi f))_{BV}=0,
\vphantom{\Big]}
\nonumber
\\
&(-1)^{(-n+|w|+|h|+1)(-n+|u|+|f|+1)}(\mu(u\,\varPhi f),(\mu(v\,\varPhi g),\mu(w\,\varPhi h)))_{BV}
\vphantom{\Big)_2}
\label{bvmc2c1a3}
\\
&\hspace{.49cm}+(-1)^{(-n+|u|+|f|+1)(-n+|v|+|g|+1)}(\mu(v\,\varPhi g),(\mu(w\,\varPhi h),\mu(u\,\varPhi f)))_{BV}
\vphantom{\Big]}
\nonumber
\\
&\hspace{.98cm}+(-1)^{(-n+|v|+|g|+1)(-n+|w|+|h|+1)}(\mu(w\,\varPhi h),(\mu(u\,\varPhi f),\mu(v\,\varPhi g)))_{BV}=0,
\vphantom{\Big]}
\nonumber
\end{align}  
for $u,v,w\in C_1$, $f,g,h\in C_2$, showing that the brackets \ceqref{bvmc2c1} enjoy the 
properties \ceqref{pgh1}, \ceqref{pgh2}. The statement follows. 
\end{proof}

\noindent
Hence, $C^\infty(M(C_2,C_1))\otimes G_{\mathbb{R}}$ is equipped with degree $1$ Gerstenhaber brackets
$(-,-)_{BV}$ induced naturally by the Poisson-Gerstenhaber brackets of $C_2$. These are in fact the BV brackets in our AKSZ 
formulation of higher Chern--Simons gauge theory.  %, as suggested by the notation. 

Using $K_1$ and $S_2$ as basic building blocks, 
we can construct a degree $0$ Hamiltonian $S_{BV}$ for the nilpotent degree $1$ defect vector field
of $M(C_2,C_1)$ (cf. subsect. \cref{subsec:geodef}),  here expressed as the map 
$\mu(u\,\varPhi f)\rightarrow \mu(u\,F_\varPhi f)$ with $u\in C_1$, $f\in C_2$: 
\begin{equation}
S_{BV}(\varPhi)=(-1)^n\mu(K_1(\varPhi)+\varPhi S_2). %\vphantom{\bigg]}
\label{sbvs1pluss2}
\end{equation}
Indeed, $S_{BV}$ satisfies the master equation 
\begin{equation}
(S_{BV}(\varPhi),S_{BV}(\varPhi))_{BV}=0 %\vphantom{\bigg]}
\label{mastersbv}
\end{equation}
and moreover one has \pagebreak 
\begin{equation}
\mu(u\,F_\varPhi f)=-(-1)^{-n+|u|}(S_{BV}(\varPhi),\mu(u\,\varPhi f))_{BV} \vphantom{\bigg]}
\label{hamiltsbv}
\end{equation}
for $u\in C_1$, $f\in C_2$. 

\begin{proof} 
By \ceqref{nat}, \ceqref{bvmc2c1}, for  $u\in C_1$, $f\in C_2$ one has 
\begin{equation}
(\mu(u\,\varPhi f),\varPhi)_{BV}=(-1)^{n(|u|+|f|)} u\,\varPhi(f,-)_2.
\label{hamilts0}
\end{equation}

Exploiting \ceqref{stokes}, \ceqref{s1local}, \ceqref{s1varpois} and \ceqref{hamilts0}, we find
\begin{align}
(\mu(K_1(&\varPhi)),\mu(u\,\varPhi f))_{BV}
\vphantom{\Big]}
\label{hamilts1}
\\
&=(-1)^{|u|+|f|-n}(\mu(u\,\varPhi f),\mu(K_1(\varPhi)))_{BV}
\vphantom{\Big]}
\nonumber
\\
&=(-1)^{(n+1)|u|+|f|-n}
\mu(TK_1(\varPhi)((\mu(u\,\varPhi f),\varPhi)_{BV}))
\vphantom{\Big]}
\nonumber
\\
&=(-1)^{|u|+|f|-n}\mu(TK_1(\varPhi)(u\,\varPhi(f,-)_2))
\vphantom{\Big]}
\nonumber
\\
&=(-1)^{|u|+|f|-n}\mu(Q_1B_1(\varPhi)(u\,\varPhi(f,-)_2)
\vphantom{\Big]}
\nonumber
\\
&\hspace{3cm}
+u(TK_1(\varPhi)(\varPhi(f,-)_2)-Q_1B_1(\varPhi)(\varPhi(f,-)_2)))
\vphantom{\Big]}
\nonumber
\\
%&=(-1)^{|u|+|f|-n}\mu(uTK_1(\varPhi)(\varPhi(f,-)_2))
%\vphantom{\Big]}
%\nonumber
%\\
&=-(-1)^{|u|}\mu(u\,Q_1\varPhi f), 
\vphantom{\Big]}
\nonumber
\end{align}
%This shows that 
%\begin{equation}
%(\mu(K_1(\varPhi)),\mu(u\,\varPhi f))_{BV}=-(-1)^{|u|}\mu(u\,Q_1\varPhi f), 
%\label{}
%\end{equation} 
%\ceqref{hamilts1} is a first basic relation. By \ceqref{hamilts1}, we have 
from which on account of \ceqref{nat} it follows readily that 
\begin{equation}
(\mu(K_1(\varPhi)),\varPhi)_{BV}=(-1)^{n-1}Q_1\varPhi.
\label{hamilts2}
\end{equation} 
Using \ceqref{hamilts2} together with \ceqref{stokes}, we obtain 
\begin{align}
(\mu(K_1(\varPhi)),\mu(K_1(\varPhi)))_{BV}&=(-1)^n\mu(TK_1(\varPhi)((\mu(K_1(\varPhi)),\varPhi)_{BV}))
\vphantom{\Big]}
\label{masters1}
\\
&=-\mu(TK_1(\varPhi)(Q_1\varPhi))=-\mu(Q_1K_1(\varPhi))=0.
\vphantom{\Big]}
\nonumber
\end{align}
\ceqref{masters1} is a first basic relation. 
%This shows that 
%\begin{equation}
%(\mu(K_1(\varPhi)),\mu(K_1(\varPhi)))_{BV}=0.
%\label{masters1}
%\end{equation}
Employing \ceqref{hamilts1} and \ceqref{stokes}, one has further
\begin{equation}
(\mu(K_1(\varPhi)),\mu(\varPhi S_2))_{BV}=-\mu(Q_1\varPhi S_2)=0.
\label{masters2}
\end{equation}
\ceqref{masters2} is a second basic relation. 
By \ceqref{bvmc2c1} and \ceqref{hams2bv}, one has 
\begin{equation}
(\mu(\varPhi S_2),\mu(u\,\varPhi f))_{BV}
=(-1)^{|u|}\mu(u\varPhi(S_2,f)_2)=(-1)^{|u|}\mu(u\varPhi Q_2f)
\label{hamiltsbvb3}
\end{equation}
for $u\in C_1$, $f\in C_2$. By \ceqref{clmaster} and \ceqref{hamiltsbvb3}, we have then 
\begin{equation}
(\mu(\varPhi S_2),\mu(\varPhi S_2))_{BV}=\mu(\varPhi(S_2,S_2)_2)=0.
\label{hamiltsbvb3/1}
\end{equation}
\ceqref{hamiltsbvb3/1} is a third basic relation. 

On account of \ceqref{sbvs1pluss2}, using \ceqref{masters2}, \ceqref{masters2} and \ceqref{hamiltsbvb3/1}, we find
\begin{align}
&(S_{BV}(\varPhi),S_{BV}(\varPhi))_{BV}=(\mu(K_1(\varPhi)),\mu(K_1(\varPhi)))_{BV} \hspace{2cm}
\vphantom{\Big]}
\label{mastersbvb1}
%\\
\end{align}
\begin{align}
&\hspace{2cm}+2(\mu(K_1(\varPhi)),\mu(\varPhi S_2))_{BV}+(\mu(\varPhi S_2),\mu(\varPhi S_2))_{BV}=0.
\vphantom{\Big]}
\nonumber
\end{align}
\ceqref{mastersbv} is thus proven. %\eject

Finally, by \ceqref{defect}, \ceqref{sbvs1pluss2}, \ceqref{hamilts1} and \ceqref{hamiltsbvb3}, we have 
\begin{align}
&-(-1)^{-n+|u|}(S_{BV}(\varPhi),\mu(u\,\varPhi f))_{BV}
\vphantom{\Big]}
\label{hamiltsbvb1}
\\
&\hspace{1cm}=-(-1)^{|u|}\big[(\mu(K_1(\varPhi)),\mu(u\,\varPhi f))_{BV}
+(\mu(\varPhi S_2),\mu(u\,\varPhi f))_{BV}\big].
\vphantom{\Big]}
\nonumber
\\
&\hspace{2cm}=\mu(u\,Q_1\varPhi f)-\mu(u\,\varPhi Q_2 f)=\mu(u\,F_\varPhi f).
\vphantom{\Big]}
\nonumber
\end{align}
\ceqref{hamiltsbv} is thus shown. 
\end{proof}

By virtue of \ceqref{hamiltsbv}, $S_{BV}$ is the Hamiltonian function for the canonical 
determination $s_Q$ of the BRST variation $s$. Indeed, 
\begin{equation}
s_Q\mu(u\,\varPhi f)=(S_{BV},\mu(u\,\varPhi f))_{BV}
\label{sqtaubvtau}
\end{equation}
for $u\in C_1$, $f\in C_2$. 

\begin{proof} One has 
\begin{equation}
s_Q\mu(u\,\varPhi f)=(-1)^{-n+|u|}\mu(u\,s_Q\varPhi f)=-(-1)^{-n+|u|}\mu(u\,F_\varPhi f). 
\label{}
\end{equation}
The statement follows then immediately from \ceqref{hamiltsbv}.
\end{proof}

\noindent
For this reason, the variation operation $s_Q$ is just the BV differential  $\delta_{BV}$. 

It should be now apparent that our analysis does indeed provide an abstract higher gauge 
theoretic generalization of Chern--Simons gauge theory, as we anticipated. The BV action 
$S_{BV}$ given by \ceqref{sbvs1pluss2} has the standard form of that of Chern--Simons gauge theory
upon viewing the contributions of $K_1$ and $S_2$ as the action's
kinetic and interaction terms, respectively.  
Furthermore, the Euler--Lagrange equation yielded by $S_{BV}$, equivalent to the vanishing of the BV variation 
$\delta_{BV}\varPhi$ of $\varPhi$, also have by \ceqref{hamiltsbv}, \ceqref{sqtaubvtau} the standard form of 
those of Chern--Simons gauge theory. We have worked out in this way a general formal framework suitable 
for the construction of Chern--Simons type higher gauge theoretic models. 

In the next section, we shall provide examples of higher gauge and gauge sigma models 
that can be constructed using the axiomatic framework developed in 
subsects. \cref{subsec:bv}, \cref{subsec:sigmamod}.

\vfil\eject

\section{\textcolor{blue}{\sffamily Higher gauge theory and gauged sigma models}}\label{sec:hisigma}

In this section, we shall work out a formulation of higher gauge theory and gauged sigma models  
relying on the formal framework developed in  sect. \cref{sec:higau}. 
The characterizing point of our construction is the realization that for a full--fledged 
BV formulation incorporating ghost degrees of freedom
higher gauge fields viewed as smooth functions on the shifted tangent bundle 
of a space time manifold are not sufficient. A more general kind of gauge fields
are required, which are internal rather than simply ordinary smooth functions. 

The mathematical framework appropriate for the attainment of this goal is furnished by 
graded differential geometry, which we review to set our terminology and notation and highlight 
those points which are relevant in the subsequent analysis. In particular, we go through the 
theory of internal graded manifold morphisms. We also dwell on the theory 
NQ--manifold and $L_\infty$ algebroids and the closely related theory  
of Lie quasi--groupoid differentiation. With this graded geometrical set--up in place,
the beauty and naturalness of our formulation of higher gauge theory and gauged sigma models
fully emerges.

%\vfil\eject

\subsection{\textcolor{blue}{\sffamily Graded manifolds and manifold morphisms}}\label{subsec:graman} 

%\vspace{.33mm}
In this subsection, we review the basic notions of graded geometry relevant in the following.
We follow mainly ref. \ccite{Roytenberg:0203110}. See  also ref. \ccite{Cattaneo:2010re}
for a readable updated introduction. 
%see for instance \ccite{Cattaneo:2010re} and references therein for a review).

%\vspace{.33mm}
A {\it graded manifold} $M$ is a locally ringed space $(M_0,\mathcal{O}_M)$, where $M_0$ is 
a smooth manifold %topological space 
and $\mathcal{O}_M$ is a sheaf of graded commutative algebras 
over $M_0$ which is locally isomorphic to $C_{\mathbb{R}^{\dim M_0}}{}^\infty\otimes S(V)$
for some fixed finite-dimensional graded real vector space $V$
of vanishing degree $0$  component $V_0$ 
\footnote{$\vphantom{\dot{\dot{\dot{x}}}}$ %$\vphantom{\dot{\dot{\dot{\dot{x}}}}}$ 
Here, $C_N{}^\infty$ is the sheaf of smooth realvalued functions on a smooth manifold 
$N$ and $S(E)$ is the graded symmetric algebra of a graded vector space $E$. 
See the cited references for further details.  
%$\vphantom{\ul{\ul{\ul{\ul{\ul{\ul{x}}}}}}}$
}. 
$M_0$ and $\mathcal{O}_M$ are called 
respectively the {\it body} and the {\it structure sheaf} of $M$. 
A {\it morphism} $\varphi:M_1\rightarrow M_2$ of two graded manifolds $M_1$, $M_2$ is a morphism 
$\varphi:(M_{10},\mathcal{O}_{M_1})\rightarrow (M_{20},\mathcal{O}_{M_2})$ of their associated 
locally ringed spaces. The associated  ordinary manifold morphism $\varphi_0:M_{10}\rightarrow M_{20}$
%of the body $M_{10}$ of $M_1$ into the body $M_{20}$ of $M_2$
and sheaf morphism $\varphi^*:\mathcal{O}_{M_2}\rightarrow \varphi_{0*}\mathcal{O}_{M_1}$ 
%of the structure sheaf $\mathcal{O}_{M_2}$  of $M_2$ into the direct image by $\varphi_0$ of the structure 
%sheaf $\mathcal{O}_{M_1}$ of $M_1$ 
are called respectively the {\it body} of and the {\it pull-back} by
 $\varphi$. Graded manifolds and graded manifold morphisms form a category $\mathsans{grMf}$. 

\vfil
An {\it N--manifold} is a graded manifold $M$ for which the vector space $V$ entering 
the definition of the local model is non negatively graded \ccite{Roytenberg:0203110}.
N--manifolds form a full subcategory $\mathsans{N\text{--}Mf}$ of $\mathsans{grMf}$. 
In this paper, we consider mainly though non exclusively N--manifolds. 

\vfil
The structure sheaf $\mathcal{O}_M$ of a graded manifold $M$ decomposes according to degree as a direct 
sum $\mathcal{O}_M=\bigoplus_k\mathcal{O}_M{}^k$. For each $k$, $\mathcal{O}_M{}^k$ is a sheaf of 
$C_{M_0}{}^\infty$ modules. Letting $\mathcal{O}_{Mk}$ be the subsheaf of graded commutative algebras generated by 
$\mathcal{O}_{Mk}=\bigoplus_{l\leq k}\mathcal{O}_M{}^l$ for $k\in\mathbb{Z}$,
one has a filtration of sheaves $\ldots \subset\mathcal{O}_{Mk}\subset\mathcal{O}_{Mk+1}\subset\dots$.
For a N--manifold $M$, $\mathcal{O}_M{}^k=\mathcal{O}_{Mk}=0$ for $k<0$ and 
$\mathcal{O}_M{}^0=\mathcal{O}_{M0}=C_{M_0}{}^\infty$. Further, for $k\geq 0$,
$(M_0,\mathcal{O}_{Mk})$ is a locally ringed space defining a graded manifold $M_k$.
The $M_k$ fit in a  sequence of fibrations $M_0\leftarrow M_1\leftarrow M_2\leftarrow\dots$
\ccite{Roytenberg:0203110}. $M$ is the projective limit of the sequence.
If $M=M_n$ for some $n$, $M$ is said of finite {\it degree} $n$. Else, $M$ is said of infinite degree. 
The N--manifolds considered in this paper are mostly %though non exclusively
finite degree. 

\vfil
In a graded manifold $M$, the local isomorphism between the structure sheaf $\mathcal{O}_M$ 
and the local model $C_{\mathbb{R}^{\dim M_0}}{}^\infty\otimes S(V)$ defines a set 
of local coordinates of $M$: if $\xi^i=(\eta_0{}^a,\zeta^r)$ are a set of coordinates of 
$\mathbb{R}^{\dim M_0}\times V$, respectively, then their preimages $x^i=(y_0{}^a,z^r)$ constitute a 
set of local coordinates of $M$. The $y_0{}^a$ are local coordinates for the body $M_0$ of $M$ and are called 
thus body coordinates. They all have $0$ degree. The $z^r$ are called vector coordinates. They can have any degree,
but for $M$ an N--manifold they are all non negatively graded.

\vfil
We consider next the space of morphisms of two graded manifolds because of its importance 
in the analysis of later subsections. Before doing that, however, the following remarks are in order. 
In geometry, one often encounters spaces of maps between manifolds and one would like to handle
these as manifolds of some sort. This is not straightforward at all because of their
infinite dimensional nature. The best approach to the subject combining mathematical 
rigour and practical usability is by employing functional diffeology (see e. g. \ccite{zemmour;2016} for background). 
The same problem arises when dealing with spaces of morphisms between graded manifolds.
In order to treat these as infinite dimensional graded manifold, one has to resort to a graded 
generalization of functional diffeology. Following this path, however, would bring us to far afield.
Here, we prefer to proceed at a more modest heuristic level by formally enlarging the categories
$\mathsans{Mf}$, $\mathsans{grMf}$ in such a way to include (graded) functional difffeological spaces. 

\vspace{.33mm} %\vfil 
The morphism set $\Hom_{\mathsans{grMf}}(M_1,M_2)$ of two graded manifolds $M_1$, $M_2$ 
%has a structure of 
is an infinite-dimensional manifold called the {\it hom manifold} of $M_1$, $M_2$.
As such, $\Hom_{\mathsans{grMf}}(M_1,M_2)$ contains $\Hom_{\mathsans{Mf}}(M_{10},M_{20})$
as a submanifold. 
%Here, it is assumed that the space $\Hom_{\mathsans{Mf}}(M_{10},M_{20})$ 
%of morphisms of any two  manifolds $M_{10}$, $M_{20}$ 
%has a structure of infinite-dimensional smooth manifold.
The construction of a special set of local coordinates of $\Hom_{\mathsans{grMf}}(M_1,M_2)$ 
illustrates this. %property. %can be obtained as follows.
Let $\varphi\in\Hom_{\mathsans{grMf}}(M_1,M_2)$ be a morphism
of $M_1$, $M_2$. Then, $\varphi$ is described locally
by the pull--backs $\varphi^*x_2{}^{i_2}$ of the local coordinates $x_2{}^{i_2}$ 
of $M_2$. These in turn can be expressed as functions
of the local coordinates $x_1{}^{i_1}$ of $M_1$ as %\hphantom{xxxxxxxxxxxxxx}
\begin{equation}
\varphi^*x_2{}^{i_2}(x_1)=\sss_{R_1,|z_1{}^{R_1}|=|x_2{}^{i_2}|}\varphi^{i_2}{}_{R_1}(y_{10})z_1{}^{R_1},
\label{graman1}
\end{equation}
where $R_1$ denotes a multi--index $r_{11}\dots r_{1h}$, $z_1{}^{R_1}$ stands for the product 
$z_1{}^{r_{11}}\cdots z_1{}^{r_{1h}}$ of the corresponding vector coordinates $z_1{}^{r_1}$ 
of $M_1$ and $\varphi^{i_2}{}_{R_1}(y_{10})$ is a degree $0$ smooth function
of the body coordinates $y_{10}{}^{a_1}$ of $M_1$ and 
$\varphi^{i_2}{}_{R_1}\neq 0$ for finitely many values of $R_1$. 
For varying $\varphi\in\Hom_{\mathsans{grMf}}(M_1,M_2)$, 
the $\varphi^{i_2}{}_{R_1}$ constitute a set of local coordinates 
of $\Hom_{\mathsans{grMf}}(M_1,M_2)$. Since the $\varphi^{i_2}{}_{R_1}$ 
are all functions and have all degree $0$, $\Hom_{\mathsans{grMf}}(M_1,M_2)$ 
is an infinite dimensional manifold as stated. 
The submanifold of $\Hom_{\mathsans{grMf}}(M_1,M_2)$  defined by the conditions 
$\varphi^{i_2}{}_{R_1}=0$ for either $i_2=r_2$ or $i_2=a_2$ and $R_1\neq\emptyset$ clearly 
is $\Hom_{\mathsans{Mf}}(M_{10},M_{20})$. 

\vspace{.33mm} %\vfil 
For two fixed graded manifolds $M_1$, $M_2$, the functor from $\mathsans{grMf}^{\mathrm{op}}$ to $\mathsans{Mf}$ 
defined by the assignment $N\rightarrow \Hom_{\mathsans{grMf}}(M_1 \times N,M_2)$ is representable, 
so that  there exists a graded manifold $\mathsans{Hom}_{\mathsans{grMf}}(M_1,M_2)$, unique up to a unique 
isomorphism, having the property that
$\Hom_{\mathsans{grMf}}(M_1 \times N,M_2)=\Hom_{\mathsans{grMf}}(N,\mathsans{Hom}_{\mathsans{grMf}}(M_1,M_2))$
\ccite{Roytenberg:0203110}. 
$\mathsans{Hom}_{\mathsans{grMf}}(M_1,M_2)$ is called the {\it internal hom manifold} of  $M_1$, $M_2$. 
Its body $\mathsans{Hom}_{\mathsans{grMf}}(M_1,M_2)_0$ is just $\Hom_{\mathsans{grMf}}(M_1,M_2)$. 
The construction of a suitable set of local coordinates of $\mathsans{Hom}_{\mathsans{grMf}}(M_1,M_2)$ 
highlights the difference between the ordinary hom manifold $\Hom_{\mathsans{grMf}}(M_1,M_2)$ 
considered above and the internal 
hom manifold $\mathsans{Hom}_{\mathsans{grMf}}(M_1,M_2)$. Consider a morphism 
$\varphi\in\Hom_{\mathsans{grMf}}(M_1\times N,M_2)$ of $M_1\times N$, $M_2$. 
Then, similarly to before, $\varphi$ is described locally by 
%local sections of the structure sheaf $\mathcal{O}_{M_1\times N}$ of $M_1\times N$ of the form 
the pull--backs $\varphi^*x_2{}^{i_2}$ of the local 
coordinates $x_2{}^{i_2}$ of $M_2$. These are functions
of the local coordinates $x_1{}^{i_1}$ of $M_1$ and $u^\kappa$ of $N$ of the form 
\begin{equation}
\varphi^*x_2{}^{i_2}(x_1,u)=\sss_{R_1,|\varphi^{i_2}{}_{R_1}|+|z_1{}^{R_1}|=|x_2{}^{i_2}|}\varphi^{i_2}{}_{R_1}(y_{10},u)z_1{}^{R_1},
\label{graman2}
\end{equation}
where $R_1$ and $z_1{}^{R_1}$ are defined as before and $\varphi^{i_2}{}_{R_1}(y_{10},u)$ is a smooth function
of the body coordinates $y_{10}{}^{a_1}$ of $M_1$ depending on the $u^\kappa$ of generally non zero degree 
and non vanishing for finitely many values of $R_1$. Now, we can regard $\varphi$ as a graded 
manifold morphism $\phi$ from $N$ to $\mathsans{Hom}_{\mathsans{grMf}}(M_1,M_2)$ 
by setting formally $\phi(u)^\#x_2{}^{i_2}(x_1)=\varphi^*x_2{}^{i_2}(x_1,u)$. 
The $\varphi^{i_2}{}_{R_1}(-,u)$ then are the local coordinates of $\phi(u)$ in $\mathsans{Hom}_{\mathsans{grMf}}(M_1,M_2)$.
This shows that in $\mathsans{Hom}_{\mathsans{grMf}}(M_1,M_2)$ coordinates are smooth functions
of the $y_{10}{}^{a_1}$ as earlier but with degrees varying in certain possibly infinite
ranges including $0$. $\mathsans{Hom}_{\mathsans{grMf}}(M_1,M_2)$ is so  
as an infinite dimensional graded manifold with body $\mathsans{Hom}_{\mathsans{grMf}}(M_1,M_2)_0=
\Hom_{\mathsans{grMf}}(M_1,M_2)$.  

The algebra of smooth functions of a graded manifold $M$ is the graded commutative algebra 
$C^\infty(M)=\mathcal{O}_M(M_0)$ 
of global sections of the structure sheaf $\mathcal{O}_M$. 
$C^\infty(M)$ is an infinite-dimensional manifold called 
the {\it graded smooth function manifold} of $M$. As such, $C^\infty(M)$ 
contains $C^\infty(M_0)$ as a submanifold. A set of local 
coordinates of $C^\infty(M)$ showing this 
can be constructed as follows. A function $f\in C^\infty(M)$ 
reads as a function of the local coordinates $x^i$ of $M$ as
\begin{equation}
f(x)=\sss_{R,|z^R|=|f|}f_R(y)z^R  \vphantom{\bigg]}
\label{graman3}
\end{equation}
analogously to \ceqref{graman1}, where $f_R(y)$ is a degree $0$ smooth function
of the body coordinates $y^a$ of $M$ and 
$f_R\neq 0$ for finitely many values of $R$. 
The $f_R$ are then the local coordinates of $f$ in $C^\infty(M)$. Since they all are degree $0$ 
functions, $C^\infty(M)$  is an infinite dimensional ordinary manifold as claimed. 
The submanifold $C^\infty(M_0)$ is defined by the conditions $f_R(y)=0$ for $R\neq\emptyset$. 
The above discussion shows further that $C^\infty(M)=\Hom_{\mathsans{grMf}}(M,G_{\mathbb{R}})$,  
where $G_{\mathbb{R}}$ is the graded algebra defined in \ceqref{gradreal}. 
analogously to ordinary differential geometry. 

A morphism $\varphi:M_1\rightarrow M_2$ %\in\Hom_{\mathsans{grMf}}(M_1,M_2)$ 
of two graded manifolds $M_1$, $M_2$ induces a morphism 
$\varphi^*:C^\infty(M_2)\rightarrow C^\infty(M_1)$ of their associated graded commutative 
algebras of graded smooth functions via its sheaf theoretic pull-back, called {\it pull-back}
of $\varphi$. $\varphi^*$ fully characterizes $\varphi$.

For a fixed graded manifolds $M$, the functor from $\mathsans{grMf}^{\mathrm{op}}$ to $\mathsans{Mf}$ 
defined by the assignment $N\rightarrow C^\infty(M \times N)$ is representable, 
so that  there exists a graded manifold $\mathsans{C}^\infty(M)$, unique up to a unique 
isomorphism, having the property that $C^\infty(M\times N)=\Hom_{\mathsans{grMf}}(N,\mathsans{C}^\infty(M))$. 
$\mathsans{C}^\infty(M)$ is called the {\it internal graded smooth function manifold} of $M$. 
Its body $\mathsans{C}^\infty(M)_0$ is just $C^\infty(M)$.
A natural set of local coordinates of $\mathsans{C}^\infty(M)$ elucidating this 
can be constructed as follows. A function $f\in C^\infty(M\times N)$ 
reads as a function of the local coordinates $x^i$ of $M$ as
\begin{equation}
f(x,u)=\sss_{R,|f_R|+|z^R|=|f|}f_R(y,u)z^R, \vphantom{\bigg]}
\label{graman4}
\end{equation}
analogously to \ceqref{graman2}, where $f_R(y,u)$ is a smooth function
of the body coordinates $y^a$ of $M$ depending on the coordinates $u^\kappa$ of $N$ of 
generally non zero degree and non vanishing for finitely many values of $R$. Now, we can regard 
$f$ as a graded manifold morphism $\phi_f$ from $N$ to $\mathsans{C}^\infty(M)$ 
by setting formally $\phi_f(u)(x)=f(x,u)$. 
The $f_R(-,u)$ then are the local coordinates of $\phi_f(u)$ in $\mathsans{C}^\infty(M)$.
This shows that in $\mathsans{C}^\infty(M)$ coordinates are smooth functions
of the $y^a$ as earlier but with degrees varying in certain possibly infinite
ranges including $0$. $\mathsans{C}^\infty(M)$ is in this way 
an infinite dimensional graded manifold with body $\mathsans{C}^\infty(M)_0=C^\infty(M)$. 
The above discussion shows further that $\mathsans{C}^\infty(M)=\mathsans{Hom}_{\mathsans{grMf}}(M,G_{\mathbb{R}})$, 
as expected. Note that $\mathsans{C}^\infty(M)$ is also a graded algebra. In fact, one has 
\begin{equation}
\mathsans{C}^\infty(M)\simeq C^\infty(M)\otimes G_{\mathbb{R}}.
\label{graman5}
\end{equation}
% where $G_{\mathbb{R}}$ is the graded algebra defined in \ceqref{gradreal}. 
% isomorphic to $C^\infty(M)\otimes G_{\mathbb{R}}$,  

\vspace{.33mm}
A internal morphism $\phi:M_1\rightarrow M_2$ %\in\Hom_{\mathsans{grMf}}(M_1,M_2)$ 
of two graded manifolds $M_1$, $M_2$ induces a morphism 
$\phi^\#:C^\infty(M_2)\rightarrow \mathsans{C}^\infty(M_1)$ of the
graded commutative algebras %$C^\infty(M_2)$, $\mathsans{C}^\infty(M_1)$ 
of graded smooth functions of $M_2$ and internal graded smooth functions of $M_1$ 
called the {\it internal pull--back} of $\phi$. To see this more explicitly, consider a graded manifold $N$ and
a graded manifold morphism $\varphi:M_1\times N\rightarrow M_2$ which we may view equivalently as a morphism
$\phi:N\rightarrow \mathsans{Hom}_{\mathsans{grMf}}(M_1,M_2)$ into the internal hom manifold
of $M_1$, $M_2$. %For $f\in C^\infty(M_2)$, %$\varphi^*f\in C^\infty(M_1\times N)$ can be regarded
%equivalently as a graded manifold morphism $\phi^\#f:N\rightarrow \mathsans{C}^\infty(M_1)$. 
For $f\in C^\infty(M_2)$, $\varphi^*f\in C^\infty(M_1\times N)$ can be regarded in 
equivalent fashion as a graded manifold morphism $\phi^\#f$ 
from $N$ to $\mathsans{C}^\infty(M_1)$ by setting formally $\phi(u)^\#f(x)=\varphi^*f(x_1,u)$. 
This defines the required internal pull--back of $f$ by $\phi$.

\vspace{.33mm}
A graded manifold 
$M$ is a locally ringed space $(M_0,\mathcal{O}_{M})$ with a special local model. 
The rings attached by the structure sheaf $\mathcal{O}_{M}$ to the open sets of the body $M_0$ are 
not in general rings of functions of a certain kind. 
For this reason, for a morphism $\varphi:M_1\rightarrow M_2$ of two graded manifolds $M_1$, $M_2$,
which is a morphism of the underlying locally ringed spaces $(M_{10},\mathcal{O}_{M_1})$, $(M_{20},\mathcal{O}_{M_2})$,
the associated pull--back $\varphi^*:C^\infty(M_2)\rightarrow C^\infty(M_1)$ is not given in general 
by a straightforward generalization of the well--known expression of elementary differential geometry. 
Let us examine this point in more detail. Suppose that a graded function $f\in C^\infty(M_2)$ is given 
by an expansion of the form \ceqref{graman3} as a function $f(x_2)$ of the local coordinates $x_2{}^{i_2}$ of $M_2$. 
Suppose further that the components $\varphi^{i_2}$ of $\varphi$ are given by expansions of the form 
\ceqref{graman1} as functions $\varphi^{i_2}(x_1)$ of the local coordinates $x_1{}^{i_1}$ of $M_1$. Then, as a function of the
$x_1{}^{i_1}$, $\varphi^*f(x_1)$ is not given by $f(x_2)|_{x_2{}^{i_2}=\varphi^{i_2}(x_1)}$ in general.
In fact, such an object may not be polynomial in the vector coordinates $z_1{}^{r_1}$ of $M_1$ as it should
even if the $\varphi^{i_2}(x_1)$ are. In particular, the assignment of the components $\varphi^{i_2}(x_1)$ cannot by itself 
specify the pull--back operation $\varphi^*$ by setting $\varphi^*f(x_1)=f(x_2)|_{x_2{}^{i_2}=\varphi^{i_2}(x_1)}$.
In two instances this however can be done, namely when either $a)$ $M_1$ is an N--manifold
or $b)$ the vector coordinates $z_1{}^{r_1}$ of $M_1$ are all odd, since then the aforementioned 
problems with polynomiality 
do not arise. Similar conclusions are reached for an internal morphism $\phi:M_1\rightarrow M_2$ 
and the associated internal pull--back operation $\phi^\#$. The assignment of the components $\phi^{i_2}(x_1)$ cannot by itself 
specify the pull--back operation $\phi^\#$ by setting $\phi^\#f(x_1)=f(x_2)|_{x_2{}^{i_2}=\phi^{i_2}(x_1)}$, but it can when 
the vector coordinates $z_1{}^{r_1}$ of $M_1$ are all odd. 
Happily, these are precisely the cases which will occur in the following analysis, where the pull--back operators 
$\varphi^*$ or $\phi^\#$ will always be tacitly assumed to be defined in the above standard fashion.

%on $\mathbb{R}^{\dim M_{10}}$
%The reason is that morphisms
%are degree preserving and so cannot allow for non trivial ghost degree. 

\begin{exa} \label{exa:nman1}
{\rm If $E\rightarrow X$ is a vector bundle over a manifold $X$, then the $1$--shifted vector bundle 
$M=E[1]$ is an N--manifold with $M_0=X$ and $M_1=E[1]$, hence of degree $1$. 
Conversely, every degree $1$ N--manifold is of the form $E[1]$ for some vector bundle $E$.}
\end{exa}

\begin{exa} \label{exa:tan1inthom}
{\rm The $-1$--shifted real line $\mathbb{R}[-1]$ is a graded manifold 
with point body concentrated in degree $-1$.  
For any manifold $M$, the internal hom manifold 
$\mathsans{Hom}_{\mathsans{grMf}}(\mathbb{R}[-1],M)$ is isomorphic to 
the $1$--shifted tangent bundle $T[1]M$ of $M$ and hence is a degree $1$ N--manifold
(see eg. \cref{exa:nman1}). More generally, the internal hom manifold 
$\mathsans{Hom}_{\mathsans{grMf}}(\mathbb{R}^q[-1],M)$ is isomorphic to 
the $q$-fold $1$--shifted tangent bundle $T[1]^qM$ of $M$.  
We shall refer to this remarkable fact in later subsections.}
\end{exa}

\begin{exa} \label{exa:intcinf}
{\rm The $1$--shifted real line $\mathbb{R}[1]$ is a point body degree $1$ N--manifold.  
For this, the graded smooth function manifold $C^\infty(\mathbb{R}[1])$ is isomorphic to $\mathbb{R}^2$ 
while the internal graded smooth function manifold $\mathsans{C}^\infty(\mathbb{R}[1])$ is isomorphic
to $G_{\mathbb{R}}{}^2$. As graded commutative algebras, $C^\infty(\mathbb{R}[1])$ and $\mathsans{C}^\infty(\mathbb{R}[1])$ 
are respectively $\mathbb{R}\oplus\vartheta\mathbb{R}$ and $G_{\mathbb{R}}\oplus\vartheta G_{\mathbb{R}}$, where $\vartheta$ 
is a formal degree $1$ parameter.}
\end{exa}

\vspace{.33mm}
In the BGKS formulation of higher gauge theory reviewed in subsect. \cref{subsec:bgkstheory}, 
higher gauge fields are morphisms of suitable graded manifolds.
In the BRST formulation of the theory, graded manifold morphisms are not sufficient 
for a complete description of BRST gauge superfields. 
To incorporate such superfields, internal morphisms are required. Indeed, as shown in
the analysis carried out above, while local coordinates for $\Hom_{\mathsans{grMf}}(M_1,M_2)$
are degree $0$ smooth functions of the body coordinates of $M_1$, 
local coordinates for $\mathsans{Hom}_{\mathsans{grMf}}(M_1,M_2)$ are 
smooth functions of the body coordinates of all possible degrees. 
Since in BRST theory fields  
of non zero ghost degree are involved, 
$\mathsans{Hom}_{\mathsans{grMf}}(M_1,M_2)$ rather %\pagebreak
than $\Hom_{\mathsans{grMf}}(M_1,M_2)$ is the natural functional manifold for describing 
BRST fields in a BRST extension of BGKS theory.

%\vfil\eject

\subsection{\textcolor{blue}{\sffamily NQ--manifolds}}\label{subsec:lgraman}

In this subsection, we review briefly the theory of NQ--manifolds because these are the kind of N--manifolds 
which are the natural target spaces of higher gauge and gauged sigma models. 
Our discussion will be kept at the level of local graded geometry focusing on those properties that are \pagebreak
most relevant in the following. See again ref. \ccite{Cattaneo:2010re} for further background
and complete referencing. 

A degree $p$ vector field $X$ on a graded manifold $M$ is a degree $p$ derivation of the graded 
commutative algebra $C^\infty(M)$. It has a local coordinate expression 
\begin{equation}
X=X^i\partial_i.
\label{lgraman1}
\end{equation}
Here, the local functions $X^i$ are the components of $X$. They are homogeneous of degree
$|X^i|=|x^i|+p$. The vector fields of all integer degrees form a graded Lie algebra
$\mathfrak{X}(M)$. 
%The Lie btackets are defined by 
%\begin{equation}
%[X,Y]=X  Y-(-1)^{pq}Y  X
%\label{}
%\end{equation}
% 
In local coordinates, the Lie brackets of two homogeneous vector fields 
$X,Y\in\mathfrak{X}(M)$ of degrees $p$, $q$ are given by 
\begin{equation}
[X,Y]=(X^j\partial_jY^i-(-1)^{pq}Y^j\partial_jX^i)\partial_i.
\label{lgraman2}
\end{equation}

A NQ--manifold $M$ is an N--manifold equipped with a {\it homological vector field} $Q$,
that is a degree $1$ vector field on $M$ such that 
\begin{equation}
Q^2=[Q,Q]/2=0.
\label{lgraman3}
\end{equation} 
It is useful to write \ceqref{lgraman3} in local coordinates using \ceqref{lgraman2}, 
\begin{equation}
Q^j\partial_jQ^i\partial_i=0.
\label{lgraman4}
\end{equation}
For a NQ--manifold $M$, the graded commutative algebra $C^\infty(M)$ of smooth functions of $M$ 
is differential with differential $Q$. The algebraic theory of these type of algebras expounded in subsect. 
\cref{subsec:algdef} thus applies. In particular, $M$ is characterized by the $Q$--cohomology of $C^\infty(M)$. 

A morphism $\varphi:M_1\rightarrow M_2$ of NQ--manifolds %$(M_1,Q_1)$ into another $(M_2,Q_2)$% 
is a morphism of graded manifolds with the property that the associated morphism 
$\varphi^*:C^\infty(M_2)\rightarrow C^\infty(M_1)$ of graded commutative algebras satisfies
\begin{equation}
Q_1 \varphi^*-\varphi^* Q_2=0
\label{lgraman5}
\end{equation}
that is that $\varphi^*:C^\infty(M_2)\rightarrow C^\infty(M_1)$ is a morphism of 
differential graded commutative algebras. 
In local coordinates, \ceqref{lgraman5} takes the form 
\begin{equation}
(Q_1{}^j\partial_{1j}\varphi^r-Q_2{}^r\circ\varphi)\partial_{2r}|_{\varphi}=0. 
\label{lgraman6}
\end{equation}
 
Suppose that $M_1$, $M_2$ are NQ--manifolds and that $\varphi:M_1\rightarrow M_2$
is a morphism of graded not necessarily NQ manifolds. Then, $\varphi$ is characterized by the defect of 
the algebra morphism $\varphi^*$ 
\begin{equation}
F_\varphi=Q_1 \varphi^*-\varphi^* Q_2
\label{lgraman7}
\end{equation}
satisfying the defect identity \hphantom{xxxxxxxxxxxxx}
\begin{equation}
Q_1F_\varphi+F_\varphi Q_2=0
\label{lgraman8}
\end{equation}
(cf. eq. \ceqref{defect}, \ceqref{bianchi}). 
In local coordinates, $F_\varphi$ is given by %the expression
\begin{equation}
F_{\varphi^*}=(Q_1{}^j\partial_{1j}\varphi^r-Q_2{}^r\circ\varphi)\partial_{2r}|_{\varphi}=0
\label{lgraman9}
\end{equation}
and the defect identity takes the form 
\begin{align}
&(Q_1{}^i\partial_{1i}F_{\varphi^*}{}^r+F_{\varphi^*}{}^s\partial_{2s}Q_2{}^r\circ\varphi)\partial_{2r}|_{\varphi}
\vphantom{\Big]}
\label{lgraman10}
\\
&\hspace{2cm}-(-1)^{|x_2{}^s|}(F_{\varphi^*}{}^sQ_1{}^i\partial_{1i}\varphi^r
+Q_2{}^s\circ\varphi F_{\varphi^*}{}^r)\partial_{2r}\partial_{2s}|_{\varphi}=0.
\vphantom{\Big]}
\nonumber
\end{align}

%Let us examine a few basic examples of NQ--manifolds.  

\begin{exa} \label{exa:tm}
{\rm The $1$--shifted tangent bundle $T[1]N$ of a manifold $N$ is an NQ--manifold.
As a graded manifold, $T[1]N$ is described locally by degree $0$ base coordinates $x^i$ 
and degree $1$ fiber coordinates $\xi^i$. The graded commutative algebra $C^\infty(T[1]N)$ consists of the functions 
of the form 
\begin{equation}
\alpha=\sss_{h\geq 0}\frac{1}{h!}\alpha_{i_1\ldots i_h}(x)\xi^{i_1}\cdots\xi^{i_h}
\label{t1mfunc}
\end{equation}
and is isomorphic to the graded commutative algebra $\Omega^*(N)$ of differential forms
$\alpha=\sum_{h\geq 0}\frac{1}{h!}\alpha_{i_1\ldots i_h}(x)dx_{N}{}^{i_1}\cdots dx_{N}{}^{i_h}$. 
The homological vector field of $T[1]N$ is 
\begin{equation}
d=\xi^i\partial_{xi}
\label{t1mq}
\end{equation}
%\vspace{-.7cm}\eject\noindent
Under the isomorphism $C^\infty(T[1]N)\simeq\Omega^*(N)$, 
$d$ answers to the de Rham diffe\-rential $d_{N}=d_{N}x^i\partial_{xi}$. The cohomology of the cochain
complex $(C^\infty(T[1]N),d)$ is therefore isomorphic to the familiar de Rham cohomology 
of the complex $(\Omega^*(N),d_{N})$.  }
\end{exa} 

\begin{exa} \label{exa:t*m}
{\rm The $1$--shifted cotangent bundle $T^*[1]N$ of a manifold $N$ \pagebreak is an N--manifold.
$T^*[1]N$  is described locally by degree $0$ base coordinates $x^i$ 
and degree $1$ fiber coordinates $\xi_i$. The graded algebra $C^\infty(T^*[1]N)$ consists therefore 
of the functions of the form
\begin{equation}
U=\sss_{h\geq 0}\frac{1}{h!}U^{i_1\ldots i_h}(x)\xi_{i_1}\cdots\xi_{i_h}
\label{t*1mfunc}
\end{equation}
and so is isomorphic to the graded commutative algebra $\mathfrak{X}^*(N)$ of multivectors 
$U=\sum_{h\geq 0}\frac{1}{h!}U^{i_1\ldots i_h}(x)\partial_{i_1}\cdots \partial_{i_h}$.

$T^*[1]N$ is equipped with the canonical degree $1$ symplectic form 
\begin{equation}
\omega=d\xi_idx^i.
\label{t*1omg}
\end{equation}
Under the isomorphism $C^\infty(T^*[1]N)\simeq \mathfrak{X}^*(N)$, the Poisson brackets 
$\{-,-\}$ as\-sociated with $\omega$
correspond to the classical Schouten brackets $[-,-]_S$. 

Let $N$ be a Poisson manifold.  %  with Poisson bivector $P\in C^\infty(N,\www^2 TN)$, 
Then, the graded vector bundle $T^*[1]N$ is an NQ--manifold. Indeed, $T^*[1]N$ is equipped with the degree $1$ 
vector field 
\begin{equation}
\delta=P^{ij}(x)\xi_j\partial_{xi}+ \tfrac{1}{2}\partial_i P^{jk}(x)\xi_j\xi_k\partial_\xi{}^i,
\label{t*1mq}
\end{equation}
where the $P^{ij}$ are the Poisson structure functions. Further, 
the relations the $P^{ij}$ satisfy are equivalent to $\delta$ 
being homological. Remarkably, under the isomorphism $C^\infty(T^*[1]N)\simeq \mathfrak{X}^*(N)$, 
$\delta$ corresponds to the Poisson--Lichnerowicz 
differential $\delta_{PL}=[P,-]_S$. The cohomology of the cochain
complex $(C^\infty(T^*[1]N),\delta)$ is therefore isomorphic to the Poisson--Lichnerowicz cohomology 
of the complex $(\mathfrak{X}^*(N),\delta_{PL})$. 

The homological vector field $\delta$ turns out to be symplectic actually Hamiltonian: one has $Q=\{S,-\}$,
where $S$ is the degree $2$ function 
\begin{equation}
S=\tfrac{1}{2}P^{ij}(x)\xi_i\xi_j.
\label{t*1ham}
\end{equation}
satsifying the master equation 
\begin{equation}
\{S,S\}=0.
\label{t*mmaster}
\end{equation}
A degree $k$ PQ--manifold is a degree $k$ NQ--manifold \pagebreak equipped with a degree $k$ symplectic $2$--form
with respect to which the homological vector field is symplectic. 
$T^*[1]N$ is therefore a degree $1$ PQ manifold. 
It can be shown that the most general degree $1$ PQ--manifold is of the form $T^*[1]N$ for some Poisson manifold
$N$ \ccite{Schwarz:1992nx}.
}
\end{exa}

\begin{exa} \label{exa:liealgoid}
{\rm A {\it Lie algebroid} is a vector bundle $A\rightarrow N$ over a manifold $N$, with a structure 
of NQ--manifold on the $1$--shifted bundle $A[1]$ necessarily of degree $1$ (cf. eg. \cref{exa:nman1}). 
If we denote by $x^i$ and $\xi^a$ the base and fiber coordinates of $A[1]$, the homological
vector field $Q$ of $A[1]$ has the form 
\begin{equation}
Q=\rho^i{}_a(x)\xi^a\partial_{xi}-\tfrac{1}{2}f^a{}_{bc}(x)\xi^b\xi^c\partial_{\xi a}.
\label{liealgoidq}
\end{equation}
The coefficients $\rho^i{}_a$ are the local coordinate representation of a bundle map $\rho:A\rightarrow TN$ called the 
{\it anchor} of $A$, viz
\begin{equation}
\rho^i(e)=\rho^i{}_ae^a
\label{liealgoida}
\end{equation}
for $e\in A$.
The coefficients $f^a{}_{bc}$ together with the $\rho^i{}_a$ define an antisymmetric bracket structure $[-,-]_A$ 
on $\varGamma(A)$ %of the sections of $A$
\footnote{$\vphantom{\dot{\dot{\dot{x}}}}$ Here and in the following, we denote by 
$\varGamma(V)$ the space of sections of a vector bundle $V$.}, explicitly 
\begin{equation}
[s,t]^a=\rho^i{}_bs^b\partial_it^a-\rho^i{}_bt^b\partial_is^a+f^a{}_{bc}s^bt^c
\label{liealgoidb}
\end{equation}
with $s,t\in\varGamma(A)$. 
The nilpotence of $Q$ ensures that the brackets $[-,-]_A$ are Lie and that the linear 
map $\varGamma(A)\rightarrow \mathfrak{X}^1(N)$ induced by $\rho$ is Lie. Examples \cref{exa:tm} and \cref{exa:t*m}
show that the tangent bundle of a manifold and the cotangent bundle of a Poisson manifold are naturally Lie algebroids.
Indeed, every degree $1$ NQ--manifold is of the form $A[1]$ for some Lie algebroid $A$. } 
\end{exa}

\noindent 
The above are all examples of $L_\infty$--algebroids, a broad class of NQ--manifolds to which the next subsection
is devoted.

%\vfil\eject

\subsection{\textcolor{blue}{\sffamily L${}_\infty$--algebras and algebroids}}\label{subsec:linf}

$L_\infty$--algebroids are NQ--manifolds representing a far reaching generalization of Lie algebroids.
They encode the symmetry of higher gauge theory \pagebreak 
and constitute the natural target spaces of higher gauged sigma models as will 
be shown later below. We illustrate their theory in this subsection. 

A graded vector bundle over an ordinary manifold $N$ is a vector bundle with a direct sum decomposition 
of the form  
\begin{equation}
E=\ddd_{p=0}^mE_p
\label{grvec}
\end{equation}
with $m\in\mathbb{N}$. The subbundles $E_p$ are conventionally assigned degree $p$. More general gradings 
are possible but they will not be considered here. 
An $L_\infty$--{\it al\-gebroid} is a graded vector bundle $E$ with an assignment of a homological vector field $Q_E$
on the $1$--shifted bundle $E[1]$,
\begin{equation}
Q_E{}^2=0.
\label{linfq}
\end{equation}
If $m=0$, $E$ is called a Lie algebroid. %For $m=2$, $E$ is called a Lie $2$--algebroid or a $2$--term $L_\infty$--algebroid.
For generic $m$, $E$ is called a Lie $m+1$--algebroid or a $m+1$--term $L_\infty$--algebroid. When $N$ is a point, one uses the 
term algebra instead of algebroid. In particular, for $m=0$ we have an ordinary Lie algebra
%, for $m=2$ a Lie $2$--algebra or $2$--term $L_\infty$--algebra 
and for a generic $m$ a Lie $m+1$--algebra or a $m+1$--term $L_\infty$--algebra. 
The nilpotence relation \ceqref{linfq} encodes a very rich geometrical structure on $E$, which we shall make more explicit next. 

The shifted bundle $E[1]$ is described locally by base coordinates $x^i$ and fiber coordinates $\xi^a$.
The $x^i$ form a vector $x\in\mathbb{R}^d$, where $d=\dim N$. The $\xi^a$ form a vector $\xi$ of the 
$1$--shift $V_E[1]$ of the graded vector space 
\begin{equation}
V_E=\ddd_{p=0}^mV_{Ep}\qquad\quad V_{Ep}=\mathbb{R}^{r_p}[p],
\label{fbcoor}
\end{equation}
where $r_p=\rank E_p$. $\xi$ decomposes accordingly in degree $p+1$ components %vectors 
$\xi_p$ $\in\mathbb{R}^{r_p}[p+1]$ with $p=0$, $\dots,m$. 

On a trivializing neighbourhood $U\subset N$, $Q_E$ has the following structure,
\begin{equation}
Q_E=\rho^i(\xi_0)\partial_{xi}+%\sss_{p=0}^n
\bigg\langle\partial\xi-\sss_{\kappa\ge 2}\frac{(-1)^\kappa}{\kappa!}[\xi,\ldots,\xi]_\kappa,\partial_{\xi}\bigg\rangle. %_{V_E}. 
\label{qexpr}
\end{equation}
In this expression, $\rho\in C^\infty(U,\Hom(V_{E0},\mathbb{R}^d))$, 
$\partial\in C^\infty(U,\Hom_{-1}(V_{E},V_E))$ and $[-,\ldots,-]_\kappa\in C^\infty(U,\Hom_{\kappa-2}(\,\www^\kappa V_{E},V_E))$,
where $\www^\lambda V_{E}$ denotes the graded $\lambda $--th exterior power of $V_E$
\footnote{$\vphantom{\dot{\dot{\dot{x}}}}$ For any integer $k$ and any two graded vector spaces $V$, $W$, 
$\Hom_k(V,W)$ is the set of all degree $k$ linear mappings $T:V\rightarrow W$. Notice that $\Hom_k(V,W)=\Hom_0(V,W[-k])$. 
Usually, one writes $\Hom(V,W)=\Hom_0(V,W)$.}. 
$\langle-,-\rangle$ denotes the $V_E\,$--$\,V_E{}^\vee$ duality pairing
\footnote{$\vphantom{\dot{\dot{\dot{x}}}}$ Recall that for a graded vector space $V$, $V^\vee$ is a graded 
vector space with $V^\vee{}_k=V_{-k}{}^\vee$. Further, the duality pairing $\langle-,-\rangle$ of $V$, $V^\vee$
pairs $V_k$ and $V^\vee{}_{-k}$.}. 
For a morphism $\varphi\in\Hom_\lambda(\,\www^\lambda V_{E},V_E)$, the notation $\langle\varphi(\xi,\ldots,\xi),\xi^*\rangle$ 
is a shorthand for $\varphi^a{}_{a_1\ldots a_\kappa}\xi^{a_1}\cdots\xi^{a_\kappa }\xi^*{}_a$, 
where the reals $\varphi^a{}_{a_1\ldots a_\kappa}$ are the components of $\varphi$ with respect to the canonical basis 
of $V_E$. The defining properties of $\rho$, $\partial$ and the $[-,\ldots,-]_\kappa$ ensure that $Q_E$ 
has degree $1$ as required. The expansion \ceqref{qexpr} can be cast in a more explicit form as  
\begin{align}
Q_E&=\rho^i(\xi_0)\partial_{xi}+\langle\partial\xi_1-\tfrac{1}{2}[\xi_0,\xi_0],\partial_{\xi_0}\rangle
\vphantom{\Big]}
\label{qexpl}
\\
&\hspace{-.5cm}+\langle\partial\xi_2-[\xi_0,\xi_1]+\tfrac{1}{6}[\xi_0,\xi_0,\xi_0],\partial_{\xi_1}\rangle
\vphantom{\Big]}
\nonumber
\\
&\hspace{-1cm}+\langle\partial\xi_3-[\xi_0,\xi_2]-\tfrac{1}{2}[\xi_1,\xi_1]+\tfrac{1}{2}[\xi_0,\xi_0,\xi_1]
-\tfrac{1}{24}[\xi_0,\xi_0,\xi_0,\xi_0],\partial_{\xi_2}\rangle+\ldots,
\vphantom{\Big]}
\nonumber
\end{align}
where we have suppressed the suffixes $\kappa $ form the brackets $[-,\ldots,-]_\kappa $ since its value is evident from 
the number of its arguments. 

Enforcing the nilpotence relation \ceqref{linfq} yields a host of relations involving
$\rho$, $\partial$ and the $[-,\ldots,-]_\kappa $,  
\begin{align}
&2\rho^j(\xi_0)\partial_{xj}\rho^i(\xi_0)-\rho^i([\xi_0,\xi_0])=0,
\vphantom{\Big]}
\label{q201}
\\
&\rho^i(\partial\xi_1)=0,
\vphantom{\Big]}
\label{q202}
\\
&\bigg(\rho^j(\xi_0)\partial_{xj}
+\bigg\langle\partial\xi-\sss_{\lambda \ge 2}\frac{(-1)^\lambda }{\lambda !}[\xi,\ldots,\xi]_\kappa ,\partial_{\xi}\bigg\rangle\bigg)
\vphantom{\Big]}
\label{q203}
\\
&\hspace{5cm} 
\bigg(\partial\xi-\sss_{\kappa \ge 2}\frac{(-1)^\kappa }{\kappa !}[\xi,\ldots,\xi]_\kappa \bigg)=0.
\vphantom{\Big]}
\nonumber
\end{align}
The first two relations %have been sinlged out because they 
express specific properties of the local function $\rho^i$. The third one summarizes the algebraic 
relations of the local functions $[-,\ldots,-]_\kappa $. These can be made more explicit  in terms 
of the components $\xi_p$ of $\xi$, %\pagebreak 
\begin{align}
&3\rho^i(\xi_0)\partial_{xi}[\xi_0,\xi_0]+3[\xi_0,[\xi_0,\xi_0]]-\partial[\xi_0,\xi_0,\xi_0]=0,
\vphantom{\ul{\ul{\ul{\ul{g}}}}}
\vphantom{\Big]}
\label{linfbrpr1}
%\\
\end{align}
\begin{align}
&\rho^i(\xi_0)\partial_{xi}\partial\xi_1+[\xi_0,\partial\xi_1]-\partial[\xi_0,\xi_1]-0, 
\vphantom{\Big]}
\label{linfbrpr2}
\\
&\partial\partial\xi_2=0, 
\vphantom{\Big]}
\label{linfbrpr3}
\\
&2\rho^i(\xi_0)\partial_{xi}[\xi_0,\xi_1]+2[\xi_0,[\xi_0,\xi_1]]-[[\xi_0,\xi_0],\xi_1] \hspace{3cm}
\vphantom{\Big]}
\label{linfbrpr4}
\\
&\hspace{6.4cm}-[\xi_0,\xi_0,\partial\xi_1]
-\partial[\xi_0,\xi_0,\xi_1]=0, 
\vphantom{\Big]}
\nonumber
\\
&4\rho^i(\xi_0)\partial_{xi}[\xi_0,\xi_0,\xi_0]+4[\xi_0,[\xi_0,\xi_0,\xi_0]]
\vphantom{\Big]}
\label{linfbrpr5}
\\
&\hspace{5cm}-6[\xi_0,\xi_0,[\xi_0,\xi_0]]-\partial[\xi_0,\xi_0,\xi_0,\xi_0]=0, 
\vphantom{\Big]}
\nonumber
\\
&2[\partial\xi_1,\xi_1]+\partial[\xi_1,\xi_1]=0, 
\vphantom{\Big]}
\label{linfbrpr6}
\\
&\rho^i(\xi_0)\partial_{xi}\partial\xi_2+[\xi_0,\partial\xi_2]-\partial[\xi_0,\xi_2]=0,
\vphantom{\Big]}
\label{linfbrpr7}
\\
&\partial\partial\xi_3=0,\qquad \qquad \qquad \ldots.
\vphantom{\Big]}
\label{linfbrpr8}
\end{align}

Up to this point, we have adhered to the usual convention that the fiber coordinates $\xi_p$ of the 
subbundles $E_p[1]$ are those induced by a choice of a local frame of $E_p[-p]$ and so of degree $p+1$. 
An equivalent description can be achieved if we use instead the fiber coordinates $\bar\xi_p$ induced by 
a choice of frame of $E_p[1]$ and thus of degree $0$. We denote by $\bar E$ and $\bar E_p$ the vector bundles 
$E$ and $E_p$ when the latter coordinate convention is adopted. 

On any trivializing neighbourhood $U\subset N$, 
a section $s\in \varGamma(\bar E)$ is represented simply by a function $s^\sim\in C^\infty(U,\mathbb{R}^r)$, 
where $r=\rank E$. Similarly, a section $s\in \varGamma(\bar E_p)$ is represented by a function 
$s^\sim\in C^\infty(U,\mathbb{R}^{r_p})$ with $r_p=\rank E_p$. 

The function $\rho$ entering the local expansion \ceqref{qexpr} of $Q_E$
is the local representation of a bundle morphism $\rho_E:\bar E_0\rightarrow TN$, 
the {\it anchor} of $\bar E$.  At the section level, 
$\rho_E$ induces a linear map $\rho_E:\varGamma(\bar E_0)\rightarrow \mathfrak{X}^1(N)$ defined by 
\begin{equation}
\rho_E{}^i(s)^\sim=\rho^i(s^\sim) \vphantom{\bigg]}
\label{lalgoid5}
\end{equation}
with $s\in\varGamma(\bar E_0)$. 

The functions $\partial$ and $[-,\ldots,-]_\kappa $ also appearing in  the local expansion \ceqref{qexpr} of $Q_E$
define a linear {\it boundary map} $\partial_E:\varGamma(\bar E)\rightarrow
\varGamma(\bar E)$ and a collection of multilinear {\it brackets} 
$[-,\ldots,-]_{E\kappa }:\ooo^\kappa \varGamma(\bar E)\rightarrow\varGamma(\bar E)$
of $\bar E$ with $\kappa \ge 2$, respectively. Explicit local expressions for these can be written down: \pagebreak 
\begin{equation}
(\partial_Es)^\sim=\partial s^\sim
\label{lalgoid2}
\end{equation}
and 
\begin{align}
&([s_1,s_2]_E)^\sim=\rho^i(\pi_0(s_1)^\sim)\partial_{xi}s_2{}^\sim
-\rho^i(\pi_0(s_2)^\sim)\partial_{xi}\hat s_1{}^\sim+[s_1{}^\sim,s_2{}^\sim],
\vphantom{\Big]}
\label{lalgoid3}
\\
&([s_1,s_2,\ldots,s_\kappa ]_E)^\sim=[s_1{}^\sim,s_2{}^\sim,\ldots,s_\kappa {}^\sim], \qquad \kappa \ge 3,
\vphantom{\Big]}
\label{lalgoid4}
\end{align}
with any $s,s_1,\ldots,s_\kappa \in\varGamma(\bar E)$. Above, $\pi_0:\varGamma(\bar E)\rightarrow\varGamma(\bar E_0)$
is the obvious projection. Furthermore, for $s\in\varGamma(\bar E)$ expressed as $s=\ddd_{p=0}^ms_p$
with $s_p\in\varGamma(\bar E_p)$, $\hat s\in\varGamma(\bar E)$ is defined through $\hat s=\ddd_{p=0}^m(-1)^ps_p$. 
$\partial_E$ and the $[-,\ldots,-]_{E\kappa }$ are all compatible with the gradation of $\bar E$ 
in the sense that $\partial_E:\varGamma(\bar E_p)\rightarrow
\varGamma(\bar E_{p-1})$ and $[-,\ldots,-]_{E\kappa }:\varGamma(\bar E_{p_1})\otimes\cdots\otimes\varGamma(\bar E_{p_\kappa })
\rightarrow \varGamma(\bar E_{\,\Sigma_ip_i+\kappa -2})$. The gradation determines also the symmetry properties of the 
$[-,\ldots,-]_{E\kappa }$. In particular, one has $[-,-]_E:\www^2 \varGamma(\bar E_0)\rightarrow\varGamma(\bar E_0)$. 
Note however that $[-,-]_E$ is not a Lie bracket in general, since the Jacobi identity 
may not be satisfied. 

The relations \ceqref{q201}--\ceqref{q203} %or \ceqref{linfbrpr1}--\ceqref{linfbrpr8} 
determine various structural properties of and relations obeyed by the anchor $\rho_E$, 
the boundary $\partial_E$ and the brackets $[-,\ldots,-]_{E\kappa }$ of $\bar E$. We mention only the most 
basic ones. The anchor $\rho_E$ has the property that for any two sections $s,t\in\varGamma(\bar E_0)$ 
\begin{equation}
[\rho_E(s),\rho_E(t)]_{\mathfrak{X}^1(N)}=\rho_E([s,t]_E), 
\label{lalgoid7}
\end{equation}
where the brackets in the left hand side are the Lie brackets of $\mathfrak{X}^1(N)$
analogously to Lie algebroids. Further, the boundary map $\partial_E$ is nilpotent 
\begin{equation}
\partial_E{}^2=0.
\label{lalgoid6}
\end{equation}
$\varGamma(\bar E)$ is a thus chain complex with boundary operator $\partial_E$, 
justifying the name given to this latter. Many other relations involving the anchor, boundary and brackets of 
$\bar E$ generalizing the classical Lie theoretic Jacobi identity follow from 
\ceqref{linfbrpr1}--\ceqref{linfbrpr8}. 

\vspace{.5mm}
\begin{exa}
{\rm Let $\mathfrak{g}$ be a simple Lie algebra of compact type and $c\in\mathbb{R}$.
The string Lie $2$--algebra $\mathfrak{string}_c(\mathfrak{g})$ is defined as follows.
As a graded vector space, \pagebreak 
\begin{equation}
\mathfrak{string}_c(\mathfrak{g})=\mathfrak{g}\oplus\mathbb{R}[1]
\label{}
\end{equation}
The boundary $\partial:\mathbb{R}[1]\rightarrow \mathfrak{g}$ vanishes. The brackets are 
\begin{align}
&[x,y]=[x,y]_{\mathfrak{g}},
\vphantom{\Big]}
\label{}
\\
&[x,y,z]=c\,\langle x,[y,z]_{\mathfrak{g}}\rangle,
\vphantom{\Big]}
\label{}
\end{align}
for $x,y,z\in\mathfrak{g}$ and vanish in all other cases, where 
$[-,-]_{\mathfrak{g}}$ and $\langle-,-\rangle$ are the Lie brackets and 
a suitably normalized invariant symmetric non singular bilinear form  
of $\mathfrak{g}$. $\mathfrak{string}_c(\mathfrak{so}(n))$ is relevant in string theory. 
}
\end{exa}

\begin{exa} \label{exa:allliealgoid}
{\rm The tangent bundle $TN$ of a manifold $N$ (cf. eg. \ceqref{exa:tm}), the  
cotangent bundle $TN$ of a Poisson manifold $N$ (cf. eg. \ceqref{exa:t*m})
and in general any Lie algebroid $A$ (cf. eg. \cref{exa:liealgoid}) are all examples of 
$L_\infty$--algebroids.
}
\end{exa}

\begin{exa} \label{exa:couralgoid}
{\rm Let $V\rightarrow N$ be a metric vector bundle. Then, with $V$ there is associated a degree $2$ symplectic N--manifold
$L\,$ as follows. Consider the $2$--shifted cotangent bundle $T^*[2]V[1]$ of the $1$--shifted bundle $V[1]$. 
Then, $T^*[2]V[1]$ is a degree $2$ symplectic N--manifold. Indeed,  
$T^*[2]V[1]$ is described locally by degree $0$ base coordinates $x^i$ and degree $1$ fiber coordinates $\xi^a$ of $V[1]$ 
and by corresponding cotangent degree $2$ base coordinates $p_i$ and degree $1$ fiber coordinates $\eta_a$.
Further, $T^*[2]V[1]$ is equipped with the canonical degree $2$ symplectic $2$--form 
$\omega_0=dp_idx^i+d\eta_a d\xi^a$. Assume  for simplicity that the chosen local trivializations of $V$ 
are such that the coefficients $g_{ab}$ of the metric of $V$ are constant. 
Then, the covariant constraint $\eta_a=\frac{1}{2}g_{ab}\xi^b$ defines 
a submanifold $M$ of $T^*[2]V[1]$. $M$ is a degree $2$ symplectic N--manifold. Indeed, $M$ 
is described by the degree $0$, $1$, $2$ coordinates $x^i$, $\xi^a$, $p_i$ and is equipped with the degree $2$ 
symplectic $2$--form %\hphantom{xxxxxxxxxxxx}
\begin{equation}
\omega=dp_idx^i+\tfrac{1}{2}d\xi^ag_{ab}d\xi^b 
\label{roytom}
\end{equation}
yielded by the pull-back of $\omega_0$ by the embedding $M\rightarrow T^*[2]V[1]$. 
It can be shown that conversely every degree $2$ symplectic N--manifold $M$ stems from a metric vector bundle $V$
by the above construction \ccite{Roytenberg:0203110}. 

%\vspace{.33mm}
Since the constraint defining the embedding $M$ into $T^*[2]V[1]$ is linear, \pagebreak 
$M$ can be identified with the $1$--shift $L[1]$ of a graded vector bundle $L$ over $N$. 
The identification is non canonical, depending 
as it does on an arbitrary choice of a metric connection of $V$. Below, we assume that 
a choice has been made.  

\vspace{.33mm}
The metric vector bundle $V$ is a Courant algebroid if the graded vector bundle $L$ is an $L_\infty$--algebroid 
with the homological vector field $Q_L$ of $L[1]$ Hamiltonian with respect to the Poisson bracket structure
associated with the symplectic form $\omega$ \ccite{Roytenberg:0203110}. In that case, $Q_L$ 
can be shown to read as 
\begin{align}
&Q_L=\rho^i{}_a(x)\xi^a\partial_{xi}+(-\partial_{xi}\rho^j{}_a(x)\xi^ap_j
\vphantom{\Big]}
\label{roytq}
\\
&\hspace{2.75cm}
+\tfrac{1}{6}\partial_{xi}f_{abc}(x)\xi^a\xi^b\xi^c)\partial_{p}{}^i
+g^{ad}(-\rho^i{}_d(x)p_i+\tfrac{1}{2}f_{dbc}(x)\xi^b\xi^c)\partial_{\xi a}
\vphantom{\Big]}
\nonumber
\end{align}
for certain local functions $\rho^i{}_a$ and $f_{abc}$. Indeed, since $Q_L$ is Hamiltonian, $Q_L=\{S,-\}$ 
for some degree $3$ function $S$ on $L[1]$ locally of the form 
\begin{equation}
S=-\rho^i{}_a(x)\xi^ap_i+\tfrac{1}{6}f_{abc}(x)\xi^a\xi^b\xi^c,
\label{royts}
\end{equation}
leading to \ceqref{roytq}. The nilpotence of $Q_L$ is equivalent to the master equation 
\begin{equation}
\{S,S\}=0. 
\label{roytmaster}
\end{equation} 
The structure functions $\rho^i{}_a$ and $f_{abc}$ define the Courant anchor and brackets of $V$, 
respectively. They obey a number of distinguished relations consequent to \ceqref{roytmaster}. 

For a Courant algebroid $V$, $L[1]$ is therefore a degree $2$ PQ--manifold.
In \ccite{Roytenberg:0203110}, it is shown that conversely every degree $2$  
PQ--manifold stems from a Courant algebroid $V$ through the above construction.
}
\end{exa} 

\vspace{.0mm}

%The formulation of the theory of $L_\infty$--algebras and algebroids provided above
%is essentially finite dimensional. It is possible to extend it removing this restriction.
%A graded vector space $V$ of possibly infinite dimension equipped with a nilpotent degree $1$
%differential $Q$ is endowed with an $L_\infty$--algebra structure. This holds in particular 
%when $V$ is the classical BV algebra $A_{\mathscr{F}}$ of functionals on a field space $\mathscr{F}$
%of a field theory and $Q$ is the BV differential $\delta_{\mathscr{F}}$. 
%%Since $(A,\delta)$ is in particular a graded vector space equipped with a nilpotent differential,
%%$A$ has also a structure of $L_\infty$ algebra. See subsect. \cref{subsec:linf} for further details. 

%\vfil\eject  

\subsection{\textcolor{blue}{\sffamily Lie quasi--groupoid and L${}_\infty$--algebroids}}\label{subsec:simplex}

Lie quasi--groupoids are groupoid--like geometrical structures yielding
$L_\infty$--alge\-broids via Lie differentiation much as Lie group yield Lie algebras. 
For this reason, they have attracted much interest 
in recent years \ccite{Jurco:2016qwv,Severa:2006aa}. Lie quasi--groupoids are Kan
simplicial manifolds and so belong to the realm of simplicial theory. \pagebreak 
We give now a concise review of the theory of Lie quasi--groupoid focused on Lie differentiation
referring the interested reader to the above papers for a more comprehensive treatment. 
%We then explain how this can be related to the higher gauge theoretic framework developed in this paper. 
Some new results are also presented. 

A {\it simplicial set} $\mathscr{X}$ is a non negatively graded set equipped a collection of degree $-1$
{\it face maps } $f_{\mathscr{X} i}:\mathscr{X}\rightarrow \mathscr{X}$ and degree $1$ {\it degeneracy maps} 
$d_{\mathscr{X} i}:\mathscr{X}\rightarrow \mathscr{X}$ indexed by $i\geq 0$ and satisfying distinguished simplicial 
identities. More ex\-plicitly, a simplicial set $\mathscr{X}$ is a collection of sets $\mathscr{X}_p$ 
indexed by $p\geq 0$ together with maps
$f^p_{\mathscr{X} i}:\mathscr{X}_p\rightarrow \mathscr{X}_{p-1}$ with $p\geq i\geq 0$, $p>0$ 
and $d^p_{\mathscr{X} i}:\mathscr{X}_p\rightarrow \mathscr{X}_{p+1}$ with $p\geq i\geq 0$ 
obeying the relations 
\begin{align}
% \begin{gathered}
&f^{p-1}_{\mathscr{X} i}\circ f^{p}_{\mathscr{X} j}=f^{p-1}_{\mathscr{X} j-1}\circ f^{p}_{\mathscr{X} i}
~~\text{for}~~ i<j,~p>0,
\vphantom{\Big]}
\label{simplex1}
\\
&d^{p+1}_{\mathscr{X} i}\circ d^{p}_{\mathscr{X} j}=d^{p+1}_{\mathscr{X} j+1}\circ d^{p}_{\mathscr{X} i}
~~\text{for}~~ i\leq j,
\vphantom{\Big]}
\label{simplex2}
\\
&f^{p+1}_{\mathscr{X} i}\circ d^{p}_{\mathscr{X} j}=d^{p-1}_{\mathscr{X} j-1}\circ f^{p}_{\mathscr{X} i}
~~\text{for}~~ i<j,~p>0,
\vphantom{\Big]}
\label{simplex3}
\\
&f^{p+1}_{\mathscr{X} i}\circ d^{p}_{\mathscr{X} j}=d^{p-1}_{\mathscr{X} j}\circ f^{p}_{\mathscr{X} i-1}
~~\text{for}~~ i>j+1,~p>0,
\vphantom{\Big]}
\label{simplex4}
\\
&f^{p+1}_{\mathscr{X} i}\circ d^{p}_{\mathscr{X} i}=\id^p_{\mathscr{X}}
=f^{p+1}_{\mathscr{X} i+1}\circ d^{p}_{\mathscr{X} i}.
\vphantom{\Big]}
\label{simplex5}
\end{align}
%In these identities, the grade index $p$ in $f^p_{\mathscr{X} i}$ and $d^p_{\mathscr{X} i}$ is customarily 
%understood for the sake of brevity, when its value is clear from context. 
An element $s_p\in\mathscr{X}_p$ is called a $p$--{\it simplex} of $\mathscr{X}$. 
A {\it simplicial set morphism} $g$ of the simplicial sets $\mathscr{X}$, $\mathscr{Y}$
is a degree $0$ map $g:\mathscr{X}\rightarrow \mathscr{Y}$ 
compatible  with the face and degeneracy maps $f_{\mathscr{X} i}$, $d_{\mathscr{X} i}$, 
$f_{\mathscr{Y} i}$, $d_{\mathscr{Y} i}$ of $\mathscr{X}$, $\mathscr{Y}$. 
Stated explicitly, a simplicial morphism $g$ of the simplicial sets $\mathscr{X}$, $\mathscr{Y}$
is a collection of maps $g^p:\mathscr{X}_p\rightarrow \mathscr{Y}_p$ indexed by $p\geq 0$ such that 
\begin{align}
&g^{p-1}\circ f^{p}_{\mathscr{X}i}= f^{p}_{\mathscr{Y}i}\circ g^p ~~\text{for}~~ p\geq i\geq 0, ~p>0,
\vphantom{\Big]}
\label{simplex6}
\\
&g^{p+1} \circ d^{p}_{\mathscr{X}i}= d^{p}_{\mathscr{Y}i}\circ g^p ~~\text{for}~~ p\geq i\geq 0. 
\vphantom{\Big]}
\label{simplex7}
\end{align}
%Again, the grade index $p$ in $g^p$ is understood when its value is clear from context.  
Simplicial sets and morphisms form a category $\mathsans{sSet}$. 

The simplicial set category $\mathsans{sSet}$ is Cartesian 
closed. Thus, for any two simplicial sets $\mathscr{X}$, $\mathscr{Y}$, there is a simplicial
set $\mathsans{Hom}_{\mathsans{sSet}}(\mathscr{X},\mathscr{Y})$, unique up to unique
simplicial isomorphism, such that, for any simplicial set $\mathscr{Z}$, 
$\Hom(\mathscr{X}\times\mathscr{Z},\mathscr{Y})\simeq\Hom(\mathscr{Z},
\mathsans{Hom}_{\mathsans{sSet}}(\mathscr{X},\mathscr{Y}))$. 
$\mathsans{Hom}_{\mathsans{sSet}}(\mathscr{X},\mathscr{Y})$ is called the 
{\it internal} simplicial hom set of $\mathscr{X},\mathscr{Y}$. 

A simplicial \pagebreak set $\mathscr{K}$ is called {\it Kan} if every horn has a simplex that fills it, i. e. 
if for every  $p>0$ and $k$ with $p\geq k\geq 0$ and every collection of 
$p-1$--simplexes $s_0,\ldots,s_{k-1},s_{k+1},\ldots,s_p\in\mathscr{K}_{p-1}$ satisfying the Kan compatibility condition
\begin{equation}
f^{p-1}_{\mathscr{K} i}(s_j)=f^{p-1}_{\mathscr{K} j-1}(s_i)~~\text{for}~~ i<j,~i\neq k,~j\neq k, 
\label{simplex8}
\end{equation}
when $p>1$, there exists a $p$--simplex $s\in \mathscr{K}_p$ such that 
\begin{equation}
s_i=f^p_{\mathscr{K} i}(s)~~\text{for}~~i\neq k. 
\label{simplex9}
\end{equation}
If $s$ is unique for $p>q$, $\mathscr{K}$ is called a $q$ Kan simplicial set. 

\vspace{.33mm}
It is possible to define a simplicial object in a category $\mathsans{C}$ by replacing sets and maps by objects and morphisms
of $\mathsans{C}$ in the above definitions. One obtains in this way a simplicial category 
conventionally denoted by $\mathsans{sC}$. In particular, one can define simplicial groups, manifolds etc.
as well as their graded counterparts. All these are simplicial sets with extra structures. 
However, Cartesian closedness does not hold in general.

\vspace{.33mm}
A ($q$) Kan simplicial set $\mathscr{G}$ is called a ($q$) {\it quasi--groupoid}. Similarly, a
($q$) Kan simplicial manifold is called a ($q$) {\it Lie quasi--groupoid}. These are the kind 
of geometrical structures yielding infinitesimally $L_\infty$--algebras and algebroids
as we show next. 

\vspace{.33mm}
Let $\mathsans{N}\mathbb{R}[-1]$ denote the nerve of the pair groupoid of the $-1$--shifted real 
line $\mathbb{R}[-1]$. Concretely, $\mathsans{N}\mathbb{R}[-1]$ is the simplicial graded manifold 
which in degree $p$ features the graded manifold  \hphantom{xxxxxxxxxxxx} %the Cartesian product  
\begin{equation}
\mathsans{N}\mathbb{R}[-1]_p=\mathbb{R}^{p+1}[-1]
\label{simplex10}
\end{equation}
and whose face and degeneracy maps are given by 
\begin{align}
&f^p_{\mathsans{N}\mathbb{R}[-1]i}(\theta_0,\ldots,\theta_p)=(\theta_0,\ldots,\widehat{\theta_i},\ldots,\theta_p),
\vphantom{\Big]}
\label{simplex11}
\\
&d^p_{\mathsans{N}\mathbb{R}[-1]i}(\theta_0,\ldots,\theta_p)=(\theta_0,\ldots,\theta_i,\theta_i,\ldots,\theta_p).
\vphantom{\Big]}
\label{simplex12}
\end{align}
For $k\geq 0$, let $\mathsans{N}\mathbb{R}[-1](k)$ be the simplicial graded submanifold of $\mathsans{N}\mathbb{R}[-1]$ 
generated by $\mathbb{R}^k[-1]\times\{0\}$ in degree $k$. 
As $f^{k+1}_{\mathsans{N}\mathbb{R}[-1]k}(\mathbb{R}^{k+1}[-1]\times \{0\})
=\mathbb{R}^k[-1]\times \{0\}$, $\mathsans{N}\mathbb{R}[-1](k)\subset \mathsans{N}\mathbb{R}[-1](k+1)$. 
%Setting $\mathsans{N}\mathbb{R}[-1](\infty)=\mathsans{N}\mathbb{R}[-1]$, 
We have so a sequence $\mathsans{N}\mathbb{R}[-1](k)$ \pagebreak of simplicial graded manifolds 
indexed by $k\geq 0$ organized in a filtration 
\begin{equation}
\mathsans{N}\mathbb{R}[-1](0)\subset \mathsans{N}\mathbb{R}[-1](1)\subset 
\mathsans{N}\mathbb{R}[-1](2)\subset\ldots\subset \mathsans{N}\mathbb{R}[-1]. %
\label{simplex13}
\end{equation}

Let $Z$ be any graded manifold. With $Z$ there is associated the constant simp\-licial graded manifold $\mathscr{C}_Z$.
Concretely, $\mathscr{C}_Z$ is the simplicial graded manifold 
which in every degree $p$ exhibits the graded manifold $Z$ itself %\hphantom{xxxxxxxxxxxx} %the Cartesian product  
\begin{equation}
\mathscr{C}_{Zp}=Z
\label{simplex10/c}
\end{equation}
and whose face and degeneracy maps are all the identity map $\id_Z$
\begin{align}
&f^p_{\mathscr{C}_Zi}(z)=z,
\vphantom{\Big]}
\label{simplex11/c}
\\
&d^p_{\mathscr{C}_Zi}(z)=z. 
\vphantom{\Big]}
\label{simplex12/c}
\end{align}

%Consider the set $\Hom_{\mathsans{sgrMf}}(\mathsans{N}\mathbb{R}[-1]\times\mathscr{C}_Z,\mathscr{X})$ of simplicial 
%graded manifold morphisms from $\mathsans{N}\mathbb{R}[-1]\times\mathscr{C}_Z$ to some $\mathscr{X}$. 
%$\Hom_{\mathsans{sgrMf}}(\mathsans{N}\mathbb{R}[-1]\times\mathscr{C}_Z,\mathscr{X})$ can be 

For a graded manifold $Z$ and simplicial graded manifold $\mathscr{X}$, the 
simplicial graded manifold hom set $\Hom_{\mathsans{sgrMf}}(\mathsans{N}\mathbb{R}[-1]\times\mathscr{C}_Z,\mathscr{X})$
can be identified with a submanifold of the infinite dimensional manifold 
\begin{equation}
\mathrm{H}(Z,\mathscr{X})
=\ppp_{p=0}^\infty\Hom_{\mathsans{grMf}}(\mathsans{N}\mathbb{R}[-1]_p\times \mathscr{C}_{Zp},\mathscr{X}_p)
\label{simplex11i1}
\end{equation}
and so it is itself a manifold. 
Each of the factors in the right hand side of \ceqref{simplex11i1} can be written in terms of the internal hom manifold
$\mathsans{Hom}_{\mathsans{grMf}}(\mathsans{N}\mathbb{R}[-1]_p,\mathscr{X}_p)$ of $\mathsans{N}\mathbb{R}[-1]_p$, $\mathscr{X}_p$,
\begin{equation}
\Hom_{\mathsans{grMf}}(\mathsans{N}\mathbb{R}[-1]_p\times \mathscr{C}_{Zp},\mathscr{X}_p)
=\Hom_{\mathsans{grMf}}(\mathscr{C}_{Zp},
\mathsans{Hom}_{\mathsans{grMf}}(\mathsans{N}\mathbb{R}[-1]_p,\mathscr{X}_p))\label{simplex11i2}
\end{equation}
(cf. subsect. \cref{subsec:graman}). This suggests defining a {\it simplicial internal 
graded manifold hom set} $\mathsans{Hom}^*_{\mathsans{sgrMf}}(\mathsans{N}\mathbb{R}[-1],\mathscr{X})$
\footnote{$\vphantom{\dot{\dot{\dot{x}}}}$ \label{foot:intsimpl}
If the graded manifold category $\mathsans{grMf}$ is suitably enlarged to accommodate graded functional diffeological
spaces as explained in subsect. \cref{subsec:graman}, then one can construct the internal simplicial graded manifold hom 
manifold $\mathsans{Hom}_{\mathsans{sgrMf}}(\mathsans{N}\mathbb{R}[-1],\mathscr{X})$. This 
is generally distinct from %and should not be confused with 
the hom manifold defined here, which we call simplicial internal for distinctiveness. \vspace{-4mm}}:  
an internal simplicial morphism is defined as an ordinary 
morphism but with an internal rather than an ordinary graded manifold morphism 
$\mathsans{N}\mathbb{R}[-1]_p\rightarrow\mathscr{X}_p$ at degree $p$. 
$\mathsans{Hom}^*_{\mathsans{sgrMf}}(\mathsans{N}\mathbb{R}[-1],\mathscr{X})$
is identifiable with a graded submanifold of the infinite dimensional graded manifold \pagebreak 
\begin{equation}
\mathsans{H}(\mathscr{X})
=\ppp_{p=0}^\infty\mathsans{Hom}_{\mathsans{grMf}}(\mathsans{N}\mathbb{R}[-1]_p,\mathscr{X}_p) %\vphantom{\ul{\ul{g}}}
\label{simplex11i3}
\end{equation}
and so it is itself a graded manifold. 
%As it is appropriate, one has 
%\begin{equation}
%\Hom_{\mathsans{sgrMf}}(\mathsans{N}\mathbb{R}[-1]\times\mathscr{C}_Z,\mathscr{X})
%=\Hom_{\mathsans{sgrMf}}(\mathscr{C}_Z,\mathsans{Hom}^*_{\mathsans{sgrMf}}(\mathsans{N}\mathbb{R}[-1],\mathscr{X})).
%\label{simplex11i4}
%\end{equation} 
In this way, by the triviality of $\mathscr{C}_Z$, the study of 
$\Hom_{\mathsans{sgrMf}}(\mathsans{N}\mathbb{R}[-1]\times\mathscr{C}_Z,\mathscr{X})$
can be %effectively 
reduced to that of %the {\it simplicial internal hom manifold} 
$\mathsans{Hom}^*_{\mathsans{sgrMf}}(\mathsans{N}\mathbb{R}[-1],\mathscr{X})$.
A similar analysis clearly can be carried out also for the $\mathsans{N}\mathbb{R}[-1](k)$.
%when $\mathsans{N}\mathbb{R}[-1]$ is replaced by one of the $\mathsans{N}\mathbb{R}[-1](k)$

%\vspace{.33mm}
%We notice that, for any $k$, every simplicial internal graded manifold morphism 
%$g(k):\mathsans{N}\mathbb{R}[-1](k)\rightarrow \mathscr{X}$ 
%%into a simplicial manifold $\mathscr{X}$ 
%is determined uniquely by the restriction 
%%$g(k)_k\big|_{\mathbb{R}^k[-1]\times \{0\}}\times Z$ 
%of its component $g(k)_k$ to $\mathbb{R}^k[-1]\times \{0\}$.

%\vspace{.5mm} %%abcd 
Let $\mathscr{G}$ be a $q$ Lie quasi--groupoid viewed as a simplicial graded manifold.
The following theorem, originally  proven % stated andby \v Severa
in ref. \ccite{Severa:2006aa} and rederived %by a different method 
in ref. \ccite{Jurco:2016qwv}, holds. 
Let $g(k)\in\mathsans{Hom}^*_{\mathsans{sgrMf}}(\mathsans{N}\mathbb{R}[-1](k),\mathscr{G})$ 
be a simplicial internal graded manifold morphism. Then, $g(k)$ can be extended to a sequence 
$g(l)\in\mathsans{Hom}^*_{\mathsans{sgrMf}}(\mathsans{N}\mathbb{R}[-1](l),\mathscr{G})$ of internal 
simplicial morphisms indexed by $l\geq 0$ 
such that 
\begin{equation}
g(l)=g(m)\big|_{\mathsans{N}\mathbb{R}[-1](l)}
\label{simplex14}
\end{equation}
for $l\leq m$. Furthermore, if $k\geq q$, the sequence $g(l)$ is unique. The proof relies 
in an essential way 
on the property that $\mathscr{G}$ is Kan and uniquely Kan in degree $q$ or larger. 
%Using this result, it can be shown 
It follows from the theorem that every simplicial internal graded manifold morphism 
$g(q)\in\mathsans{Hom}^*_{\mathsans{sgrMf}}(\mathsans{N}\mathbb{R}[-1](q),\mathscr{G})$ 
determines a unique simplicial internal morphism 
$g\in\mathsans{Hom}^*_{\mathsans{sgrMf}}(\mathsans{N}\mathbb{R}[-1],\mathscr{G})$ such that %\hphantom{xxxxxxxxxx}
\begin{equation}
g(q)=g\big|_{\mathsans{N}\mathbb{R}[-1](q)}.
\label{simplex15}
\end{equation}
$g$ is completely defined in terms of the unique sequence of simplicial internal morphisms 
$g(l)\in\mathsans{Hom}^*_{\mathsans{sgrMf}}(\mathsans{N}\mathbb{R}[-1](l),\mathscr{G})$ extending $g(q)$ by 
\begin{equation}
g\big|_{\mathsans{N}\mathbb{R}[-1](l)}=g(l).
\label{simplex16}
\end{equation}
Each simplicial internal graded manifold morphism 
$g\in\mathsans{Hom}^*_{\mathsans{sgrMf}}(\mathsans{N}\mathbb{R}[-1],\mathscr{G})$,
conversely, is determined by the restriction %$g\big|_{\mathsans{N}\mathbb{R}[-1](q)}$ 
of $g$ to $\mathsans{N}\mathbb{R}[-1](q)$
and, consequently, by the restriction %$g^q\big|_{\mathbb{R}^q[-1]\times \{0\}}$ 
of the component $g^q$ of $g$ to $\mathbb{R}^q[-1]\times \{0\}$. The internal hom manifold
$\mathsans{Hom}_{\mathsans{grMf}}(\mathbb{R}^q[-1]\times \{0\},\mathscr{G}_q)$ 
of $\mathbb{R}^q[-1]\times \{0\}$, $\mathscr{G}_q$
is isomorphic to the to\-tal space of the graded vector bundle $T[1]^q\mathscr{G}_q\rightarrow \mathscr{G}_q$. 
So, the graded manifold 
$\mathsans{Hom}^*_{\mathsans{sgrMf}}(\mathsans{N}\mathbb{R}[-1],\mathscr{G})$
can be identified with a graded submanifold of this latter. 
%of internal sim\-plicial %graded manifold 
%morphisms $g:\mathsans{N}\mathbb{R}[-1]\rightarrow\mathscr{G}$. 

%\vspace{.5mm}
A point $\gamma\in\mathscr{G}_0$ defines a mapping $g(0)_0:\{0\} \rightarrow \mathscr{G}_0$ 
by setting $g(0)_0(0)=\gamma$ and therefore \pagebreak 
determines a unique simplicial internal graded manifold morphism 
$g(0)\in\mathsans{Hom}^*_{\mathsans{sgrMf}}(\mathsans{N}\mathbb{R}[-1](0),\mathscr{G})$. This can be 
extended non uniquely to a sequence $g(l)\in\mathsans{Hom}^*_{\mathsans{sgrMf}}(\mathsans{N}\mathbb{R}[-1](l),\mathscr{G})$
of simplicial internal morphisms which determines in turn a simplicial internal morphism  
$g\in\mathsans{Hom}^*_{\mathsans{sgrMf}}(\mathsans{N}\mathbb{R}[-1],\mathscr{G})$ via \ceqref{simplex16}. 
The non uniqueness of the sequence $g(l)$ entails that of the resulting morphism $g$.
In \ccite{Jurco:2016qwv}, it is shown that the simplicial morphisms $g$ so yielded %in this way 
are parametrized by the total space of the $1$--shift $\mathsans{Lie}\,\mathscr{G}[1]$ of the graded vector bundle 
%defined by 
\begin{equation}
\mathsans{Lie}\,\mathscr{G}=\ddd_{p=1}^{q}\nnn_{i=0}^{p-1}\,\varsigma^p_{\mathscr{G}}{}^*\ker f^p_{\mathscr{G}i*}[p-1]
\rightarrow \mathscr{G}_0,
\label{simplex17}
\end{equation}
where $\varsigma^p_{\mathscr{G}}=d^{p-1}_{\mathscr{G}0}\circ\ldots\circ d^0_{\mathscr{G}0}:\mathscr{G}_0\rightarrow \mathscr{G}_p$
with $\varsigma^0_{\mathscr{G}}=\id_{\mathscr{G}_0}$ %by convention 
and here and below $f_{*\ldots *|m}$ with $k$--fold 
$*$ denotes the $k$--fold tangent map of a map $f$ evaluated at a point $m$ of its domain. 
Notice that $\mathsans{Lie}\,\mathscr{G}\subset \ddd_{p=1}^{q}\,\varsigma^p_{\mathscr{G}}{}^*T[p-1]\mathscr{G}_p$.
%The problem we have to tackle next is to give 
Remarkably, a rather explicit expression  of this parametrization can be furnished. 
Its elaboration is a bit lengthy but in fact  completely algorithmic. One writes a general ansatz for the 
maps $g^p(\theta_0,\ldots,\theta_{p-1},0)$ of the form
\begin{equation}
g^p(\theta_0,\ldots,\theta_{p-1},0)=\gamma^p{}_{/\!\!\!\hskip.6pt 0}+\sss_{r=0}^{p-1}\sss_{0\leq i_0<\ldots<i_r\leq p-1}
\theta_{i_0}\cdots \theta_{i_r}\gamma^p{}_{i_0\ldots i_r},
\label{simplex17/0}
\end{equation}
where $\gamma^p{}_{/\!\!\!\hskip.6pt 0}$ is a degree $0$ and the $\gamma^p{}_{i_1\ldots i_r}$ are degree $r$ local functions on $\mathscr{G}_0,$ 
and derives all re\-lations $\gamma^p{}_{/\!\!\!\hskip.6pt 0}$ and the $\gamma^p{}_{i_1\ldots i_r}$ must obey in order the $g^p$ to be the 
components of a simplicial morphism $g\in\mathsans{Hom}^*_{\mathsans{sgrMf}}(\mathsans{N}\mathbb{R}[-1],\mathscr{G})$
starting from $g^0$. %We were able to compute $g^p$ up to $p=2$. The result of this calculation is shown next. 

$g^0$ is parametrized by a point $\gamma\in\mathscr{G}_0$ and a vector $\xi_0$ of the fiber 
$T[1]_{\varsigma^1_{\mathscr{G}}(\gamma)}\mathscr{G}_1$ of the vector bundle $\varsigma^1_{\mathscr{G}}{}^*T[1]\mathscr{G}_1$ 
satisfying
\begin{equation}
f^1_{\mathscr{G}0*|\varsigma^1_{\mathscr{G}}(\gamma)}(\xi_0)=0.
\label{simplex18}
\end{equation}
$\xi_0$ can hence be identified with a vector $\alpha_1$ of the fiber 
$\ker f^1_{\mathscr{G}0*|\varsigma^1_{\mathscr{G}}(\gamma)}[1]$ of the vector subbundle $\varsigma^1_{\mathscr{G}}{}^*\ker f^1_{\mathscr{G}0*}[1]
\subset\mathsans{Lie}\,\mathscr{G}[1]$. In terms of $\gamma$, $\xi_0$, $g^0$ reads 
\begin{equation}
g^0(\theta_0)=\varsigma^0_{\mathscr{G}}(\gamma)+\theta_0\rho_{\mathscr{G}\gamma}(\xi_0),
\label{simplex19}
\end{equation}
where $\rho_{\mathscr{G}\gamma}(\xi_0)$ is given by  \hphantom{xxxxxxxxxxxxxxxx}
\begin{equation}
\rho_{\mathscr{G}\gamma}(\xi_0)=f^1_{\mathscr{G}1*|\varsigma^1_{\mathscr{G}}(\gamma)}(\xi_0) \vphantom{\ul{\ul{\ul{\ul{x}}}}}
\label{simplex20}
\end{equation}
$g^1$ is parametrized by $\gamma$, $\xi_0$ and the degree $2$ component $\xi_1$ of a vector of the fiber 
$T[1]^2{}_{\varsigma^2_{\mathscr{G}}(\gamma)}\mathscr{G}_2$ of the vector bundle 
$\varsigma^2_{\mathscr{G}}{}^*T[1]^2\mathscr{G}_2$ satisfying
\begin{equation}
f^2_{\mathscr{G}i*|\varsigma^2_{\mathscr{G}}(\gamma)}(\xi_1)
+f^2_{\mathscr{G}i**|\varsigma^2_{\mathscr{G}}(\gamma)}
(d^1_{\mathscr{G}0*|\varsigma^1_{\mathscr{G}}(\gamma)}(\xi_0), d^1_{\mathscr{G}1*|\varsigma^1_{\mathscr{G}}(\gamma)}(\xi_0))=0 
\label{simplex21}
\end{equation}
for $i=0,1$, the lowest degree components of which are determined by $\xi_0$. 
The Kan nature of $\mathscr{G}$
ensures that \ceqref{simplex21} has a solution. The solution is unique up to the degree $2$ component of a vector 
$\alpha_2$ of the fiber $\bigcap_{i=0}^1\ker f^2_{\mathscr{G}i*|\varsigma^2_{\mathscr{G}}(\gamma)}[2]$ 
of the vector subbundle $\bigcap_{i=0}^1\varsigma^2_{\mathscr{G}}{}^*\ker f^2_{\mathscr{G}i*}[2]\subset \mathsans{Lie}\,\mathscr{G}[1]$. 
In terms of $\gamma$, $\xi_0$, $\xi_1$, $g^1$ %reads %
has the expansion %\pagebreak 
\begin{align}
&g^1(\theta_0,\theta_1)=\varsigma^1_{\mathscr{G}}(\gamma)-(\theta_1-\theta_0)\xi_0
\vphantom{\Big]}
\label{simplex22}
\\
&\hspace{2cm}+\theta_1d^0_{\mathscr{G}0*|\varsigma^0_{\mathscr{G}}(\gamma)}(\rho_{\mathscr{G}\gamma}(\xi_0))
+\theta_0\theta_1\big(\partial_{\mathscr{G}\gamma}\xi_1-\tfrac{1}{2}[\xi_0,\xi_0]_{\mathscr{G}\gamma}\big),
\vphantom{\Big]}
\nonumber
\end{align}
where $\partial_{\mathscr{G}\gamma}\xi_1$, $[\xi_0,\xi_0]_{\mathscr{G}\gamma}$ are given by
\begin{align}
&\partial_{\mathscr{G}\gamma}\xi_1=f^2_{\mathscr{G}2*|\varsigma^2_{\mathscr{G}}(\gamma)}(\xi_1),
\vphantom{\Big]}
\label{simplex23}
\\
&\tfrac{1}{2}[\xi_0,\xi_0]_{\mathscr{G}\gamma}=-f^2_{\mathscr{G}2**|\varsigma^2_{\mathscr{G}}(\gamma)}
(d^1_{\mathscr{G}0*|\varsigma^1_{\mathscr{G}}(\gamma)}(\xi_0), d^1_{\mathscr{G}1*|\varsigma^1_{\mathscr{G}}(\gamma)}(\xi_0)).
\vphantom{\Big]}
\label{simplex23/1}
\end{align}
$g^2$ is parametrized by $\gamma$, $\xi_0$, $\xi_1$ and 
the degree $3$ component $\xi_2$ of a vector of the fiber 
$T[1]^3{}_{\varsigma^3_{\mathscr{G}}(\gamma)}\mathscr{G}_3$ 
of the vector bundle $\varsigma^3_{\mathscr{G}}{}^*T[1]^3\mathscr{G}_3$ satisfying
\begin{align}
&\hphantom{\,+\,}f^3_{\mathscr{G}i*|\varsigma^3_{\mathscr{G}}(\gamma)}(\xi_2)
\vphantom{\Big]}
\label{simplex24}
\\
&+f^3_{\mathscr{G}i**|\varsigma^3_{\mathscr{G}}(\gamma)}
((d^2_{\mathscr{G}1}\circ d^1_{\mathscr{G}1})_{*|\varsigma^1_{\mathscr{G}}(\gamma)}(\xi_0), d^2_{\mathscr{G}0*|\varsigma^2_{\mathscr{G}}(\gamma)}(\xi_1))
\nonumber
\vphantom{\Big]}
\\
&-f^3_{\mathscr{G}i**|\varsigma^3_{\mathscr{G}}(\gamma)}
((d^2_{\mathscr{G}0}\circ d^1_{\mathscr{G}1})_{*|\varsigma^1_{\mathscr{G}}(\gamma)}(\xi_0), d^2_{\mathscr{G}1*|\varsigma^2_{\mathscr{G}}(\gamma)}(\xi_1))
\vphantom{\Big]}
\nonumber
\\
&+f^3_{\mathscr{G}i**|\varsigma^3_{\mathscr{G}}(\gamma)}
((d^2_{\mathscr{G}0}\circ d^1_{\mathscr{G}0})_{*|\varsigma^1_{\mathscr{G}}(\gamma)}(\xi_0), d^2_{\mathscr{G}2*|\varsigma^2_{\mathscr{G}}(\gamma)}(\xi_1))
\vphantom{\Big]}
\nonumber  
\\  
&+f^3_{\mathscr{G}i**|\varsigma^3_{\mathscr{G}}(\gamma)}
((d^2_{\mathscr{G}1}\circ d^1_{\mathscr{G}1})_{*|\varsigma^1_{\mathscr{G}}(\gamma)}(\xi_0), 
\vphantom{\Big]}
\nonumber
\\  
&\hspace{4.5cm}
d^2_{\mathscr{G}0**|\varsigma^2_{\mathscr{G}}(\gamma)}(d^1_{\mathscr{G}0*|\varsigma^1_{\mathscr{G}}(\gamma)}(\xi_0), 
d^1_{\mathscr{G}1*|\varsigma^1_{\mathscr{G}}(\gamma)}(\xi_0)))
\vphantom{\Big]}
\nonumber
\\
&-f^3_{\mathscr{G}i**|\varsigma^3_{\mathscr{G}}(\gamma)}
((d^2_{\mathscr{G}0}\circ d^1_{\mathscr{G}1})_{*|\varsigma^1_{\mathscr{G}}(\gamma)}(\xi_0), 
\vphantom{\Big]}
\nonumber
\\
&\hspace{4.5cm}
d^2_{\mathscr{G}1**|\varsigma^2_{\mathscr{G}}(\gamma)}(d^1_{\mathscr{G}0*|\varsigma^1_{\mathscr{G}}(\gamma)}(\xi_0), 
d^1_{\mathscr{G}1*|\varsigma^1_{\mathscr{G}}(\gamma)}(\xi_0)))
\vphantom{\Big]}
\nonumber
\\
&+f^3_{\mathscr{G}i**|\varsigma^3_{\mathscr{G}}(\gamma)}
((d^2_{\mathscr{G}0}\circ d^1_{\mathscr{G}0})_{*|\varsigma^1_{\mathscr{G}}(\gamma)}(\xi_0), 
\vphantom{f^f}
\nonumber
%\\  
\end{align}  
\begin{align}  
&\hspace{4.5cm}
d^2_{\mathscr{G}2**|\varsigma^2_{\mathscr{G}}(\gamma)}(d^1_{\mathscr{G}0*|\varsigma^1_{\mathscr{G}}(\gamma)}(\xi_0), 
d^1_{\mathscr{G}1*|\varsigma^1_{\mathscr{G}}(\gamma)}(\xi_0)))
\vphantom{\Big]}
\nonumber
\\
&-f^3_{\mathscr{G}i***|\varsigma^3_{\mathscr{G}}(\gamma)}
((d^2_{\mathscr{G}1}\circ d^1_{\mathscr{G}1})_{*|\varsigma^1_{\mathscr{G}}(\gamma)}(\xi_0), 
\nonumber
\\
&\hspace{4.4cm}
(d^2_{\mathscr{G}0}\circ d^1_{\mathscr{G}1})_{*|\varsigma^1_{\mathscr{G}}(\gamma)}(\xi_0), 
(d^2_{\mathscr{G}0}\circ d^1_{\mathscr{G}0})_{*|\varsigma^1_{\mathscr{G}}(\gamma)}(\xi_0))=0.
\vphantom{\Big]}
\nonumber
\end{align}  
for $i=0,1,2$, the lowest degree components of which are determined by $\xi_0$, $\xi_1$. Again, the Kan nature of 
$\mathscr{G}$ ensures that \ceqref{simplex23} has a solution unique up to the degree $3$
component $\alpha_3$ of a vector of the fiber $\bigcap_{i=0}^2\ker f^3_{\mathscr{G}i*|\varsigma^3_{\mathscr{G}}(\gamma)}[3]$
of the vector subbundle $\bigcap_{i=0}^2\varsigma^3_{\mathscr{G}}{}^*\ker f^3_{\mathscr{G}i*}[3]\subset \mathsans{Lie}\,\mathscr{G}[1]$. 
%Explicitly, i
In terms of $\gamma$, $\xi_0$, $\xi_1$, $\xi_2$, $g^2$ reads %takes the form
\begin{align}
&g^2(\theta_0,\theta_1,\theta_2)
=\varsigma^2_{\mathscr{G}}(\gamma)-(\theta_1-\theta_0)d^1_{\mathscr{G}1*|\varsigma^1_{\mathscr{G}}(\gamma)}(\xi_0)
-(\theta_2-\theta_1)d^1_{\mathscr{G}0*|\varsigma^1_{\mathscr{G}}(\gamma)}(\xi_0)
\vphantom{\Big]}
\label{simplex25}
\\
&%-(\theta_2-\theta_1)d^1_{\mathscr{G}0*|\varsigma^1_{\mathscr{G}}(\gamma)}(\xi_0)
+\theta_2(d^1_{\mathscr{G}0}\circ d^0_{\mathscr{G}0})_{*|\varsigma^0_{\mathscr{G}}(\gamma)}(\rho_{\mathscr{G}\gamma}(\xi_0))
+(\theta_1\theta_2- \theta_0\theta_2+\theta_0 \theta_1)\xi_1
%\vphantom{\ul{\ul{\ul{g}}}}
\vphantom{\Big]}
\nonumber
\\
&-(\theta_1\theta_2- \theta_0\theta_2)
\big[d^1_{\mathscr{G}1*|\varsigma^1_{\mathscr{G}}(\gamma)}
(\partial_{\mathscr{G}\gamma}\xi_1-\tfrac{1}{2}[\xi_0,\xi_0]_{\mathscr{G}\gamma})
+d^1_{\mathscr{G}1**|\varsigma^1_{\mathscr{G}}(\gamma)}(d^0_{\mathscr{G}0*}(\rho_{\mathscr{G}\gamma}(\xi_0)),\xi_0)\big]
\vphantom{\Big]}
\nonumber
\\
%&\hspace{.74cm}\times \big[d^1_{\mathscr{G}1*|\varsigma^1_{\mathscr{G}}(\gamma)}
%(\partial_{\mathscr{G}\gamma}\xi_1-\tfrac{1}{2}[\xi_0,\xi_0]_{\mathscr{G}\gamma})
%+d^1_{\mathscr{G}1**|\varsigma^1_{\mathscr{G}}(\gamma)}(\varsigma^1_{\mathscr{G}*}(\rho_{\mathscr{G}\gamma}(\xi_0)),\xi_0)\big]
%\vphantom{\Big]}
%\nonumber
%\\
&+\theta_1\theta_2\big[d^1_{\mathscr{G}0*|\varsigma^1_{\mathscr{G}}(\gamma)}
(\partial_{\mathscr{G}\gamma}\xi_1-\tfrac{1}{2}[\xi_0,\xi_0]_{\mathscr{G}\gamma})
+d^1_{\mathscr{G}0**|\varsigma^1_{\mathscr{G}}(\gamma)}(d^0_{\mathscr{G}0*}(\rho_{\mathscr{G}\gamma}(\xi_0)),\xi_0)\big]
\vphantom{\Big]}
\nonumber
\\
&+\theta_0 \theta_1\theta_2\big(\partial_{\mathscr{G}\gamma}\xi_2-[\xi_0,\xi_1]_{\mathscr{G}\gamma}
+\tfrac{1}{6}[\xi_0,\xi_0,\xi_0]_{\mathscr{G}\gamma}\big),
\vphantom{\Big]}
\nonumber
\end{align}
where $\partial_{\mathscr{G}\gamma}\xi_1$, $[\xi_0,\xi_1]_{\mathscr{G}\gamma}$, $[\xi_0,\xi_0,\xi_0]_{\mathscr{G}\gamma}$ are given by
\begin{align}
&\partial_{\mathscr{G}\gamma}\xi_2=f^3_{\mathscr{G}3*|\varsigma^3_{\mathscr{G}}(\gamma)}(\xi_2),
\vphantom{\Big]}
\label{simplex26}
\\
&[\xi_0,\xi_1]_{\mathscr{G}\gamma}=-f^3_{\mathscr{G}3**|\varsigma^3_{\mathscr{G}}(\gamma)}
((d^2_{\mathscr{G}1}\circ d^1_{\mathscr{G}1})_{*|\varsigma^1_{\mathscr{G}}(\gamma)}(\xi_0), 
d^2_{\mathscr{G}0*|\varsigma^2_{\mathscr{G}}(\gamma)}(\xi_1))
\label{simplex27}
\vphantom{\Big]}
\\
&\hphantom{[\xi_0,\xi_1]_{\mathscr{G}\gamma}=}
+f^3_{\mathscr{G}3**|\varsigma^3_{\mathscr{G}}(\gamma)}
((d^2_{\mathscr{G}0}\circ d^1_{\mathscr{G}1})_{*|\varsigma^1_{\mathscr{G}}(\gamma)}(\xi_0), 
d^2_{\mathscr{G}1*|\varsigma^2_{\mathscr{G}}(\gamma)}(\xi_1))
\vphantom{\Big]}
\nonumber
\\
&\hphantom{[\xi_0,\xi_1]_{\mathscr{G}\gamma}=}
-f^3_{\mathscr{G}3**|\varsigma^3_{\mathscr{G}}(\gamma)}
((d^2_{\mathscr{G}0}\circ d^1_{\mathscr{G}0})_{*|\varsigma^1_{\mathscr{G}}(\gamma)}(\xi_0), 
d^2_{\mathscr{G}2*|\varsigma^2_{\mathscr{G}}(\gamma)}(\xi_1)),
\vphantom{\Big]}
\nonumber
\\
&\tfrac{1}{6}[\xi_0,\xi_0,\xi_0]_{\mathscr{G}\gamma}
\vphantom{\Big]}
\label{simplex28}
\\
&=+f^3_{\mathscr{G}3**|\varsigma^3_{\mathscr{G}}(\gamma)}
((d^2_{\mathscr{G}1}\circ d^1_{\mathscr{G}1})_{*|\varsigma^1_{\mathscr{G}}(\gamma)}(\xi_0), 
\vphantom{\Big]}
\nonumber
\\
&\hphantom{=\,}\hspace{4.5cm}
d^2_{\mathscr{G}0**|\varsigma^2_{\mathscr{G}}(\gamma)}(d^1_{\mathscr{G}0*|\varsigma^1_{\mathscr{G}}(\gamma)}(\xi_0), 
d^1_{\mathscr{G}1*|\varsigma^1_{\mathscr{G}}(\gamma)}(\xi_0)))
\vphantom{\Big]}
\nonumber
\\
&\hphantom{=\,}
-f^3_{\mathscr{G}3**|\varsigma^3_{\mathscr{G}}(\gamma)}
((d^2_{\mathscr{G}0}\circ d^1_{\mathscr{G}1})_{*|\varsigma^1_{\mathscr{G}}(\gamma)}(\xi_0), 
\vphantom{\Big]}
\nonumber
\\
&\hphantom{=\,}\hspace{4.5cm}
d^2_{\mathscr{G}1**|\varsigma^2_{\mathscr{G}}(\gamma)}(d^1_{\mathscr{G}0*|\varsigma^1_{\mathscr{G}}(\gamma)}(\xi_0), 
d^1_{\mathscr{G}1*|\varsigma^1_{\mathscr{G}}(\gamma)}(\xi_0)))
\vphantom{\Big]}
\nonumber
\\
&\hphantom{=\,}
+f^3_{\mathscr{G}3**|\varsigma^3_{\mathscr{G}}(\gamma)}
((d^2_{\mathscr{G}0}\circ d^1_{\mathscr{G}0})_{*|\varsigma^1_{\mathscr{G}}(\gamma)}(\xi_0), 
\vphantom{\Big]}
\nonumber
\\
&\hphantom{=\,}\hspace{4.5cm}
d^2_{\mathscr{G}2**|\varsigma^2_{\mathscr{G}}(\gamma)}(d^1_{\mathscr{G}0*|\varsigma^1_{\mathscr{G}}(\gamma)}(\xi_0), 
d^1_{\mathscr{G}1*|\varsigma^1_{\mathscr{G}}(\gamma)}(\xi_0)))
\vphantom{f^f}
\nonumber
%\\  
\end{align}  
\begin{align}  
&\hphantom{=\,}-f^3_{\mathscr{G}3***|\varsigma^3_{\mathscr{G}}(\gamma)}
((d^2_{\mathscr{G}1}\circ d^1_{\mathscr{G}1})_{*|\varsigma^1_{\mathscr{G}}(\gamma)}(\xi_0), 
\nonumber
\\
&\hspace{4.6cm}
(d^2_{\mathscr{G}0}\circ d^1_{\mathscr{G}1})_{*|\varsigma^1_{\mathscr{G}}(\gamma)}(\xi_0), 
(d^2_{\mathscr{G}0}\circ d^1_{\mathscr{G}0})_{*|\varsigma^1_{\mathscr{G}}(\gamma)}(\xi_0)).
\vphantom{\Big]}
\nonumber
\end{align}
The components $g^p$ of $g$ with $p\geq 3$ can be obtained in the same way, although the amount of computation 
required increases very rapidly with $p$. 

Via the degeneracy maps of $\mathscr{G}$, the data $\gamma$ and $\xi_p$ with $0\leq p\leq q-1$ 
coordinatize a graded submanifold of the graded vector bundle $T[1]^q\mathscr{G}_q\rightarrow\mathscr{G}_q$. 
Conversely, via the face maps of $\mathscr{G}$, 
a point of the submanifold gives rise to a set of data $\gamma$ and $\xi_p$. At the same time 
the data $\gamma$ and $\xi_p$ coordinatize the vector bundle $\mathsans{Lie}\,\mathscr{G}[1]$.

In ref. \ccite{Jurco:2016qwv}, it is shown $\mathsans{Lie}\,\mathscr{G}$ is an $L_\infty$--algebroid.
It is not difficult to see the reason why. The parametrization of the simplicial internal graded manifold hom manifold 
$\mathsans{Hom}^*_{\mathsans{sgrMf}}(\mathsans{N}\mathbb{R}[-1],\mathscr{G})$
described above can be thought of as defining %defines 
a higher exponential map $\exp_{\mathscr{G}}:\mathsans{Lie}\,\mathscr{G}[1]
\rightarrow \mathsans{Hom}^*_{\mathsans{sgrMf}}(\mathsans{N}\mathbb{R}[-1],\mathscr{G})$, 
which with the data $\gamma$ and $\xi_p$ %with $0\leq p\leq q-1$ 
associates the simplicial internal morphism 
\begin{equation}
g=\exp_{\mathscr{G}}(\gamma,\xi_0,\ldots,\xi_{q-1}).
\label{simplex29}
\end{equation}
The Chevalley--Eilenberg differential 
$Q_{\mathsans{Lie}\,\mathscr{G}}$ of $\mathsans{Lie}\,\mathscr{G}$ is defined implicitly by
\begin{equation}
Q_{\mathsans{Lie}\,\mathscr{G}}\exp_{\mathscr{G}}(\gamma,\xi_0,\ldots,\xi_{q-1})=D\exp_{\mathscr{G}}(\gamma,\xi_0,\ldots,\xi_{q-1}),
\label{simplex30}
\end{equation}
where $D$ is the formal operator acting as 
\begin{equation}
Du_p(\theta_0,\ldots,\theta_p)
=\frac{d}{d\epsilon}u_p(\theta_0+\epsilon,\ldots,\theta_p+\epsilon).
\label{simplex31}
\end{equation}
on a simplicial morphism $u\in\mathsans{Hom}^*_{\mathsans{sgrMf}}(\mathsans{N}\mathbb{R}[-1],\mathscr{G})$. 
As $D^2=0$, $Q_{\mathsans{Lie}\,\mathscr{G}}$ is nilpotent by construction. 
It is straightforward to verify that $Q_{\mathsans{Lie}\,\mathscr{G}}$ 
is given by an expression of the form \ceqref{qexpl}, with $\rho_{\mathscr{G}}$, $\partial_{\mathscr{G}}$,
$[-,-]_{\mathscr{G}}$, $[-,-,-]_{\mathscr{G}}$, ... being precisely the anchor, the boundary and the brackets 
of an $L_\infty$--algebroid structure on  $\mathsans{Lie}\,\mathscr{G}$. 
In this sense, $\mathsans{Lie}\,\mathscr{G}$ describes $\mathscr{G}$ infinitesimally
and is the quasi Lie groupoid analog of the Lie algebra of a Lie group. The classic Lie case is recovered by 
considering the 1 Lie quasi--groupoid $\mathscr{G}$ that is the nerve of the single object delooping groupoid
$\mathsans{B}G$ of an ordinary Lie group $G$. 

In ref. \ccite{Jurco:2016qwv} \pagebreak it is claimed that higher gauge symmetry can be described mathematically 
in terms of simplicial homotopies of simplicial maps. For this reason, we briefly review this topic next.
A {\it simplicial homotopy} of two simplicial sets $\mathscr{X}$, $\mathscr{Y}$ is simply 
a simplicial set morphism $h\in \Hom_{\mathsans{sSet}}(\mathscr{X}\times\varDelta^1,\mathscr{Y})$, 
where $\varDelta^1$ is the so--called standard simplicial $1$--simplex, a simplicial set that is a 
combinatorial analogue of the geometric $1$--simplex. 
There are source and target maps $\sigma,\tau:\Hom_{\mathsans{sSet}}(\mathscr{X}\times\varDelta^1,\mathscr{Y})
\rightarrow \Hom_{\mathsans{sSet}}(\mathscr{X},\mathscr{Y})$ so that a homotopy $h$ of $\mathscr{X}$, $\mathscr{Y}$ 
connects two simplicial maps $\sigma(h), \tau(h)\in  \Hom_{\mathsans{sSet}}(\mathscr{X},\mathscr{Y})$, 
$h:\sigma(h)\Rightarrow \tau(h)$. 

%For any two simplicial sets $\mathscr{D}$, $\mathscr{Y}$, there is a simplicial set
%$\mathsans{Hom}_{\mathsans{sSet}}(\mathscr{D},\mathscr{Y})$, called {\it internal simplicial hom set} 
%of $\mathscr{D}$, $\mathscr{Y}$ with the following property. For any simplicial set
%$\mathscr{X}$, $\Hom_{\mathsans{sSet}}(\mathscr{X}\times \mathscr{D},\mathscr{Y})
%=\Hom_{\mathsans{sSet}}(\mathscr{X},\mathsans{Hom}_{\mathsans{sSet}}(\mathscr{D},\mathscr{Y}))$.
As $\Hom_{\mathsans{sSet}}(\mathscr{X}\times\varDelta^1,\mathscr{Y})=
\Hom_{\mathsans{sSet}}(\mathscr{X},\mathsans{Hom}_{\mathsans{sSet}}(\varDelta^1,\mathscr{Y}))$
for any pair of simplicial sets $\mathscr{X}$, $\mathscr{Y}$, 
the study of simplicial homotopies can be reduced to that of the internal simplicial hom set 
$\mathsans{Hom}_{\mathsans{sSet}}(\varDelta^1,\mathscr{Y})$. As we shall explain shortly, 
$\mathsans{Hom}_{\mathsans{sSet}}(\varDelta^1,\mathscr{Y})$ admits a rather explicit description. 
Furthermore, it has an important property: when $\mathscr{G}$ is a quasi--groupoid,
$\mathsans{Hom}_{\mathsans{sSet}}(\varDelta^1,\mathscr{G})$ also is. 

Let $\mathscr{G}$ be a quasi--groupoid. 
For $p\ge 0$, an element $w\in\mathsans{Hom}_{\mathsans{sSet}}(\varDelta^1,\mathscr{G})_p$
can be encoded in a $p+1$-tuple $w_{i}\in\mathscr{G}_{p+1}$, $0\le i\le p$, satisfying 
\begin{equation}
f^{p+1}_{\mathscr{G}i}(w_{i})=f^{p+1}_{\mathscr{G}i}(w_{i-1})\qquad \text{for $0<i\le p$}.
\label{homo}
\end{equation}
The face and degeneracy maps of $\mathsans{Hom}_{\mathsans{sSet}}(\varDelta^1,\mathscr{G})$ are defined as
\begin{align}
f^p_{\mathsans{Hom}_{\mathsans{sSet}}(\varDelta^1,\mathscr{G})i}(w)_j=&\,f^{p+1}_{\mathscr{G}i}(w_{j+1}) &&\text{for $i\le j$,}
\vphantom{\Big]}
\label{homface}
\\
&\,f^{p+1}_{\mathscr{G}i+1}(w_{j})&&\text{for $i>j$, with $0\le i\le p$, $0\le j\le p-1$,}
\vphantom{\Big]}
\nonumber
\\
d^p_{\mathsans{Hom}_{\mathsans{sSet}}(\varDelta^1,\mathscr{G})i}(w)_j=&\,d^{p+1}_{\mathscr{G}i}(w_{j-1})&&\text{for $i< j$},
\vphantom{\Big]}
\label{homdeg}
\\
&\,d^{p+1}_{\mathscr{G}i+1}(w_{j})&&\text{for $i\ge j$, with $0\le i\le p$, $0\le j\le p+1$,}
\vphantom{\Big]}
\nonumber
\end{align}
where $w\in\mathsans{Hom}_{\mathsans{sSet}}(\varDelta^1,\mathscr{G})_p$. 

{\it Source} and {\it target} simplicial maps $\sigma,\tau
\in\Hom_{\mathsans{sSet}}(\mathsans{Hom}_{\mathsans{sSet}}(\varDelta^1,\mathscr{G}),
\mathscr{G})$ exist. The maps $\sigma^p,\tau^p\in
\Hom_{\mathsans{Mf}}(\mathsans{Hom}_{\mathsans{sSet}}(\varDelta^1,\mathscr{G})_p,\mathscr{G}_p)$ 
describing $\sigma$, $\tau$ at degree $p$ are %given by 
\begin{align}
&\sigma^p(w)=f^{p+1}_{\mathscr{G}0}(w_{0}),
\vphantom{\Big]}
\label{homsour}
\\
&\tau^p(w)=f^{p+1}_{\mathscr{G}p+1}(w_{p})
\vphantom{\Big]}
\label{homtarg}
\end{align}
with $w\in\mathsans{Hom}_{\mathsans{sSet}}(\varDelta^1,\mathscr{G})_p$. 

The above consideration extend essentially unchanged to simplicial objects in relevant categories 
such as $\mathsans{Mf}$ and $\mathsans{grMf}$. 

It turns out that, when $\mathscr{G}$ is a Lie quasi--groupoid,
$\mathsans{Hom}_{\mathsans{sSet}}(\varDelta^1,\mathscr{G})$ is as well. 
However, when $\mathscr{G}$ is a $q$ Lie quasi--groupoid for some $q$, 
$\mathsans{Hom}_{\mathsans{sSet}}(\varDelta^1,\mathscr{G})$ is not in general a $q^1$ Lie quasi--groupoid 
for any finite $q^1$.

Let $\mathscr{G}$ be a $q$ Lie quasi--groupoid. 
Then, the exponential parametrization 
of the simplicial internal graded manifold hom manifold $\mathsans{Hom}_{\mathsans{sSet}}(\varDelta^1,\mathscr{K})$
associated with an $r$ Lie quasi--groupoid $\mathscr{K}$ 
we described earlier in this subsection can be extended to $\mathsans{Hom}^*_{\mathsans{sgrMf}}(\mathsans{N}\mathbb{R}[-1],
\mathsans{Hom}_{\mathsans{sSet}}(\varDelta^1,\mathscr{G}))$
upon replacing $\mathscr{K}$ with $\mathsans{Hom}_{\mathsans{sSet}}(\varDelta^1,\mathscr{G})$. 
Since $\mathsans{Hom}_{\mathsans{sSet}}(\varDelta^1,\mathscr{G})$ may not 
be a finite $q^1$ Lie quasi--groupoid, the parametrization is generally infinte dimensional. 
Let us spell this  out in some detail. 

By \ceqref{simplex17}, $\mathsans{Hom}^*_{\mathsans{sgrMf}}(\mathsans{N}\mathbb{R}[-1],
\mathsans{Hom}_{\mathsans{sSet}}(\varDelta^1,\mathscr{G}))$ is parametrized  
by the total space of the $1$--shift $\mathsans{Lie}\,\mathsans{Hom}_{\mathsans{sSet}}(\varDelta^1,\mathscr{G})[1]$ 
of the graded vector bundle %defined by 
\begin{equation}
\mathsans{Lie}\,\mathsans{Hom}_{\mathsans{sSet}}(\varDelta^1,\mathscr{G})
=\ddd_{p=1}^{\infty}\nnn_{0\le i,j\le p, i\neq j,j+1}\,\varsigma^{p+1}_{\mathscr{G}l_{pj}}{}^*\ker f^{p+1}_{\mathscr{G}i*}
[p-1]\rightarrow \mathscr{G}_1.
\label{htpsimplex17}
%&\mathsans{Lie}\,\mathsans{Hom}_{\mathsans{sSet}}(\varDelta^1,\mathscr{G})
%=\ddd_{p=1}^{\infty}\Big(\nnn_{0\le i,j\le p-1, j<i}\,\varsigma^{p+1}_{\mathscr{G}s_j}{}^*\ker f^{p+1}_{\mathscr{G}i+1*}
%\vphantom{\Big]}
%\label{htpsimplex17}
%\\
%&\hspace{5cm}
%\cap \nnn_{0\le i,j\le p-1, j\ge i}\,\varsigma^{p+1}_{\mathscr{G}s_{j+1}}{}^*\ker f^{p+1}_{\mathscr{G}i*}\Big)[p-1]\rightarrow \mathscr{G}_1.
%\vphantom{\Big]}
%\nonumber
\end{equation}
In the above expression and the following, 
$\varsigma^p_{\mathscr{G}i_{p-1}\cdots i_1}=d^{p-1}_{\mathscr{G}i_{p-1}}\circ\cdots \circ d^1_{\mathscr{G}i_1}:
\mathscr{G}_1\rightarrow\mathscr{G}_p$ with $\varsigma^1_{\mathscr{G}/\!\!\!\hskip.6pt 0}=\id_{\mathscr{G}_1}$ by convention.
$l_{pj}$ is a notational shorthand for the index string $0,\ldots,0,1,\ldots,1$ with $0$, $1$ 
occurring $j$ and $p-j$ times, respectively. 

Let $h\in\mathsans{Hom}^*_{\mathsans{sgrMf}}(\mathsans{N}\mathbb{R}[-1],\mathsans{Hom}_{\mathsans{sSet}}(\varDelta^1,\mathscr{G}))$
be a simplicial internal graded manifold morphism. 
Then, by our earlier analysis, %the component 
$h^0$ is parametrized by a point $\beta\in\mathscr{G}_1$ and two
vectors $\chi^1{}_{0}$, $\chi^1{}_{1}$ lying in  the fibers $T[1]_{\varsigma^2_{\mathscr{G}1}(\beta)}\mathscr{G}_2$, 
$T[1]_{\varsigma^2_{\mathscr{G}0}(\beta)}\mathscr{G}_2$ 
of the vector bundles $\varsigma^2_{\mathscr{G}1}{}^*T[1]\mathscr{G}_2$, $\varsigma^2_{\mathscr{G}0}{}^*T[1]\mathscr{G}_2$, 
respectively,  satisfying by \ceqref{homo} %the relation
\begin{equation}
f^2_{\mathscr{G}1*|\varsigma^2_{\mathscr{G}0}(\beta)}(\chi^1{}_{1})=f^2_{\mathscr{G}1*|\varsigma^2_{\mathscr{G}1}(\beta)}(\chi^1{}_{0}). \vphantom{\bigg]}
\label{htpsimplex18}
\end{equation}
Owing to \ceqref{simplex18}, using \ceqref{homface}, \ceqref{homdeg}, we have further
\begin{equation}
f^2_{\mathscr{G}0*|\varsigma^2_{\mathscr{G}0}(\beta)}(\chi^1{}_{1})=0. \vphantom{\bigg]}
\label{htpsimplex19}
\end{equation}
From \ceqref{simplex19}, \ceqref{simplex20}, using again \ceqref{homface}, \ceqref{homdeg}, we find that 
\begin{equation}
h^0{}_{0}(\theta_0)=\varsigma^1_{\mathscr{G}/\!\!\!\hskip.6pt 0}(\beta)+\theta_0f^2_{\mathscr{G}2*|\varsigma^2_{\mathscr{G}1}(\beta)}(\chi^1{}_{0}), 
\label{htpsimplex20}
\end{equation}
$h^1$ is parametrized by $\beta$, $\chi^1{}_{0}$, $\chi^1{}_{1}$ and the degree $2$ components 
$\chi^2{}_{0}$, $\chi^2{}_{1}$, $\chi^2{}_{2}$ of three vectors belonging to the fibers 
$T[1]^2{}_{\varsigma^3_{\mathscr{G}11}(\beta)}\mathscr{G}_3$, $T[1]^2{}_{\varsigma^3_{\mathscr{G}01}(\beta)}\mathscr{G}_3$, 
$T[1]^2{}_{\varsigma^3_{\mathscr{G}00}(\beta)}\mathscr{G}_3$ of the vector bundles 
$\varsigma^3_{\mathscr{G}11}{}^*T[1]^2\mathscr{G}_3$, $\varsigma^3_{\mathscr{G}01}{}^*T[1]^2\mathscr{G}_3$, 
$\varsigma^3_{\mathscr{G}00}{}^*T[1]^2\mathscr{G}_3$, respectively, satisfying on account of \ceqref{homo}
%\begin{align}
%&f^3_{\mathscr{G}1*|\varsigma^3_{\mathscr{G}01}(\beta)}(\chi^2{}_{1})
%+f^3_{\mathscr{G}1**|\varsigma^3_{\mathscr{G}01}(\beta)}
%(d^2_{\mathscr{G}0*|\varsigma^2_{\mathscr{G}1}(\beta)}(\chi^1{}_{0}),d^2_{\mathscr{G}2*|\varsigma^2_{\mathscr{G}0}(\beta)}(\chi^1{}_{1}))
%\vphantom{\Big]}
%\label{htpsimplex}
%\\
%&\qquad=f^3_{\mathscr{G}1*|\varsigma^3_{\mathscr{G}11}(\beta)}(\chi^2{}_{0})
%+f^3_{\mathscr{G}1**|\varsigma^3_{\mathscr{G}11}(\beta)}
%(d^2_{\mathscr{G}1*|\varsigma^2_{\mathscr{G}1}(\beta)}(\chi^1{}_{0}),d^2_{\mathscr{G}2*|\varsigma^2_{\mathscr{G}1}(\beta)}(\chi^1{}_{0})),
%\vphantom{\Big]}
%\nonumber
%\\
%&f^3_{\mathscr{G}2*|\varsigma^3_{\mathscr{G}00}(\beta)}(\chi^2{}_{2})
%+f^3_{\mathscr{G}2**|\varsigma^3_{\mathscr{G}00}(\beta)}
%(d^2_{\mathscr{G}0*|\varsigma^2_{\mathscr{G}0}(\beta)}(\chi^1{}_{1}),d^2_{\mathscr{G}1*|\varsigma^2_{\mathscr{G}0}(\beta)}(\chi^1{}_{1}))
%\vphantom{\Big]}
%\label{htpsimplex}
%\\
%&\qquad=f^3_{\mathscr{G}2*|\varsigma^3_{\mathscr{G}01}(\beta)}(\chi^2{}_{1})
%+f^3_{\mathscr{G}2**|\varsigma^3_{\mathscr{G}01}(\beta)}
%(d^2_{\mathscr{G}0*|\varsigma^2_{\mathscr{G}1}(\beta)}(\chi^1{}_{0}),d^2_{\mathscr{G}2*|\varsigma^2_{\mathscr{G}0}(\beta)}(\chi^1{}_{1})).
%\vphantom{\Big]}
%\nonumber
%\end{align}
\begin{align}
&f^3_{\mathscr{G}i*|\varsigma^3_{\mathscr{G}jk}(\beta)}(\chi^2{}_{2-j-k})
\vphantom{\Big]}
\label{htpsimplex21}
\\
&\hspace{1.5cm}+f^3_{\mathscr{G}i**|\varsigma^3_{\mathscr{G}jk}(\beta)}
(d^2_{\mathscr{G}j*|\varsigma^2_{\mathscr{G}k}(\beta)}(\chi^1{}_{1-k}),d^2_{\mathscr{G}k+1*|\varsigma^2_{\mathscr{G}j}(\beta)}(\chi^1{}_{1-j}))
\vphantom{\Big]}
\nonumber
\\
=\,&f^3_{\mathscr{G}i*|\varsigma^3_{\mathscr{G}kj+1}(\beta)}(\chi^2{}_{1-j-k})
\vphantom{\Big]}
\nonumber 
\\
&\hspace{1.5cm}+f^3_{\mathscr{G}i**|\varsigma^3_{\mathscr{G}kj+1}(\beta)}
(d^2_{\mathscr{G}k*|\varsigma^2_{\mathscr{G}j+1}(\beta)}(\chi^1{}_{-j}),d^2_{\mathscr{G}j+2*|\varsigma^2_{\mathscr{G}k}(\beta)}(\chi^1{}_{1-k}))
\vphantom{\Big]}
\nonumber
\end{align}
with $(i,j,k)=(1,0,1),(2,0,0)$. By \ceqref{simplex21}, on account of \ceqref{homface}, \ceqref{homdeg}, we have further
%\begin{align}
%&f^3_{\mathscr{G}0*|\varsigma^3_{\mathscr{G}01}(\beta)}(\chi^2{}_{1})
%+f^3_{\mathscr{G}0**|\varsigma^3_{\mathscr{G}01}(\beta)}
%(d^2_{\mathscr{G}0*|\varsigma^2_{\mathscr{G}1}(\beta)}(\chi^1{}_{0}),d^2_{\mathscr{G}2*|\varsigma^2_{\mathscr{G}0}(\beta)}(\chi^1{}_{1}))=0,
%\vphantom{\Big]}
%\label{htpsimplex}
%\\
%&f^3_{\mathscr{G}0*|\varsigma^3_{\mathscr{G}00}(\beta)}(\chi^2{}_{2})
%+f^3_{\mathscr{G}0**|\varsigma^3_{\mathscr{G}00}(\beta)}
%(d^2_{\mathscr{G}0*|\varsigma^2_{\mathscr{G}0}(\beta)}(\chi^1{}_{1}),d^2_{\mathscr{G}1*|\varsigma^2_{\mathscr{G}0}(\beta)}(\chi^1{}_{1}))=0,
%\vphantom{\Big]}
%\label{htpsimplex}
%\\
%&f^3_{\mathscr{G}2*|\varsigma^3_{\mathscr{G}11}(\beta)}(\chi^2{}_{0})
%+f^3_{\mathscr{G}2**|\varsigma^3_{\mathscr{G}11}(\beta)}
%(d^2_{\mathscr{G}1*|\varsigma^2_{\mathscr{G}1}(\beta)}(\chi^1{}_{0}),d^2_{\mathscr{G}2*|\varsigma^2_{\mathscr{G}1}(\beta)}(\chi^1{}_{0}))=0,
%\vphantom{\Big]}
%\label{htpsimplex}
%\\
%&f^3_{\mathscr{G}1*|\varsigma^3_{\mathscr{G}00}(\beta)}(\chi^2{}_{2})
%+f^3_{\mathscr{G}1**|\varsigma^3_{\mathscr{G}00}(\beta)}
%(d^2_{\mathscr{G}0*|\varsigma^2_{\mathscr{G}0}(\beta)}(\chi^1{}_{1}),d^2_{\mathscr{G}1*|\varsigma^2_{\mathscr{G}0}(\beta)}(\chi^1{}_{1}))=0.
%\vphantom{\Big]}
%\label{htpsimplex}
%\end{align}
\begin{align}
&f^3_{\mathscr{G}i*|\varsigma^3_{\mathscr{G}jk}(\beta)}(\chi^2{}_{2-j-k})
\vphantom{\Big]}
\label{htpsimplex22}
\\
&\hspace{1cm}+f^3_{\mathscr{G}i**|\varsigma^3_{\mathscr{G}jk}(\beta)}
(d^2_{\mathscr{G}j*|\varsigma^2_{\mathscr{G}k}(\beta)}(\chi^1{}_{1-k}),d^2_{\mathscr{G}k+1*|\varsigma^2_{\mathscr{G}j}(\beta)}(\chi^1{}_{1-j}))=0
\vphantom{\Big]}
\nonumber
\end{align}
with $(i,j,k)=(0,0,0),(0,0,1),(1,0,0),(2,1,1)$. By \ceqref{simplex22}--\ceqref{simplex24}, employing 
\ceqref{homface}, \ceqref{homdeg}, we obtain then
%\begin{align}
%&h^1{}_{0}(\theta_0,\theta_1)=\varsigma^2_{\mathscr{G}1}(\beta)-(\theta_1-\theta_0)\chi^1{}_{0}
%+\theta_1(d^1_{\mathscr{G}1} \circ %*|\varsigma^1_{\mathscr{G}/\!\!\!\hskip.6pt 0}(\beta)}(
%f^2_{\mathscr{G}2})_{*|\varsigma^2_{\mathscr{G}1}(\beta)}(\chi^1{}_{0})
%\vphantom{\Big]}
%\label{htpsimplex}
%\\
%&\hspace{.5cm}+\theta_0\theta_1[f^3_{\mathscr{G}3*|\varsigma^3_{\mathscr{G}11}(\beta)}(\chi^2{}_{0})
%+f^3_{\mathscr{G}3**|\varsigma^3_{\mathscr{G}11}(\beta)}
%(d^2_{\mathscr{G}1*|\varsigma^2_{\mathscr{G}1}(\beta)}(\chi^1{}_{0}),d^2_{\mathscr{G}2*|\varsigma^2_{\mathscr{G}1}(\beta)}(\chi^1{}_{0}))]
%\vphantom{\Big]}
%\nonumber
%\\
%&h^1{}_{1}(\theta_0,\theta_1)=\varsigma^2_{\mathscr{G}0}(\beta)-(\theta_1-\theta_0)\chi^1{}_{1}
%+\theta_1(d^1_{\mathscr{G}0} \circ %*|\varsigma^1_{\mathscr{G}/\!\!\!\hskip.6pt 0}(\beta)}(
%f^2_{\mathscr{G}2})_{*|\varsigma^2_{\mathscr{G}1}(\beta)}(\chi^1{}_{0})
%\vphantom{\Big]}
%\label{htpsimplex}
%\\
%&\hspace{.5cm}+\theta_0\theta_1[f^3_{\mathscr{G}3*|\varsigma^3_{\mathscr{G}01}(\beta)}(\chi^2{}_{1})
%+f^3_{\mathscr{G}3**|\varsigma^3_{\mathscr{G}01}(\beta)}
%(d^2_{\mathscr{G}0*|\varsigma^2_{\mathscr{G}1}(\beta)}(\chi^1{}_{0}),d^2_{\mathscr{G}2*|\varsigma^2_{\mathscr{G}0}(\beta)}(\chi^1{}_{1}))]
%\vphantom{\Big]}
%\nonumber
%\end{align}
\begin{align}
&h^1{}_{i}(\theta_0,\theta_1)=\varsigma^2_{\mathscr{G}1-i}(\beta)-(\theta_1-\theta_0)\chi^1{}_{i}
+\theta_1(d^1_{\mathscr{G}1-i} \circ %*|\varsigma^1_{\mathscr{G}/\!\!\!\hskip.6pt 0}(\beta)}(
f^2_{\mathscr{G}2})_{*|\varsigma^2_{\mathscr{G}1}(\beta)}(\chi^1{}_{0})
\vphantom{\Big]}
\label{htpsimplex23}
\\
&\hspace{.1cm}+\theta_0\theta_1[f^3_{\mathscr{G}3*|\varsigma^3_{\mathscr{G}1-i1}(\beta)}(\chi^2{}_{i})
+f^3_{\mathscr{G}3**|\varsigma^3_{\mathscr{G}1-i1}(\beta)}
(d^2_{\mathscr{G}1-i*|\varsigma^2_{\mathscr{G}1}(\beta)}(\chi^1{}_{0}),d^2_{\mathscr{G}2*|\varsigma^2_{\mathscr{G}1-i}(\beta)}(\chi^1{}_{i}))]
\vphantom{\Big]}
\nonumber
\end{align}
with $i=0,1$. The components $h^p$ of $h$ with $p\geq 2$ can be obtained in the same way with an amount of computation 
increasing very rapidly with $p$. 

For any simplicial internal graded manifold morphism
$h\in\mathsans{Hom}^*_{\mathsans{sgrMf}}(\mathsans{N}\mathbb{R}[-1]$, $\mathsans{Hom}_{\mathsans{sSet}}(\varDelta^1,\mathscr{G}))$
the source and target of $h$ are the simplicial internal %graded manifold 
morphisms
$\sigma\circ h$, $\tau\circ h\in\mathsans{Hom}^*_{\mathsans{sgrMf}}(\mathsans{N}\mathbb{R}[-1],\mathscr{G})$,
because the source and target maps $\sigma$, $\tau$ are simplicial as recalled earlier. Their components
$\sigma^p\circ h^p$, $\tau^p\circ h^p$ should so
be expressible by means of the exponential parametrization found earlier, eqs. \ceqref{simplex19}, \ceqref{simplex22},
\ceqref{simplex25} etc. This is indeed the case: there exists bundle maps 
$\dot\sigma,\dot\tau:\mathsans{Lie}\,\mathsans{Hom}_{\mathsans{sSet}}(\varDelta^1,\mathscr{G})[1]\rightarrow\mathsans{Lie}\,\mathscr{G}[1]$, 
such that 
\begin{align}
&\sigma\circ_{\mathsans{sSet}}\exp_{\mathsans{Hom}_{\mathsans{sSet}}(\varDelta^1,\mathscr{G})}=\exp_{\mathscr{G}}\circ\,\dot\sigma, 
\vphantom{\Big]}
\label{htpsimplex30}
%\\
\end{align}
\begin{align}
&\tau\circ_{\mathsans{sSet}}\exp_{\mathsans{Hom}_{\mathsans{sSet}}(\varDelta^1,\mathscr{G})}=\exp_{\mathscr{G}}\circ\,\dot\tau.
\vphantom{\Big]}
\label{htpsimplex31}
\end{align}
The source map $\dot\sigma$ is given by 
\begin{align}
&\dot\sigma^*\gamma=f^1_{\mathscr{G}0}(\beta),
\vphantom{\Big]}
\label{htpsimplex24}
\\
&\dot\sigma^*\xi_0=f^2_{\mathscr{G}0*|\varsigma^2_{\mathscr{G}1}(\beta)}(\chi^1{}_{0}),
\vphantom{\Big]}
\label{htpsimplex25}
\\
&\dot\sigma^*\xi_1=
f^3_{\mathscr{G}0*|\varsigma^3_{\mathscr{G}11}(\beta)}(\chi^2{}_{0})
+f^3_{\mathscr{G}0**|\varsigma^3_{\mathscr{G}11}(\beta)}
(d^2_{\mathscr{G}1*|\varsigma^2_{\mathscr{G}1}(\beta)}(\chi^1{}_{0}),d^2_{\mathscr{G}2*|\varsigma^2_{\mathscr{G}1}(\beta)}(\chi^1{}_{0}))
\vphantom{\Big]}
\label{htpsimplex26}
%\\
%&\ldots
%\nonumber
\end{align}
etc. Similarly, the target map $\dot\tau$ is given by 
\begin{align}
&\dot\tau^*\gamma=f^1_{\mathscr{G}1}(\beta),
\vphantom{\Big]}
\label{htpsimplex27}
\\
&\dot\tau^*\xi_0=f^2_{\mathscr{G}2*|\varsigma^2_{\mathscr{G}0}(\beta)}(\chi^1{}_{1}),
\vphantom{\Big]}
\label{htpsimplex28}
\\
&\dot\tau^*\xi_1=
f^3_{\mathscr{G}3*|\varsigma^3_{\mathscr{G}00}(\beta)}(\chi^2{}_{2})
+f^3_{\mathscr{G}3**|\varsigma^3_{\mathscr{G}00}(\beta)}
(d^2_{\mathscr{G}0*|\varsigma^2_{\mathscr{G}0}(\beta)}(\chi^1{}_{1}),d^2_{\mathscr{G}1*|\varsigma^2_{\mathscr{G}0}(\beta)}(\chi^1{}_{1}))
\vphantom{\Big]}
\label{htpsimplex29}
%\\
%&\ldots
%\nonumber
\end{align}
etc. 

There is a subtle relationship of the maps $\dot\sigma$, $\dot\tau$ and the Chevalley--Eilenberg
differential $Q_{\mathsans{Lie}\,\mathscr{G}}$ and this shows that simplicial homotopy secretly incarnates the symmetry of 
simplicial mapping. %More s
Specifically, there is %exists 
a family of graded manifold morphisms $p_t:\mathsans{Lie}\,\mathscr{G}[1]\rightarrow
\mathsans{Lie}\,\mathsans{Hom}_{\mathsans{sSet}}(\varDelta^1,\mathscr{G})[1]$ depending on an odd parameter
$t\in\mathbb{R}[-1]$ (that is a mapping $p:\mathsans{Lie}\,\mathscr{G}[1]\rightarrow T[1]
\mathsans{Lie}\,\mathsans{Hom}_{\mathsans{sSet}}(\varDelta^1,\mathscr{G})[1]$) such that 
%with the property that  \hphantom{xxxxxxxxxxx}
\begin{align}
&\dot\sigma\circ p_t=\id_{\mathsans{Lie}\,\mathscr{G}[1]},
\vphantom{\Big]}
\label{htpsimplex32}
\\
&\dot\tau\circ p_t=\id_{\mathsans{Lie}\,\mathscr{G}[1]}+tQ_{\mathsans{Lie}\,\mathscr{G}}.
\vphantom{\Big]}
\label{htpsimplex33}
\end{align}
We do not have a general proof of this fact, but we have been able to construct the map $p_t$ up to simplicial degree $2$,
\begin{align}
&p_t{}^*\beta=\varsigma^1_{\mathscr{G}}(\gamma)+t\xi_0,
\vphantom{\Big]}
\label{htpsimplex34}
\\
&p_t{}^*\chi^1{}_{i}=d^1_{\mathscr{G}i*|\varsigma^1_{\mathscr{G}}(\gamma)}(\xi_0)-(-1)^it\xi_1,
\vphantom{\Big]}
\label{htpsimplex35}
\\
&p_t{}^*\chi^2{}_{i}=d^2_{\mathscr{G}i*|\varsigma^2_{\mathscr{G}}(\gamma)}(\xi_1)
%+d^2_{\mathscr{G}i**|\varsigma^2_{\mathscr{G}}(\gamma)}
%(d^1_{\mathscr{G}0*|\varsigma^1_{\mathscr{G}}(\gamma)}(\xi_0), d^1_{\mathscr{G}1*|\varsigma^1_{\mathscr{G}}(\gamma)}(\xi_0))+(-1)^it\xi_2
\vphantom{\Big]}
\label{htpsimplex36}
\\
&\hspace{2cm}%p_t{}^*\chi^2{}_{i}=d^2_{\mathscr{G}i*|\varsigma^2_{\mathscr{G}}(\gamma)}(\xi_1)
+d^2_{\mathscr{G}i**|\varsigma^2_{\mathscr{G}}(\gamma)}
(d^1_{\mathscr{G}0*|\varsigma^1_{\mathscr{G}}(\gamma)}(\xi_0), d^1_{\mathscr{G}1*|\varsigma^1_{\mathscr{G}}(\gamma)}(\xi_0))+(-1)^it\xi_2
\vphantom{\Big]}
\nonumber
\end{align}
etc. Using \ceqref{simplex18}, \ceqref{simplex21}, \ceqref{simplex23}, \ceqref{simplex23/1}, \ceqref{simplex24}, \pagebreak 
\ceqref{simplex26}--\ceqref{simplex28}, 
it is straightforward though lengthy to verify that the conditions %\pagebreak
\ceqref{htpsimplex18}, \ceqref{htpsimplex19}, \ceqref{htpsimplex21}, \ceqref{htpsimplex22} as well as 
the basic relations \ceqref{htpsimplex32}, \ceqref{htpsimplex33} are all identically satisfied. 

%\vfil\eject

%\vspace{.5mm}

\subsection{\textcolor{blue}{\sffamily Higher gauged sigma models}}\label{subsec:hisigma}

%\vspace{.33mm}
We now expound a formulation of higher gauged sigma model theory based on the abstract algebraic framework
of sect. \cref{sec:higau}. Its basic data are therefore two differential graded commutative algebras $C_1$, $C_2$  
and higher gauged sigma model fields are modeled as non differential morphisms
from $C_2$ to $C_1$. Both $C_1$ and $C_2$ are algebras of functions on appropriate graded manifolds. While $C_2$ 
is the algebra of smooth functions on the $1$--shifted $L_\infty$--algebroid encoding the higher gauge symmetry, 
$C_1$ is the algebra of {\it internal} smooth functions of $1$--shift\-ed tangent bundle of the relevant space--time manifold.
The higher fields are in this way the pull--backs of internal graded manifold morphisms from the latter to the former.
The incorporation of ghost degrees of freedom is so possible and a complete BRST formulation is reached. 

%\vspace{.33mm}
$C_1$ is the graded commutative algebra $C^\infty(T[1]Z)\otimes G_{\mathbb{R}}$ of functions on the $1$--shifted 
tangent bundle $T[1]Z$ of an ordinary manifold $Z$ (cf. eg. \cref{exa:tm}), the source manifold, 
valued in the graded vector space $G_{\mathbb{R}}$ defined by \ceqref{gradreal}.
The differential $Q_1$ of $C_1$ is the canonical homological vector field $d$ of $T[1]Z$, as is natural. 
Describe locally $T[1]Z$ by degree $0$ base coordinates $z^\alpha$ 
and degree $1$ fiber coordinates $\zeta^\alpha$. Then, a homogeneous degree $s$ element $f\in C^\infty(T[1]Z)\otimes G_{\mathbb{R}}$
has an expansion of the form 
\begin{equation}
f=\sss_{0\le \nu\le n}\frac{1}{h!}f_{\alpha_1\ldots \alpha_h}(z)\zeta^{\alpha_1}\cdots\zeta^{\alpha_h} \vphantom{\bigg]}
\label{fexpans}
\end{equation}
(cf. eq. \ceqref{t1mfunc}), where $n=\dim Z$ and the local functions $f_{\alpha_1\ldots \alpha_h}$ have degree $s-h$.
Further, $d$ is given by given by \hphantom{xxxxxxxxxxxxxx}
\begin{equation}
d=\zeta^\alpha\partial_{z\alpha} \vphantom{\bigg]}
\label{canhom}
\end{equation}
(cf. eq. \ceqref{t1mq}). It is important to realize that the algebra $C^\infty(T[1]Z)\otimes G_{\mathbb{R}}$ 
is in fact bigraded. \pagebreak  
The two gradings are the {\it form grading},  that is the polynomial degree in the odd 
coordinates $\zeta^\alpha$, and the {\it ghost grading}, that is the $G_{\mathbb{R}}$--degree. 
The grading of $C^\infty(T[1]Z)\otimes G_{\mathbb{R}}$ is just the total grading of the bigrading. 
So, while a degree $s$ element $f\in C^\infty(T[1]Z)$ can be identified with an $s$--form of $Z$,
a degree $s$ element $f\in C^\infty(T[1]Z)\otimes G_{\mathbb{R}}$ can be viewed as a non homogeneous form--ghost
whose total form plus ghost degree is $s$,
\begin{equation}
f=\sss_{0\le h\le n}f^{(h,s-h)} \vphantom{\bigg]}
\label{fgexp}
\end{equation}
where $f^{(h,g)}$ has form--ghost bidegree $(h,g)$. 
Notice that the degree $1$ vector field $d$ has form--ghost bidegree
$(1,0)$. 

$C_2$ is the graded commutative algebra $C^\infty(E[1])$ of functions on the $1$--shifted 
bundle $E[1]$ of an $L_\infty$--algebroid $E$ over an ordinary manifold $N$ (cf. subsect. \cref{subsec:linf}). 
The differential $Q_2$ of $C_2$ is the homological vector field $Q_E$ on $E[1]$ associated with the $L_\infty$ structure of $E$
and is given by  eq. \ceqref{qexpl} in terms of the local representations $\rho$, $\partial$ and $[-,\ldots,-]_\kappa$ 
of the anchor, boundary map and multiple argument brackets of $E$, respectively, 
when $E[1]$ is described locally  by degree $0$ base coordinates $x^i$ and 
degree $1$ fiber coordinates $\xi_p$. 
%Then, $Q_E$ has the local expansion 
%\begin{align}
%Q_E&=\rho^i(\xi_0)\partial_{xi}+\langle\partial\xi_1-\tfrac{1}{2}[\xi_0,\xi_0],\partial_{\xi_0}\rangle
%\vphantom{\Big]}
%%\label{qexpl}
%\\
%&\hspace{-.5cm}+\langle\partial\xi_2-[\xi_0,\xi_1]+\tfrac{1}{6}[\xi_0,\xi_0,\xi_0],\partial_{\xi_1}\rangle
%\vphantom{\Big]}
%\nonumber
%\\
%&\hspace{-1cm}+\langle\partial\xi_3-[\xi_0,\xi_2]-\tfrac{1}{2}[\xi_1,\xi_1]+\tfrac{1}{2}[\xi_0,\xi_0,\xi_1]
%-\tfrac{1}{24}[\xi_0,\xi_0,\xi_0,\xi_0],\partial_{\xi_2}\rangle+\ldots,
%\vphantom{\Big]}
%\nonumber
%\end{align}

\vspace{.33mm}
A higher gauge field is a graded commutative algebra morphism $\varPhi:C_2\rightarrow C_1$,
its curvature is the associated defect $F_\varPhi$ defined according to \ceqref{defect},
and the Bianchi identity this obeys is the defect identity \ceqref{bianchi} as we anticipated in 
subsect. \ref{subsec:algdef}.
Adopting a somewhat more suitable notation and terminology, we define a {\it BRST higher gauged sigma model field} 
as a graded commutative algebra morphisms $A:C^\infty(E[1])\rightarrow C^\infty(T[1]Z)\otimes G_{\mathbb{R}}$.
The {\it BRST higher gauged sigma model curvature} $F_A$ of $A$ is then, \hphantom{xxxxxxxxxx}
\begin{equation}
F_A=dA-AQ_E. \vphantom{\bigg]}
\label{curvfa}
\end{equation}
The Bianchi identity correspondingly is 
\begin{equation}
dF_A+F_AQ_E=0. \vphantom{\bigg]}
\label{bianchifa}
\end{equation}
The higher gauged sigma model field components are 
\begin{align}
&\varphi^i=Ax^i,
\vphantom{\Big]}
\label{phir}
%\\
\end{align}
\vspace{-1.cm}\eject\noindent
\begin{align}
&A_p=A\xi_p.
\vphantom{\Big]}
\label{aa}
\end{align}
$\varphi$ is the customary sigma model field describing the embedding of $Z$ into $N$.
The $A_p$ are the higher gauge fields $\varphi$ couples to.  
The higher gauged sigma model curvature components are 
\begin{align}
&\upsilon^i=d\varphi^i-\rho^i(A_0),
\vphantom{\Big]}
\label{upsr}
\\
&F_p=dA_p+P_p(A_0,A_1,\ldots),
\vphantom{\Big]}
\label{fa}
\end{align}
where $P_p(A_0,A_1,\ldots)$ is a polynomial in the gauge field components  $A_0,A_1,\ldots$
constructed using the $L_\infty$--algebroid boundary and brackets. Specifically, 
\begin{align}
&F_0=dA_0+\tfrac{1}{2}[A_0,A_0]-\partial A_1, 
\vphantom{\Big]}
\label{f0}
\\
&F_1=dA_1+[A_0,A_1]-\tfrac{1}{6}[A_0,A_0,A_0]-\partial A_2, 
\vphantom{\Big]}
\label{f1}
\\
&F_2=dA_2+[A_0,A_2]+\tfrac{1}{2}[A_1,A_1]-\tfrac{1}{2}[A_0,A_0,A_1]
\vphantom{\Big]}
\label{f2}
\\
&\hspace{5cm}+\tfrac{1}{24}[A_0,A_0,A_0,A_0]-\partial A_3,\quad\ldots.
\vphantom{\Big]}
\nonumber
\end{align} 
We remind that it is tacitly understood that in \ceqref{upsr} $\rho^i$ 
and in \ceqref{fa} and \ceqref{f0}--\ceqref{f2}
the brackets $[-,\ldots,-]$ are all formally evaluated at $x^j=\varphi^j$. 
The Bianchi identities obeyed by the sigma model field components are 
\begin{align}
&d\upsilon^i+\upsilon^j\partial_{xj}\rho^i(A_0)+\rho^i(F_0)=0,
\vphantom{\Big]}
\label{biauspr}
\\
&dF_p+Q_p(A_0,A_1,\ldots;F_0,F_1,\ldots)-\upsilon^i\partial_{xi}P_p(A_0,A_1,\ldots)=0,
\vphantom{\Big]}
\label{biafa}
\end{align}
where $Q_p(A_0,A_1,\ldots;F_0,F_1,\ldots)$ is a polynomial in the gauge field 
and gauge curvature components $A_0,A_1,\ldots$ and $F_0,F_1,\ldots$ of degree $1$ in the latter
constructed again using the $L_\infty$--algebroid boundary and brackets. Explicitly,
\begin{align}
&dF_0+[A_0,F_0]+\partial F_1
-\upsilon^i\partial_{xi}(\tfrac{1}{2}[A_0,A_0]-\partial A_1)=0,
\vphantom{\Big]}
\label{biaf0}
\\
&dF_1+[A_0,F_1]-[F_0,A_1]+\tfrac{1}{2}[A_0,A_0,F_0]+\partial F_2
\vphantom{\Big]}
\label{biaf1}
\\
&\hspace{4cm}-\upsilon^i\partial_{xi}([A_0,A_1]-\tfrac{1}{6}[A_0,A_0,A_0]-\partial A_2)=0, 
\vphantom{\Big]}
\nonumber
\\
&dF_2+[A_0,F_2]-[F_0,A_2]-[A_1,F_1]-[A_0,F_0,A_1]
\vphantom{\Big]}
\label{biaf2}
%\\
\end{align}
\begin{align}
&\hspace{1.5cm}+\tfrac{1}{2}[A_0,A_0,F_1]+\tfrac{1}{6}[A_0,A_0,A_0,F_0]+\partial F_3
-\upsilon^i\partial_{xi}([A_0,A_2]
\vphantom{\Big]}
\nonumber
\\
&\hspace{.75cm}
+\tfrac{1}{2}[A_1,A_1]-\tfrac{1}{2}[A_0,A_0,A_1]+\tfrac{1}{24}[A_0,A_0,A_0,A_0]-\partial A_3)=0,\quad\ldots.
\vphantom{\Big]}
\nonumber
\end{align}
All the above relations can be made more explicit by expanding the fields in 
subcomponents of given form--ghost bidegree as in \ceqref{fgexp}. 
For the higher gauged sigma model field components, this expansion has the form
\begin{align}
&\varphi^i=\sss_{0\le h\le n}\varphi^{(h,-h)i},
\vphantom{\Big]}
\label{fgphir}
\\
&A_p=\sss_{0\le h\le n}A_p{}^{(h,p+1-h)}.
\vphantom{\Big]}
\label{fgaa}
\end{align}
For the  corresponding higher %gauged sigma model 
curvature components, the expansion reads as %\hphantom{xxxxxxxxxxxxxxx}
\begin{align}
&\upsilon^i=\sss_{0\le h\le n}\upsilon^{(h,1-h)i},
\vphantom{\Big]}
\label{fgupsr}
\\
&F_p=\sss_{0\le h\le n}F_p{}^{(h,p+2-h)}.
\vphantom{\Big]}
\label{fgfa}
\end{align}
Substituting the \ceqref{fgphir}, \ceqref{fgaa}, \ceqref{fgupsr}, \ceqref{fgfa} into the 
\ceqref{upsr}, \ceqref{fa} or \ceqref{f0}--\ceqref{f2}, \ceqref{biauspr}, \ceqref{biafa} 
or \ceqref{biaf0}--\ceqref{biaf2}, we obtain rather explicit expressions
of the curvature subcomponents and the Bianchi identities they obey in terms of the field
subcomponents. 
%This is a straightforward exercise that is left to reader. 
For a given $L_\infty$--algebroid $E$, the structure of these expressions
depends explicitly on the dimension $n$ of the base manifold $Z$. However, 
if we truncate the expansions by setting all fields of negative
ghost degree to $0$, we are left with expressions of a universal form 
independent from $n$.  
In particular, if we set all components with non zero ghost degree to zero, we recover the familiar expressions of the 
higher gauged sigma model curvatures and Bianchi identities.

\begin{exa} \label{exa:sstrhgt}
{\rm Semistrict higher gauge theory is a higher gauge theory for which the $L_\infty$ algebroid $E$ 
is a Lie 2--algebra. It has been studied by different approaches in refs. 
\ccite{Jurco:2014mva,Zucchini:2011aa}. In semistrict higher gauge theory, 
neglecting all negative ghost degree contributions,
the subcomponents of the higher gauge fields $A_0$, $A_1$ are 
\begin{align}
&A_0{}^{(0,1)}=\gamma, \qquad A_0{}^{(1,0)}=a
\vphantom{\Big]}
\label{}
\\
&A_1{}^{(0,2)}=\varGamma, \quad A_1{}^{(1,1)}=C, \quad A_1{}^{(0,1)}=B. 
\vphantom{\Big]}
\label{}
\end{align}
The subcomponents of the corresponding higher gauge curvatures $F_0$, $F_1$ are 
\begin{align}
&F_0{}^{(0,2)}=\phi=\tfrac{1}{2}[\gamma,\gamma]-\partial\varGamma,
\vphantom{\Big]}
\label{}
\\
&F_0{}^{(1,1)}=g=d\gamma+[a,\gamma]-\partial C,
\vphantom{\Big]}
\nonumber
\\
&F_0{}^{(2,0)}=f=da+\tfrac{1}{2}[a,a]-\partial B,
\vphantom{\Big]}
\nonumber
\\
&F_1{}^{(0,3)}=\varPsi=[\gamma,\varGamma]-\tfrac{1}{6}[\gamma,\gamma,\gamma],
\vphantom{\Big]}
\label{}
\\
&F_1{}^{(1,2)}=K=d\varGamma+[a,\varGamma]+[\gamma, C]-\tfrac{1}{2}[a,\gamma,\gamma],
\vphantom{\Big]}
\nonumber
\\
&F_1{}^{(2,1)}=H=dC+[a,C]+[\gamma,B]-\tfrac{1}{2}[a,a,\gamma],
\vphantom{\Big]}
\nonumber
\\
&F_1{}^{(3,0)}=G=dB+[a,B]-\tfrac{1}{2}[a,a,a].
\vphantom{\Big]}
\nonumber
\end{align}
These obey the Bianchi identities
\begin{align}
&d\phi+[a,\phi]-[g,\gamma]+\partial K=0,
\vphantom{\Big]}
\label{}
\\
&dg+[a,g]+[\gamma,f]+\partial H=0,
\vphantom{\Big]}
\nonumber
\\
&df+[a,f]+\partial G=0,
\vphantom{\Big]}
\nonumber
\\
&d\varPsi+[a,\varPsi]-[g,\varGamma]-[\phi,C]+[\gamma,K]+\tfrac{1}{2}[g,\gamma,\gamma]+[a,\gamma,\phi]=0,
\vphantom{\Big]}
\label{}
\\
&dK+[a,K]-[g,C]-[\phi,B]-[f,\varGamma]
\vphantom{\Big]}
\nonumber
\\
&\hspace{2.85cm}
+[\gamma,H]+\tfrac{1}{2}[a,a,\phi]+[g,a,\gamma]+\tfrac{1}{2}[f,\gamma,\gamma]=0,
\vphantom{\Big]}
\nonumber
\\
&dH+[a,H]-[g,B]-[f,C]+[\gamma,G]+\tfrac{1}{2}[a,a,g]+[f,a,\gamma]=0.
\vphantom{\Big]}
\end{align}
These identities were obtained by another route in \ccite{Zucchini:2011aa,Soncini:2014ara,Zucchini:2015ohw}.}
\end{exa}

%The subcomponent expansion \ceqref{fgphir}, \ceqref{fgaa} of the components $\varphi^i$, $A_p$ of the BRST higher 
%gauged sigma model field $A$ show that $A$ specifies  an internal graded manifold morphism $a:T[1]Z\rightarrow E[1]$
%for reasons explained in subsect. \cref{subsec:graman}. 

For reasons explained in subsect. \cref{subsec:graman}, the components $\varphi^i$, $A_p$ of the BRST higher 
gauged sigma model field $A$ given by the expansions \ceqref{fgphir}, \ceqref{fgaa}
specify an internal graded manifold morphism $a:T[1]Z\rightarrow E[1]$. 
Since the vector coordinates of $T[1]Z$ are all odd, a pull--back operator 
$a^\#:C^\infty(E[1])\rightarrow \mathsans{C}^\infty(T[1]Z)$ of the standard differential geometric 
form such that %by setting
\begin{align}
&a^\#x^i=\varphi^i, 
\vphantom{\Big]}
\label{phirrev}
\\
&a^\#\xi_p=A_p
\vphantom{\Big]}
\label{aarev}
\end{align} 
is defined. 
By eqs. \ceqref{phir}, \ceqref{aa}, recalling that $\mathsans{C}^\infty(T[1]Z)\simeq C^\infty(T[1]Z)\otimes G_{\mathbb{R}}$ 
(cf. eq. \ceqref{graman5}), it is apparent that $a^\#$ equals precisely the BRST field $A$ one started with.
%at least when acting on  all the polynomials of the coordinates $x^i$, $\xi_p$ of $E[1]$. 
Viceversa, an internal morphism $a:T[1]Z\rightarrow E[1]$ of components $\varphi^i$, $A_p$ specifies a BRST field $A$ 
such that 
\begin{align}
&Ax^i=\varphi^i, 
\vphantom{\Big]}
\label{phirinv}
\\
&A\xi_p=A_p.
\vphantom{\Big]}
\label{aainv}
\end{align} 
by setting $A=a^\#$. The manifold $M(C^\infty(E[1]),C^\infty(T[1]Z)\otimes G_{\mathbb{R}})$ of BRST high\-er 
gauged sigma model fields can so be identified with the internal hom manifold $\mathsans{Hom}_{\mathsans{grMf}}(T[1]Z,E[1])$
of $T[1]Z$, $E[1]$. In this sense, the present theory is a
BRST extension of the BGKS theory of refs. \ccite{Bojowald:0406445,Kotov:2007nr,Gruetzmann:2014ica}. 

The above construction applies in particular when $E$ is the $L_\infty$--algebroid 
$\mathsans{Lie}\,\mathscr{G}$ of a Lie quasi--groupoid $\mathscr{G}$ (cf. subsect. 
\cref{subsec:simplex}). As long as we are concerned with higher 
gauge field and curvature components, as we have done so far, there is not much more we can say in this particular case. 
The relationship of $\mathsans{Lie}\,\mathscr{G}$ to $\mathscr{G}$ is expected to be relevant
in a formulation of a model of higher parallel transport and in the analysis of finite higher gauge symmetry. 
While the former lies beyond the scope of the present work, the latter will be examined %in some detail 
in the next subsection.

%\vfil\eject

\subsection{\textcolor{blue}{\sffamily Higher gauged sigma model field BRST variations}}\label{subsec:brsthisigma}

In this subsection, building upon the results of subsect. \cref{subsec:brst}, 
we shall compute the canonical determination BRST 
variations of the higher gauged sigma model fields for a general $L_\infty$--algebroid.
For the $L_\infty$--algebroid of a Lie quasi--groupoid, we provide evidence that the definition of 
higher guage transformations through simplicial homotopy reproduces infinitesimally the BRST variations. 

From the general relation \ceqref{sphiq}, using \ceqref{curvfa}, we find 
\begin{equation}
s_QA=-F_A=-dA+AQ_E. \vphantom{\bigg]^f_g}
\label{sqhgs}
\end{equation} 
By \ceqref{upsr}, \ceqref{fa}, we have 
\begin{align}
&s_Q\varphi^i=-d\varphi^i+\rho^i(A_0),
\vphantom{f^f} %\vphantom{\Big]}
\label{sqphir}
%\\
\end{align}
\begin{align}
&s_QA_p=-dA_p-P_p(A_0,A_1,\ldots),
\vphantom{\Big]}
\label{sqaa}
\end{align}
The variations $s_QA_p$ can be written down more explicitly using the 
expressions \ceqref{f0}--\ceqref{f2} of the higher gauged sigma model curvatures,
\begin{align}
&s_QA_0=-dA_0-\tfrac{1}{2}[A_0,A_0]+\partial A_1, 
\vphantom{\Big]}
\label{sqa0}
\\
&s_QA_1=-dA_1-[A_0,A_1]+\tfrac{1}{6}[A_0,A_0,A_0]+\partial A_2, 
\vphantom{\Big]}
\label{sqa1}
\\
&s_QA_2=-dA_2-[A_0,A_2]-\tfrac{1}{2}[A_1,A_1]+\tfrac{1}{2}[A_0,A_0,A_1]
\vphantom{\Big]}
\label{sqa2}
\\
&\hspace{5cm}-\tfrac{1}{24}[A_0,A_0,A_0,A_0]+\partial A_3,\qquad\ldots.
\vphantom{\Big]}
\nonumber
\end{align}
Substituting the expansions \ceqref{fgphir}, \ceqref{fgaa}, \ceqref{fgupsr}, \ceqref{fgfa} 
of the fields and curvatures 
in subcomponents of given form--ghost bidegree into the \ceqref{sqphir}, \ceqref{sqaa}
and \ceqref{sqa0}--\ceqref{sqa2}, we obtain rather explicit expressions of these BRST variations. 
The non zero ghost degree components are  necessary for the nilpotence of $s_Q$ 
%the canonical determination $s_Q$ of the BRST variation operation $s$
(cf. subsect. \cref{subsec:brst}). 

\vspace{.33mm}
\begin{exa}
{\rm In semistrict higher gauge theory, introduced in eg. \cref{exa:sstrhgt}, 
neglecting all the negative ghost degree fields, the canonical determination
BRST variations are given  by 
\begin{align}
&s_Q\gamma=-\phi=-\tfrac{1}{2}[\gamma,\gamma]+\partial\varGamma,
\vphantom{\Big]}
\label{}
\\
&s_Qa=-g=-d\gamma-[a,\gamma]+\partial C,
\vphantom{\Big]}
\nonumber
\\
&s_Q\varGamma=-\varPsi=-[\gamma,\varGamma]+\tfrac{1}{6}[\gamma,\gamma,\gamma],
\vphantom{\Big]}
\nonumber
\\
&s_QC=-K=-d\varGamma-[a,\varGamma]-[\gamma, C]+\tfrac{1}{2}[a,\gamma,\gamma],
\vphantom{\Big]}
\label{}
\\
&s_QB=-H=-dC-[a,C]-[\gamma,B]+\tfrac{1}{2}[a,a,\gamma].
\vphantom{\Big]}
\nonumber
\end{align}
These expressions were also obtained in \ccite{Zucchini:2011aa,Soncini:2014ara,Zucchini:2015ohw}. 
Since they are truncated to non negative ghost degree,  nilpotence of $s_Q$ holds 
only if certain on--shell con\-ditions are satisfied. 
}
\end{exa}

When $E$ is the $L_\infty$--algebroid $\mathsans{Lie}\,\mathscr{G}$ of a Lie quasi--groupoid $\mathscr{G}$,
we can exploit the simplicial description of $\mathsans{Lie}\,\mathscr{G}$ worked out in  
subsect. \cref{subsec:simplex}. In particular, we can write he canonical determination BRST variation 
\ceqref{sqhgs} in terms of the simplicial homotopy source and target maps $\dot\sigma$,
$\dot\tau$ using \ceqref{htpsimplex32}, \ceqref{htpsimplex33},% \hphantom{xxxxxxxxxxxxxxxxxxx}
\begin{equation}
s_QA=-dA+AD_t\dot\tau\circ p_t,   %-\dot\sigma\circ p_t),
\label{sqhgslg}
\end{equation} 
where the maps $p_t:\mathsans{Lie}\,\mathscr{G}[1]\rightarrow
\mathsans{Lie}\,\mathsans{Hom}_{\mathsans{sSet}}(\varDelta^1,\mathscr{G})[1]$, $t\in\mathbb{R}[-1]$,  were introduced in subsect. 
\cref{subsec:simplex} and $D_t=d/dt$.  This shows rather explicitly 
the relation between infinitesimal higher gauge symmetry 
and simplicial homotopy %of simplicial maps 
hypothesized in ref. \ccite{Jurco:2016qwv}. 
One may wonder whether the analysis of 
subsect. \cref{subsec:simplex} can tell us %more with regard to this. %
something about finite higher gauge symmetry. 

The exponential map $\exp_{\mathscr{G}}$ of $\mathscr{G}$ is a graded manifold morphism from 
$\mathsans{Lie}\,\mathscr{G}[1]$ to $\mathsans{Hom}^*_{\mathsans{sgrMf}}(\mathsans{N}\mathbb{R}[-1],\mathscr{G})$.
By virtue of this, a higher gauged sigma model field 
$a:T[1]Z\rightarrow \mathsans{Lie}\,\mathscr{G}[1]$, viewed here as an internal graded manifold morphism, 
can be equivalently encoded in the composition  
\begin{equation}
\exp_{\mathscr{G}}{} \circ a:T[1]Z \rightarrow \mathsans{Hom}^*_{\mathsans{sgrMf}}(\mathsans{N}\mathbb{R}[-1],\mathscr{G}). 
\vphantom{\bigg]}
\label{jns1}
\end{equation} 
Substituting the Lie quasi--groupoid $\mathscr{G}$ with the simplicial homotopy Lie quasi--groupoid
$\mathsans{Hom}_{\mathsans{sSet}}(\varDelta^1,\mathscr{G})[1])$, we can 
view a finite higher gauge transformation as a higher gauged sigma model field 
$\gamma:T[1]Z\rightarrow \mathsans{Lie}\,\mathsans{Hom}_{\mathsans{sSet}}(\varDelta^1,\mathscr{G})[1]$
encoded in the composition 
\begin{equation}
\exp_{\mathsans{Hom}_{\mathsans{sSet}}(\varDelta^1,\mathscr{G})}\circ \,\gamma:T[1]Z \rightarrow 
\mathsans{Hom}^*_{\mathsans{sgrMf}}(\mathsans{N}\mathbb{R}[-1],\mathsans{Hom}_{\mathsans{sSet}}(\varDelta^1,\mathscr{G})). 
\vphantom{\bigg]}
\label{jns2}
\end{equation} 
As shown in subsect. \cref{subsec:simplex}, 
the source and target maps $\sigma$, $\tau$ are simplicial manifold morphisms of 
$\mathsans{Hom}_{\mathsans{sSet}}(\varDelta^1,\mathscr{G})$ into $\mathscr{G}$. They thus induce
internal graded manifold morphisms
\begin{equation}
\omega\circ_{\mathsans{sSet}}\exp_{\mathsans{Hom}_{\mathsans{sSet}}(\varDelta^1,\mathscr{G})}\circ \,\gamma:T[1]Z \rightarrow 
\mathsans{Hom}^*_{\mathsans{sgrMf}}(\mathsans{N}\mathbb{R}[-1],\mathscr{G}), \quad \omega=\sigma,\tau, \vphantom{\bigg]}
\label{jns3/0}
\end{equation}
by simplicial composition. By relations \ceqref{htpsimplex30}, \ceqref{htpsimplex31}, these can be cast as 
\begin{equation}
\omega\circ_{\mathsans{sSet}}\exp_{\mathsans{Hom}_{\mathsans{sSet}}(\varDelta^1,\mathscr{G})}\circ \,\gamma=
\exp_{\mathscr{G}}\circ\,\dot\omega\circ \,\gamma. \vphantom{\bigg]}
\label{jns3}
\end{equation} 
%$\omega=\sigma,\tau$. Similar relations hold also for the source map $\sigma$. 
A comparison of \ceqref{jns1} and \ceqref{jns3} indicates that $\gamma$ may represent a finite gauge transformation
relating the higher gauge fields \hphantom{xxxxxxxxxxxxxxxxx} \pagebreak 
\begin{equation}
a=\dot\sigma\circ\,\gamma, \qquad a'=\dot\tau\circ\,\gamma.
\label{jns3/1}
\end{equation} 
We might substantiate to some extent this claim if we were able to find a one parameter family 
$\gamma_t:T[1]Z\rightarrow \mathsans{Lie}\,\mathsans{Hom}_{\mathsans{sSet}}(\varDelta^1,\mathscr{G})[1]$,
$t\in\mathbb{R}[-1]$, of fields such that $\dot\sigma\circ\,\gamma_0=a$ and that \hphantom{xxxxxxxxxxxxxxxxx}
\begin{equation}
D_t\dot\tau\circ \,\gamma_t=s_Q a=-da+Q_{\mathsans{Lie}\,\mathscr{G}}\circ a.
\label{jns4}
\end{equation} 
Indeed, \ceqref{jns4} would precisely reproduce \ceqref{sqhgs} with $E=\mathsans{Lie}\,\mathscr{G}$ 
upon switching to the graded algebra morphism $A=a^\#:C^\infty(\mathsans{Lie}\,\mathscr{G}[1])
\rightarrow \mathsans{C}^\infty(T[1]Z)$ associated to $a$.
Relation \ceqref{htpsimplex33} suggests that $\gamma_t=p_t\,\circ\, a$ might do the job. 
This seems reasonable in the light of the relationship of simplicial homotopy to 
infinitesimal higher gauge symmetry observed earlier. However, it does not quite work because
the differential term $-da$ in \ceqref{jns4} cannot be generated in this way. To remedy for this, 
we introduce a family of translations of $T[1]Z$ in the form of one of internal graded manifold morphisms
$u_t:T[1]Z\rightarrow T[1]Z$, $t\in\mathbb{R}[-1]$, defined as 
\begin{equation}
u_t=\id_{T[1]Z}+td.   \vphantom{\bigg]}
\label{jns5}
\end{equation} 
By its means, we then set \hphantom{xxxxxxxxxxxx}
\begin{equation}
\gamma_t=p_t\circ a\circ u_{-t}.  \vphantom{\bigg]}
\label{jns6}
\end{equation} 
%\vspace{-.9cm}\pagebreak \noindent
It is now straightforward to verify that \ceqref{jns4} holds true. 
%Switching next to $A=a^\#:C^\infty(\mathsans{Lie}\,\mathscr{G}[1])\rightarrow \mathsans{C}^\infty(T[1]Z)$, 
%\ceqref{jns4} yields \pagebreak 
%\begin{equation}
%s_QA=-dA+AQ_{\mathsans{Lie}\,\mathscr{G}}
%\label{jns7}
%\end{equation} 
%
%\par\noindent
%as required. 
Further investigation is required about this issue. Our analysis
however already shows that simplicial homotopy is related to BRST symmetry in 
a deep way. 

\subsection{\textcolor{blue}{\sffamily Some explicit constructions}}\label{subsec:hgtexa}

All the main examples of higher gauge theories can be formulated within the abstract 
algebraic framework worked out in sect. \cref{sec:higau} in which higher gauge fields and 
their curvatures are modeled as non differential morphisms of differential graded commutative 
algebras and their defects. In this final subsection, we revisit in detail a few of these 
along the lines of subsects. \cref{subsec:bv} and \cref{subsec:sigmamod} to illustrate
the formalism and test its range of applicability.

All the models considered below are instances of higher gauged sigma models and so are covered by the analysis 
of subsect. \cref{subsec:hisigma}. Therefore, the differential graded commutative algebra $C_1$ is the algebra 
$\mathsans{C}^\infty(T[1]Z)$ of internal graded functions of the $1$--shifted tangent bundle 
$T[1]Z$ of an $n$--fold $Z$ and the differential $Q_1$ is the canonical homological vector field  
$d$ of $T[1]Z$ given by \ceqref{canhom}. Further, the differential graded commutative algebra $C_2$ is
the algebra $C^\infty(E[1])$ of graded functions of the $1$--shift $E[1]$ of an $L_\infty$ algebroid $E$ 
on a manifold $N$ and the differential $Q_2$ is the associated homological vector field $Q_E$ of $E[1]$ 
given by eq. \ceqref{qexpl}. Finally, the manifold $M(C_2,C_1)$ of BRST higher 
gauged sigma model fields is the internal hom manifold $\mathsans{Hom}_{\mathsans{grMf}}(T[1]Z,E[1])$.

When the manifold $Z$ is closed, as we assume henceforth, the function algebra $\mathsans{C}^\infty(T[1]Z)$ 
is characterized by the existence of the Berezin integration map
\begin{equation}
\int_{T[1]Z}\varrho\,:\mathsans{C}^\infty(T[1]Z)\rightarrow G_{\mathbb{R}}.
\label{berez1}
\end{equation}
Here, $\varrho$ denotes the Berezin integration measure locally given by
$d^nzd^n\zeta$. The Berezin map is linear, has degree $-n$, 
is non singular, that is 
\begin{equation}
\int_{T[1]Z}\varrho\,uv=0~~\text{for all $v\in\mathsans{C}^\infty(T[1]Z)$} \Rightarrow u=0
\label{berez2}
\end{equation}
for any $v\in\mathsans{C}^\infty(T[1]Z)$, and satisfies the Stokes' theorem, so that
\begin{equation}
\int_{T[1]Z}\varrho\,du=0
\label{berez3}
\end{equation}
for $u\in\mathsans{C}^\infty(T[1]Z)$. The  Berezin map allows the construction of BV master actions
as space--time integrals of suitable local Lagrangians. 

We present now an example of a higher BF gauge theory following the axiomatic approach of subsect. 
\cref{subsec:bv}. 

\begin{exa} Semistrict higher BF gauge theory. 
{\rm Semistrict higher BF gauge theory is the simplest example of higher BF gauge theory.
 
The source manifold $Z$ is a generic closed $n$--fold. The target $L_\infty$--algebroid $E$ is a cyclic Lie 
$2$--algebra. Thus, the base manifold $N$ of $E$ is a point and $E$ is just a $2$ term graded vector 
space $\mathfrak{v}=\mathfrak{v}_0\oplus\mathfrak{v}_1[1]$ equipped with a boundary map 
$\partial:\mathfrak{v}_1\rightarrow\mathfrak{v}_0$, a set of $2$-- and $3$--argument brackets
$[-,-]:\mathfrak{v}_0\wedge\mathfrak{v}_0\rightarrow\mathfrak{v}_0$, 
$[-,-]:\mathfrak{v}_0\otimes\mathfrak{v}_1\rightarrow\mathfrak{v}_1$ and 
$[-,-,-]:\mathfrak{v}_0\wedge\mathfrak{v}_0\wedge\mathfrak{v}_0\rightarrow\mathfrak{v}_1$ 
with certain properties and 
a non singular bilinear pairing $(-,-):\mathfrak{v}_0\times\mathfrak{v}_1\rightarrow\mathbb{R}$ 
satisfying 
\begin{align}
&(\partial x_1,y_1)-(\partial y_1,x_1)=0, 
\vphantom{\Big]}
\label{linftyform1}
\\
&([u_0,x_0],x_1)+(x_0,[u_0,x_1])=0,
\vphantom{\Big]}
\label{linftyform2}
\\
&(x_0,[u_0,v_0,y_0])+(y_0,[u_0,v_0,x_0])=0
\vphantom{\Big]}
\label{linftyform3}
\end{align}
for $x_0,y_0,u_0,v_0\in\mathfrak{v}_0$, $x_1,y_1\in\mathfrak{v}_1$. Note that here $\mathfrak{v}_0$, 
$\mathfrak{v}_1$ are conventionally assumed to both have degree $0$. Recall also that 
the non singularity of $(-,-)$ implies that $\dim \mathfrak{v}_0=\dim \mathfrak{v}_1=\dim \mathfrak{v}/2=r/2$. 

To construct semistrict higher BF gauge theory, it is necessary to study the geometry of higher BF gauge field 
manifold $\mathsans{Hom}_{\mathsans{grMf}}(T[1]Z,\mathfrak{v}[1])$. 
A point $a\in\mathsans{Hom}_{\mathsans{grMf}}(T[1]Z,\mathfrak{v}[1])$ is specified by 
its components  \hphantom{xxxxxxxxxx}
\begin{equation}
A_p=a^\#\xi_p,
\label{}
\end{equation}
with $p=0,1$. Albeit the $A_p$ belong to the spaces $\mathsans{C}^\infty(T[1]Z)_{p+1}\otimes\mathbb{R}^{r/2}$, it is  
possible to regard them as elements of the spaces $\mathsans{C}^\infty(T[1]Z)_{p+1}\otimes\mathfrak{v}_p$.
$a$ can therefore be identified with a pair $A_p \in \mathsans{C}^\infty(T[1]Z)_{p+1}\otimes\mathfrak{v}_p$. 
A tangent vector $\dot a\in T_a[k]\mathsans{Hom}_{\mathsans{grMf}}(T[1]Z,\mathfrak{v}[1])$ is so a pair 
$\dot A_p \in \mathsans{C}^\infty(T[1]Z)_{p+1+k}\otimes\mathfrak{v}_p$ 
and a cotangent vector $\dot a^*\in T^*{}_a[l] \mathsans{Hom}_{\mathsans{grMf}}(T[1]Z,\mathfrak{v}[1])$ a pair
$\dot A^*{}_p \in \mathsans{C}^\infty(T[1]Z)_{n-p-1+l}$ $\otimes\,\mathfrak{v}_{1-p}$. 
The canonical cotangent--tangent pairing of $\dot a$, $\dot a^*$ is
\begin{equation}
\langle\dot a^*,\dot a\rangle
=\int_{T[1]Z}\varrho\,[(-1)^{(n+1+k)(n+1+l)}(\dot A_0,\dot A^*{}_0)+(-1)^{n(n+l)}(\dot A^*{}_1,\dot A_1)].
\label{bfexa1}
\end{equation}

As we explained in subsect. \cref{subsec:bv}, the field manifold of higher BF gauge theory is the $-1$--shifted
cotangent bundle $T^*[-1] \mathsans{Hom}_{\mathsans{grMf}}(T[1]Z,\mathfrak{v}[1])$, which we describe
through its base and fiber coordinates $A_p \in \mathsans{C}^\infty(T[1]Z)_{p+1}\otimes\mathfrak{v}_p$ and 
$B_p\in\mathsans{C}^\infty(T[1]Z)_{n-p-2}$ $\otimes\,\mathfrak{v}_{1-p}$. 
The BV symplectic form $\varOmega_{BV}$ is given by \ceqref{omegabvbf} in terms of the 
cotangent--tangent pairing. By \ceqref{bfexa1}, $\varOmega_{BV}$ reads explicitly as  
\begin{equation}
\varOmega_{BV}=\int_{T[1]Z}\varrho\,[-(\delta A_0,\delta B_0)+(\delta B_1,\delta A_1)].
\label{bfexa2}
\end{equation}
The ensuing BV antibrackets can be obtained just by a straightforward application of \ceqref{bvbra} and can be cast compactly as
\begin{align}
&\bigg(\int_{T[1]Z}\varrho\,(A_0,U_0{}^*),\int_{T[1]Z}\varrho\,(V_0,B_0)\bigg)_{BV}=\int_{T[1]Z}\varrho (V_0,U^*{}_0),
\vphantom{\Big]}
\label{bfexa8}
\\
&\bigg(\int_{T[1]Z}\varrho\,(U_1{}^*,A_1),\int_{T[1]Z}\varrho\,(B_1,V_1)\bigg)_{BV}=\int_{T[1]Z}\varrho (U^*{}_1,V_1)
\vphantom{\Big]}
\label{bfexa9}
\end{align}
with $V_p \in \mathsans{C}^\infty(T[1]Z)_{p+1}\otimes\mathfrak{v}_p$
and $U^*{}_p\in \mathsans{C}^\infty(T[1]Z)_{n-p-1}$ $\otimes\,\mathfrak{v}_{1-p}$. 
The BV master action $S_{BV}$ is given eq. \ceqref{sbfbv} in terms of cotangent--tangent pairing
and the curvature map. In the present case, the former is given by \ceqref{bfexa1}, the latter 
is specified by the pair $F_p\in\mathsans{C}^\infty(T[1]Z)_{p+2}\otimes\mathfrak{v}_p$ given by \ceqref{f0}, 
\ceqref{f1}. We find in this way
\begin{align}
&S_{BV}=-\int_{T[1]Z}\varrho\,\big[(-1)^n\big(dA_0+\tfrac{1}{2}[A_0,A_0]-\partial A_1,B_0\big)
\vphantom{\Big]}
\label{bfexa3}
\\
&\hspace{4.5cm}+\big(B_1,dA_1+[A_0,A_1]-\tfrac{1}{6}[A_0,A_0,A_0]\big)\big]
\vphantom{\Big]}.
\nonumber
\end{align}
The BV variations of the $A_p$ and $B_p$ are given by the expressions \ceqref{sphiqcot}, \ceqref{sphi*qcot}
with the canonical determination BRST variation $s_Q$ identified with the BV variation $\delta_{BV}$.
After a straightforward computation, we find 
\begin{align}
&\delta_{BV}A_0=-dA_0-\tfrac{1}{2}[A_0,A_0]+\partial A_1,
\vphantom{\Big]}
\label{bfexa4}
\\
&\delta_{BV}A_1=-dA_1-[A_0,A_1]+\tfrac{1}{6}[A_0,A_0,A_0],
\vphantom{\Big]}
\label{bfexa5}
\\
&\delta_{BV}B_0=dB_0+[A_0,B_0]+[B_1,A_1]-\tfrac{1}{2}[A_0,A_0,B_1],
\vphantom{\Big]}
\label{bfexa6}
\\
&\delta_{BV}B_1=-dB_1-[A_0,A_1]+\partial B_0.
\vphantom{\Big]}
\label{bfexa7}
\end{align}
}
\end{exa} 

\noindent 
Of course, higher BF gauge theories based on general Lie $m+1$--algebras can be formulated 
by means of an obvious generalization of the above construction. 

There exists a broad variety of higher Chern--Simons like gauged sigma models. 
They can all be elegantly formulated along the lines of subsect. \cref{subsec:sigmamod} as 
we shall show momentarily by illustrating a couple of examples. 

In all these models, the map $\mu: C_1\rightarrow G_{\mathbb{R}}$ is just the Berezin integration map \ceqref{berez1}.
This is indeed linear, has degree $-n$ and satisfies by virtue of the properties \ceqref{berez2}, \ceqref{berez3}
the conditions \ceqref{nat} and \ceqref{stokes}. \pagebreak 
The degree $-1$ Gerstenhaber brackets $(-,-)_2$ on $C_2$ and the Hamiltonian $S_2$ of the differential $Q_2$ appear respectively %concretely
as Gerstenhaber brackets $(-,-)$ on $C^\infty(E[1])$ and a Hamiltonian $S$ for the homological vector field $Q_E$ of $E[1]$.
Likewise, the kinetic and boundary maps $K_1:M(C_2,C_1)\rightarrow C_1$ and $B_1:TM(C_2,C_1)\rightarrow C_1$ 
with the properties \ceqref{s1local}, \ceqref{s1varpois} appear as 
maps $K:\mathsans{Hom}_{\mathsans{grMf}}(T[1]Z,E[1])$ $\rightarrow \mathsans{C}^\infty(T[1]Z)$ and 
$B:T\mathsans{Hom}_{\mathsans{grMf}}(T[1]Z,E[1])\rightarrow \mathsans{C}^\infty(T[1]Z)$ of the same nature. 
The definition of all these data must be provided on a case by case basis. 

\begin{exa} The Poisson sigma model.
{\rm The Poisson sigma model was introduced long ago in refs. \ccite{Ikeda:1993fh, Schaller:1994es}.
Its defining data are the following. 

The source manifold $Z$ is a closed $2$--fold. The target $L_\infty$--algebroid $E$ is the cotangent Lie algebroid
$T^*N$ of a Poisson manifold $N$ (cf. eg. \cref{exa:t*m}). The homological vector field $Q_E$ is taken to be $-\delta$, 
where $\delta$ is the Poisson--Lichnero\-wicz homological vector field of the shifted algebroid 
$T^*[1]N$ given by \ceqref{t*1mq}. The degree $-1$ Gerstenhaber brackets $(-,-)$ on $C^\infty(T^*[1]N)\,$ are the odd Pois\-son 
brackets $\{-,-\}$ associated with the canonical degree $1$ symplectic $2$--form $\omega$ of eq. \ceqref{t*1omg}. 
With respect to these, $Q_E$ is Hamiltonian with Hamiltonian $-S$, where $S$ is the function \ceqref{t*1ham} 
satisfying the master equation \ceqref{t*mmaster}, as follows from  the analogous properties enjoyed by $\delta$. 

%\vspace{.33mm}
A higher gauged sigma model field $a\in \mathsans{Hom}_{\mathsans{grMf}}(T[1]Z,T^*[1]N)$ is fully specified by the sigma model fields
\begin{align}
&\varphi^i=a^\#x^i,
\vphantom{\Big]}
\label{bvpsm2}
\\
&A_{0i}=a^\#\xi_i
\vphantom{\Big]}
\label{bvpsm3}
\end{align}
associated with the base and fiber coordinates $x^i$, $\xi_i$ of $T^*[1]N$. 
Their BV antibrackets defined according to \ceqref{bvmc2c1} can be cast as
\begin{equation}
\bigg(\int_{T[1]Z}\varrho \,u\varphi^i,\int_{T[1]Z}\varrho \,vA_{0j}\bigg)_{BV}=\delta^i{}_j \int_{T[1]Z}\varrho uv\,
\label{bvpsm3/1}
\end{equation}
for any  $u\in \mathsans{C}^\infty(T[1]Z)_2$, $v\in \mathsans{C}^\infty(T[1]Z)_0$
as follows readily from the relation  $\{x^i,\xi_j\}=\delta^i{}_j$.

In terms of $\varphi^i$ and $A_{oi}$, the kinetic map $K$ is given by \pagebreak 
\begin{equation}
K(\varphi,A_0)=-A_{0i}d\varphi^i. \vphantom{\bigg]}
\label{bvpsm4}
\end{equation}
%\vspace{-.9cm}\eject\noindent
The boundary map $B$ reads as 
\begin{equation}
B(\varphi,A_0)(\dot \varphi,\dot A_0)=\dot\varphi^iA_{0i}. \vphantom{\bigg]}
\label{bvpsm5}
\end{equation}
The fulfilment of the conditions \ceqref{s1local} and \ceqref{s1varpois}
is straightforwardly checked. 

We can now write down the model's BV master action $S_{BV}$ using expression \ceqref{sbvs1pluss2}. We obtain 
\hphantom{xxxxxxxxxxxxx}
\begin{equation}
S_{BV}%=\mu(K_1(a^\#)+a^\#S_2)
=-\int_{T[1]Z}\varrho \,\big[A_{0i}d\varphi^i+\tfrac{1}{2}P^{ij}(\varphi)A_{0i}A_{0j}\big].
\vphantom{\bigg]}
\label{bvpsm6}
\end{equation}
We have recovered in this way up to an overall sign the well--known expression of the Poisson 
sigma model BV action first obtained in \ccite{Cattaneo:1999fm}. 

Recalling that the canonical determination BRST variation $s_Q$ equals the BV variation $\delta_{BV}$, we can readily compute
the BV variations of $\varphi^i$ and $A_{0i}$ using \ceqref{sqphir} and \ceqref{sqa0}, 
\begin{align}
&\delta_{BV}\varphi^i=-d\varphi^i-P^{ij}(\varphi)A_{0j},
\vphantom{\Big]}
\label{bvpsm7}
\\
&\delta_{BV}A_{0i}=-dA_{0i}-\tfrac{1}{2}\partial_{xi}P^{jk}(\varphi)A_{0j}A_{0k}.
\vphantom{\Big]}
\label{bvpsm8}
\end{align}
They coincide with the  BV variations of the Poisson sigma model fields computed by other means in
\ccite{Cattaneo:1999fm}. 
}
\end{exa}

\vspace{.0mm}
\begin{exa} The Courant sigma model.
{\rm The Courant sigma model was first laid forward in ref. 
\ccite{Roytenberg:2006qz}. The data defining it are the following. 

The source manifold $Z$ is a closed $3$--fold. The target $L_\infty$--algebroid $E$ is 
is the $L_\infty$--algebroid $L$ associated with a Courant algebroid $V$ through the construction 
of eg. \cref{exa:couralgoid}. The homological vector field $Q_E$ %of the shifted algebroid $E[1]$  
is the homological vector field $Q_L$ of the shifted algebroid $L[1]$ given by \ceqref{roytq}.
The degree $-1$ Gerstenhaber brackets $(-,-)$ on $C^\infty(E[1])$ are the Poisson 
brackets $\{-,-\}$ associated with the degree $2$ symplectic $2$--form $\omega$ of eq. \ceqref{roytom}.
With respect to these, $Q_E$ is Hamiltonian with Hamiltonian $S$, where $S$ is the function given in 
\ceqref{royts} satisfying the master equation \ceqref{roytmaster}, reflecting the analogous properties 
enjoyed by $Q_L$. 

A higher gauged sigma model field $a\in \mathsans{Hom}_{\mathsans{grMf}}(T[1]Z,E[1])$ is fully specified by the 
sigma model fields 
\begin{align}
&\varphi^i=a^\#x^i,
\vphantom{\Big]}
\label{bvcour2}
\\
&A_0{}^a=a^\#\xi^a,
\vphantom{\Big]}
\label{bvcour3}
\\
&A_{1i}=a^\#p_i
\vphantom{\Big]}
\label{bvcour4}
\end{align}
corresponding the base and fiber coordinates $x^i$, $p_i$ $\xi^a$ of $E[1]N$. We are abusing notation a bit here,
since $p_i$ is an affine rather than a vector coordinate. $p_i$ can be turned into a vector coordinate by covariantizing
it by means of a metric connection for $V$. At the end, nothing can depend on the choice of the connection and so 
it does not make any difference working directly with $p_i$. With this granted,
the BV antibrackets defined according to \ceqref{bvmc2c1} can be cast as
\begin{align}
&\bigg(\int_{T[1]Z}\varrho \,u\varphi^i,\int_{T[1]Z}\varrho \,vA_{1j}\bigg)_{BV}=\delta^i{}_j \int_{T[1]Z}\varrho uv\,
\vphantom{\Big]}
\label{bvcour4/1}
\\
&\bigg(\int_{T[1]Z}\varrho \,wA_0{}^a,\int_{T[1]Z}\varrho \,zA_0{}^b\bigg)_{BV}=g^{ab}\int_{T[1]Z}\varrho wz\,
\vphantom{\Big]}
\label{bvcour4/2}
\end{align}
with $u\in \mathsans{C}^\infty(T[1]Z)_3$, $v\in \mathsans{C}^\infty(T[1]Z)_0$,
$w,z\in \mathsans{C}^\infty(T[1]Z)_2$, using that $\{x^i,p_j\}=\delta^i{}_j$
and $\{\xi^a,\xi^b\}=g^{ab}$.

The kinetic map $K$ is now given by 
\begin{equation}
K(\varphi,A_0,A_1)=A_{1i}d\varphi^i-\tfrac{1}{2}A_0{}^ag_{ab}dA_0{}^b
\label{bvcour5}
\end{equation}
in terms of sigma model fields. The boundary map $B$ reads
\begin{equation}
B_1(\varphi,A_0,A_1)(\dot \varphi,\dot A_0,\dot A_1)=\dot\varphi^iA_{1i}-\tfrac{1}{2}\dot A_0{}^ag_{ab}A_0{}^b.
\label{bvcour6}
\end{equation}
It is straightforward to check that \ceqref{s1local} and \ceqref{s1varpois} are fulfilled. 

We can write the model's BV action using expression \ceqref{sbvs1pluss2} 
\begin{align}
&S_{BV}%=\mu(K_1(a^\#)+a^\#S_2)
=\int_{T[1]Z}\varrho \,\big[A_{1i}d\varphi^i-\tfrac{1}{2}A_0{}^ag_{ab}dA_0{}^b
\vphantom{\Big]}
\label{bvcour7}
\\
&\hspace{6cm} -\rho^i{}_a(\varphi)A_0{}^aA_{1i}+\tfrac{1}{6}f_{abc}(\varphi)A_0{}^aA_0{}^bA_0{}^c]. 
\vphantom{\Big]}
\nonumber
\end{align}
Up to sign conventions, this is the expression found in ref. \ccite{Roytenberg:2006qz}.

Identifying again the canonical determination BRST variation $s_Q$ with the BV variation $\delta_{BV}$, we can compute
the BV variations of $\varphi^i$, $A_0{}^a$ and $A_{0i}$ using \ceqref{sqphir} and \ceqref{sqa0}, 
\begin{align}
&\delta_{BV}\varphi^i=-d\varphi^i+\rho^i{}_a(\varphi)A_0{}^a,
\vphantom{\Big]}
\label{bvcour8}
\\
&\delta_{BV}A_0{}^a=-dA_0{}^a  
+g^{ad}(-\rho^i{}_d(\varphi)A_{1i}+\tfrac{1}{2}f_{dbc}(\varphi)A_0{}^bA_0{}^c),
\vphantom{\Big]}
\label{bvcour9}
\\
&\delta_{BV}A_{1i}=-dA_{1i}-\partial_{xi}\rho^j{}_a(\varphi)A_0{}^aA_{1j}
+\tfrac{1}{6}\partial_{xi}f_{abc}(\varphi)A_0{}^aA_0{}^bA_0{}^c.
\vphantom{\Big]}
\label{bvcour10}
\end{align}
}
\end{exa}

\noindent
Other examples include the ordinary Chern--Simons gauge theory and its semistrict higher counterpart.

\vfill\eject

%\begin{small}

\vspace{.4cm}
\noindent
\textcolor{blue}{Acknowledgements.} 
The author thanks P. Ritter for useful discussions.
The author acknowledges financial support from INFN Research Agency
under the provisions of the agreement between Bologna University and INFN. 

\vfil\eject

%\end{small}


\begin{thebibliography}{99}


\bibitem{Baez:2010ya}
J.~C.~Baez and J.~Huerta,
{\it An invitation to higher gauge theory},\\
\textcolor{blue}{\href{http://dx.doi.org/10.1007/s10714-010-1070-9}
{Gen. Relativ. Gravit. {\bf 43} (2011) 2335}}
[\textcolor{blue}{\href{http://www.arxiv.org/abs/1003.4485}{\sffamily 1003.4485[hep-th]}}].
%%CITATION = ARXIV:1003.4485;%%


%\cite{Saemann:2016sis}
\bibitem{Saemann:2016sis}
  C.~Saemann,
{\it Lectures on Higher Structures in M-Theory}, \\
\textcolor{blue}{\href{https://arxiv.org/abs/1609.09815}{\sffamily  arXiv:1609.09815 [hep-th]}}.
  %%CITATION = ARXIV:1609.09815;%%





\bibitem{Baez:1998}
J. Baez and J. Dolan, 
{\it Categorification}, in {\it Higher Category Theory}, eds.
E. Getzler and M. Kapranov, \\ 
\textcolor{blue}{\href{http://www.ams.org/books/conm/230/}
{Contemp. \ Math.\ {\bf 230} AMS (1998) 1}}
[\textcolor{blue}{\href{https://arxiv.org/abs/math/9802029}
{\sffamily math.QA/9802029}}].

\bibitem{Baez:2002jn}
J.~C.~Baez,
{\it Higher Yang-Mills theory}, \\
\textcolor{blue}{\href{http://www.arxiv.org/abs/hep-th/0206130}{\sffamily hep-th/0206130}}.
%%CITATION = HEP-TH/0206130;%%


\bibitem{Baez:2004in}
J.~C.~Baez and U.~Schreiber,
{\it Higher gauge theory: 2-connections on 2-bundles},\\
\textcolor{blue}{\href{http://www.arxiv.org/abs/hep-th/0412325}{\sffamily hep-th/0412325}}.
%%CITATION = HEP-TH/0412325;%%

%\cite{Baez:2005qu}
\bibitem{Baez:2005qu}
J.~C.~Baez and U.~Schreiber,
{\it Higher gauge theory},
in {\it Categories in Algebra, Geometry and Mathematical Physics}, 
eds. A. Davydov et al., \\
\textcolor{blue}{\href{http://www.ams.org/books/conm/431/}
{Contemp.\ Math.\ {\bf 431} AMS (2007) 7}}
[\textcolor{blue}{\href{https://arxiv.org/abs/math/0511710}{\sffamily arXiv:math/0511710}}].
  %%CITATION = MATH/0511710;%%

%\cite{Sati:2008eg}
\bibitem{Sati:2008eg}
H.~Sati, U.~Schreiber and J.~Stasheff,
{\it $L_{\infty}$ algebra connections and applications to string and Chern-Simons $n$--transport},
in {\it Quantum Field Theory}, eds. B. Fauser, J. Tolksdorf and E. Zeidler, \\
\textcolor{blue}{\href{http://link.springer.com/chapter/10.1007\%2F978-3-7643-8736-5_17}
{Birkhauser (2009) %303
}}
[\textcolor{blue}{\href{https://arxiv.org/abs/0801.3480}{\sffamily arXiv:0801.3480 [math.DG]}}].
  %%Citation = ARXIV:0801.3480;%%
  %30 citations counted in INSPIRE as of 02 Apr 2014


%\cite{Fiorenza2011}
\bibitem{Fiorenza2011} 
D.~Fiorenza, U.~Schreiber and J.~Stasheff
{\it \v Cech cocycles for differential characteristic classes. 
An $\infty$--Lie theoretic construction}, \\
\textcolor{blue}{\href{https://projecteuclid.org/euclid.atmp/1358950853}
{Adv.\ Theor.\ Math.\ Phys.\  {\bf 16} (2012) 149}}
[\textcolor{blue}{\href{https://arxiv.org/abs/1011.4735}
{\sffamily arXiv:1011.4735 [math.AT]}}].
  %%CITATION = ARXIV:1011.4735;%%
  %19 citations counted in INSPIRE as of 31 Oct 2015

\bibitem{Ritter:2013wpa}
P.~Ritter and C.~Saemann,
{\it Lie 2-algebra models},\\
\textcolor{blue}{\href{http://dx.doi.org/10.1007/JHEP04(2014)066}{JHEP {\bf 1404} (2014) 066}}
[\textcolor{blue}{\href{http://www.arxiv.org/abs/1308.4892}{\sffamily 1308.4892 [hep-th]}}].
%%CITATION = ARXIV:1308.4892;%%
  

%\cite{Ritter:2015zur}
\bibitem{Ritter:2015zur}
  P.~Ritter, C.~Sämann and L.~Schmidt,
{\it Generalized Higher Gauge Theory},\\
\textcolor{blue}{\href{http://link.springer.com/article/10.1007\%2FJHEP04\%282016\%29032}
{JHEP {\bf 1604} (2016) 032}}
  [\textcolor{blue}{\href{https://arxiv.org/abs/1512.07554}{\sffamily arXiv:1512.07554 [hep-th]}}].
  %%CITATION = doi:10.1007/JHEP04(2016)032;%%
  %4 citations counted in INSPIRE as of 07 Oct 2016



\bibitem{Jurco:2014mva}
B.~Jurco, C.~Saemann, and M.~Wolf, {\it Semistrict higher gauge theory},\\
\textcolor{blue}{\href{http://dx.doi.org/10.1007/JHEP04(2015)087}{JHEP {\bf 1504} (2015) 087}}
[\textcolor{blue}{\href{http://www.arxiv.org/abs/1403.7185}{\sffamily 1403.7185 [hep-th]}}].
%%CITATION = ARXIV:1403.7185;%%

%\cite{Jurco:2016qwv}
\bibitem{Jurco:2016qwv}
B.~Jurco, C.~Saemann and M.~Wolf,
{\it Higher Groupoid Bundles, Higher Spaces, and Self-Dual Tensor Field Equations}\\
\textcolor{blue}{\href{http://onlinelibrary.wiley.com/doi/10.1002/prop.201600031/abstract}
{Fortsch.\ Phys.\  {\bf 64} (2016) 674}}
[\textcolor{blue}{\href{https://arxiv.org/abs/1604.01639}{\sffamily arXiv:1604.01639 [hep-th]}}].
  %%CITATION = doi:10.1002/prop.201600031;%%
  %1 citations counted in INSPIRE as of 06 Oct 2016

%%


%\cite{Baez5} 
\bibitem{Baez5}
J.~Baez and A.~Lauda, 
{\it Higher dimensional algebra V: 2-groups}, \\
 \textcolor{blue}{\href{http://www.tac.mta.ca/tac/volumes/12/14/12-14abs.html}{Theor.\ Appl.\ Categor.\ {\bf 12} (2004) 423}}
[\textcolor{blue}{\href{https://arxiv.org/abs/math/0307200}{\sffamily arXiv:math.0307200}}].

%\cite{Baez:2003fs}
\bibitem{Baez:2003fs}
J.~C.~Baez and A.~S.~Crans,
{\it Higher dimensional algebra VI: Lie $2$--algebras},\\
\textcolor{blue}{\href{http://www.tac.mta.ca/tac/volumes/12/15/12-15abs.html}
{Theor.\ Appl.\ Categor.\  {\bf 12} (2004) 492}}
[\textcolor{blue}{\href{https://arxiv.org/abs/math/0307263}{\sffamily arXiv:math/0307263}}].
%%CITATION = 00594,12,492;%%

\bibitem{Lada:1992wc}
T.~Lada and J.~Stasheff,
{\it Introduction to sh Lie algebras for physicists},\\
\textcolor{blue}{\href{http://dx.doi.org/10.1007/BF00671791}{Int. J. Theor. Phys. {\bf 32} (1993) 1087}}
[\textcolor{blue}{\href{http://www.arxiv.org/abs/hep-th/9209099}{\sffamily hep-th/9209099}}].
%%CITATION = HEP-TH/9209099;%%

%\cite{Lada:1994mn}
\bibitem{Lada:1994mn}
T.~Lada and M.~Markl,
{\it Strongly homotopy Lie algebras},\\
\textcolor{blue}{\href{http://www.tandfonline.com/doi/abs/10.1080/00927879508825335}
{Comm.\ Algebra {\bf 23} (1995) 2147}}
[\textcolor{blue}{\href{https://arxiv.org/abs/hep-th/9406095}{\sffamily arXiv:hep-th/9406095}}].
  %%CITATION = HEP-TH/9406095;%%

%\cite{Brylinski:1993ab}
\bibitem{Brylinski:1993ab}
  J.~L.~Brylinski,
{\it Loop spaces, characteristic classes and geometric quantization}, \\
\textcolor{blue}{\href{http://link.springer.com/book/10.1007\%2F978-0-8176-4731-5}
{Progress in mathematics, Birkhäuser (1993)}}.


%\cite{Breen:2001ie}
\bibitem{Breen:2001ie}
  L.~Breen and W.~Messing,
{\it Differential geometry of gerbes},\\
\textcolor{blue}{\href{http://www.sciencedirect.com/science/article/pii/S0001870805002513}
{Adv.\ Math.\ {\bf 198} (2005) 732}}
[\textcolor{blue}{\href{https://arxiv.org/abs/math/0106083}
{\sffamily arXiv:math/0106083}}].
%%CITATION = MATH/0106083;%%


%\cite{Zucchini:2015wba}
\bibitem{Zucchini:2015wba}
  R.~Zucchini,
{\it On higher holonomy invariants in higher gauge theory I},\\
\textcolor{blue}{\href{http://www.worldscientific.com/doi/10.1142/S0219887816500900} 
{Int.\ J.\ Geom.\ Meth.\ Mod.\ Phys.\  {\bf 13} (2016) no. 07, 1650090}} \hfill \hfill \hfill \hfill \hfill \hfill \hfill \hfill
[\textcolor{blue}{\href{https://arxiv.org/abs/1505.02121}{\sffamily arXiv:1505.02121 [hep-th]}}].
  %%CITATION = doi:10.1142/S0219887816500900;%%
  %1 citations counted in INSPIRE as of 07 Oct 2016

%\cite{Zucchini:2015xba}
\bibitem{Zucchini:2015xba}
  R.~Zucchini,
{\it On higher holonomy invariants in higher gauge theory II},\\
\textcolor{blue}{\href{http://www.worldscientific.com/doi/10.1142/S0219887816500912} 
{Int.\ J.\ Geom.\ Meth.\ Mod.\ Phys.\  {\bf 13} (2016) no. 07, 1650091}} \hfill \hfill \hfill \hfill \hfill \hfill \hfill \hfill
[\textcolor{blue}{\href{https://arxiv.org/abs/1505.02122}{\sffamily arXiv:1505.02122 [hep-th]}}].
  %%CITATION = doi:10.1142/S0219887816500912;%%
  %1 citations counted in INSPIRE as of 07 Oct 2016


%\cite{Schreiber2011}
\bibitem{Schreiber2011}
U.~Schreiber,
{\it Differential cohomology in a cohesive $\infty$--topos}, \\
\textcolor{blue}{\href{https://arxiv.org/abs/1310.7930}{\sffamily arXiv:1310.7930 [math-ph]}}.
%%CITATION = ARXIV:1310.7930;%%



%\cite{Sharpe:2015mja}
\bibitem{Sharpe:2015mja}
  E.~Sharpe,
{\it Notes on generalized global symmetries in QFT}, \\
\textcolor{blue}{\href{http://onlinelibrary.wiley.com/doi/10.1002/prop.201500048/abstract}
{Fortsch.\ Phys.\  {\bf 63} (2015) 659}}
[\textcolor{blue}{\href{https://arxiv.org/abs/1508.04770}{\sffamily arXiv:1508.04770 [hep-th]}}].
  %%CITATION = ARXIV:1508.04770;%%
  %1 citations counted in INSPIRE as of 06 Nov 2015

%%


%\cite{Palmer:2013ena}
\bibitem{Palmer:2013ena}
S.~Palmer and C.~Saemann,
{\it The ABJM Model is a Higher Gauge Theory}, \\
\textcolor{blue}{\href{http://www.worldscientific.com/doi/abs/10.1142/S0219887814500753}
{Int.\ J.\ Geom.\ Meth.\ Mod.\ Phys.\  {\bf 11} (2014) no. 08,  1450075}}
[\textcolor{blue}{\href{https://arxiv.org/abs/1311.1997}
{\sffamily arXiv:1311.1997 [hep-th]}}].
  %%CITATION = doi:10.1142/S0219887814500753;%%
  %4 citations counted in INSPIRE as of 07 Oct 2016

\bibitem{Palmer:2013pka}
S.~Palmer and C.~Saemann,
{\it Six-Dimensional (1,0) Superconformal Models and Higher Gauge Theory},\\
\textcolor{blue}{\href{https://arxiv.org/abs/1308.2622}
{J.\ Math.\ Phys.\  {\bf 54} (2013) 113509}}
[\textcolor{blue}{\href{http://scitation.aip.org/content/aip/journal/jmp/54/11/10.1063/1.4832395}
{\sffamily arXiv:1308.2622 [hep-th]}}].
  %%CITATION = doi:10.1063/1.4832395;%%
  %15 citations counted in INSPIRE as of 07 Oct 2016

%\cite{Lavau:2014iva}
\bibitem{Lavau:2014iva}
  S.~Lavau, H.~Samtleben and T.~Strobl,
{\it Hidden Q-structure and Lie 3-algebra for non-abelian superconformal models in six dimensions},\\
 \textcolor{blue}{\href{http://www.sciencedirect.com/science/article/pii/S0393044014002216}
{J.\ Geom.\ Phys.\  {\bf 86} (2014) 497}}
  [\textcolor{blue}{\href{https://arxiv.org/abs/1403.7114}{\sffamily arXiv:1403.7114 [math-ph]}}].
  %%CITATION = ARXIV:1403.7114;%%

%\cite{Zucchini:2011aa}
\bibitem{Zucchini:2011aa} 
  R.~Zucchini,
 {\it AKSZ models of semistrict higher gauge theory},\\
\textcolor{blue}{\href{http://link.springer.com/article/10.1007\%2FJHEP03\%282013\%29014}
{JHEP {\bf 1303}, 014 (2013)}}
  [\textcolor{blue}{\href{https://arxiv.org/abs/1112.2819}
{\sffamily arXiv:1112.2819 [hep-th]}}].
  %%CITATION = ARXIV:1112.2819;%%

  %\cite{Soncini:2014ara}
\bibitem{Soncini:2014ara}
  E.~Soncini and R.~Zucchini,
{\it 4-d semistrict higher Chern-Simons theory I},\\
\textcolor{blue}{\href{http://link.springer.com/article/10.1007\%2FJHEP10\%282014\%29079}
{JHEP {\bf 1410} (2014) 79}}
  [\textcolor{blue}{\href{https://arxiv.org/abs/1406.2197}
{\sffamily arXiv:1406.2197 [hep-th]}}].
  %%CITATION = ARXIV:1406.2197;%%

%\cite{Ritter:2015ymv}
\bibitem{Ritter:2015ymv}
  P.~Ritter and C.~Saemann,
{\it $L_\infty$-Algebra Models and Higher Chern-Simons Theories}, \\
\textcolor{blue}{\href{https://arxiv.org/abs/1511.08201}
{\sffamily arXiv:1511.08201 [hep-th]}}.
  %%CITATION = ARXIV:1511.08201;%%
  %3 citations counted in INSPIRE as of 07 Oct 2016

%\cite{Zucchini:2015ohw}
\bibitem{Zucchini:2015ohw}
  R.~Zucchini,
 {\it A Lie based 4–dimensional higher Chern–Simons theory},\\
 \textcolor{blue}{\href{http://scitation.aip.org/content/aip/journal/jmp/57/5/10.1063/1.4947531}
{J.\ Math.\ Phys.\  {\bf 57} (2016) no.5,  052301}}
  [\textcolor{blue}{\href{https://arxiv.org/abs/1512.05977}
{\sffamily arXiv:1512.05977 [hep-th]}}].
  %%CITATION = doi:10.1063/1.4947531;%%
  %2 citations counted in INSPIRE as of 07 Oct 2016


\bibitem{Roytenberg:0203110}
D.~Roytenberg,
{\it On the structure of graded symplectic supermanifolds and Courant algebroids},
in: {\it Quantization, Poisson Brackets and Beyond}, ed.\ Theodore Voronov, \\
\textcolor{blue}{\href{http://www.ams.org/books/conm/315/}
{Contemp. Math. {\bf 315}, Amer. Math. Soc. %, Providence
 (2002)}}
[\textcolor{blue}{\href{http://www.arxiv.org/abs/math.SG/0203110}{\sffamily math.SG/0203110}}].


\bibitem{Bojowald:0406445}
M.~Bojowald, A.~Kotov, and T.~Strobl,
{\it Lie algebroid morphisms, Poisson sigma models, and off-shell closed gauge symmetries}, \\
\textcolor{blue}{\href{http://dx.doi.org/10.1016/j.geomphys.2004.11.002}
{J. Geom. Phys. {\bf 54}  (2005) 400}}
[\textcolor{blue}{\href{http://www.arxiv.org/abs/math.DG/0406445}{\sffamily math.DG/0406445}}].

\bibitem{Kotov:2007nr}
A.~Kotov and T.~Strobl,
{\it Characteristic classes associated to $Q$-bundles}, \\
\textcolor{blue}{\href{http://dx.doi.org/10.1142/S0219887815500061}
{Int. J. Geom. Meth. Mod. Phys. {\bf 12} (2015) 1550006}}
[\textcolor{blue}{\href{http://www.arxiv.org/abs/0711.4106}{\sffamily 0711.4106 [math.DG]}}].
%%CITATION = ARXIV:0711.4106;%%

\bibitem{Gruetzmann:2014ica}
M.~Gruetzmann and T.~Strobl,
{\it General Yang-Mills type gauge theories for p-form gauge fields: From
  physics-based ideas to a mathematical framework or From Bianchi identities to
  twisted Courant algebroids}, \\
\textcolor{blue}{\href{http://dx.doi.org/10.1142/S0219887815500097}
{Int. J. Geom. Meth. Mod.Phys. {\bf 12} (2014) 1550009}}
[\textcolor{blue}{
  \href{http://www.arxiv.org/abs/1407.6759}{\sffamily 1407.6759 [hep-th]}}].
%%CITATION = ARXIV:1407.6759;%%

\bibitem{Fiorenza:2011jr}
D.~Fiorenza, C.~L.~Rogers, and U.~Schreiber,
{\it A higher Chern-Weil derivation of AKSZ $\sigma$-models}, \\
\textcolor{blue}{\href{http://dx.doi.org/10.1142/S0219887812500788}
{Int. J. Geom. Meth. Mod. Phys. {\bf 10} (2013) 1250078}}
[\textcolor{blue}{
  \href{http://www.arxiv.org/abs/1108.4378}{\sffamily 1108.4378 [math-ph]}}].
%%CITATION = ARXIV:1108.4378;%%


%\cite{Barnich:2000zw}
\bibitem{Barnich:2000zw}
G.~Barnich, F.~Brandt and M.~Henneaux,
{\it Local BRST cohomology in gauge theories},\\
\textcolor{blue}{\href{http://www.sciencedirect.com/science/article/pii/S0370157300000491}
{Phys.\ Rept.\  {\bf 338} (2000) 439}}
[\textcolor{blue}{\href{https://arxiv.org/abs/hep-th/0002245}
{\sffamily hep-th/0002245}}].
  %%CITATION = doi:10.1016/S0370-1573(00)00049-1;%%
  %325 citations counted in INSPIRE as of 03 Oct 2016

%\cite{Baulieu:1981sb}
\bibitem{Baulieu:1981sb}
  L.~Baulieu and J.~Thierry-Mieg,
{\it The principle of BRS symmetry: an alternative approach to Yang-Mills teories},\\
\textcolor{blue}{\href{http://www.sciencedirect.com/science/article/pii/0550321382904540}
{Nucl.\ Phys.\ B {\bf 197} (1982) 477}}.
%%CITATION = doi:10.1016/0550-3213(82)90454-0;%%
%195 citations counted in INSPIRE as of 03 Oct 2016

%\cite{Baulieu:1984iw}
\bibitem{Baulieu:1984iw}
L.~Baulieu, 
{\it Anomalies and gauge symmetry},\\
\textcolor{blue}{\href{http://www.sciencedirect.com/science/article/pii/0550321384900609}
{Nucl.\ Phys.\ B {\bf 241} (1984) 557}}.
%%CITATION = doi:10.1016/0550-3213(84)90060-9;%%
%88 citations counted in INSPIRE as of 03 Oct 2016

%\cite{Baulieu:1984ih}
\bibitem{Baulieu:1984ih}
L.~Baulieu, 
{\it Algebraic construction of gauge invariant theories}, \\
\textcolor{blue}{\href{http://inspirehep.net/record/965295}
{Cargese Summer Institute: Particles and Fields, Cargese (1983) 1}}.
  %%CITATION = PAR-LPTHE-84-4, C83-07-06;%%

%\cite{Bonora:1980pt}
\bibitem{Bonora:1980pt}
  L.~Bonora and M.~Tonin,
{\it Superfield formulation of extended BRS symmetry}, \\
\textcolor{blue}{\href{http://www.sciencedirect.com/science/article/pii/0370269381903658}
{Phys.\ Lett.\ B {\bf 98} (1981) 48}}.
%%CITATION = doi:10.1016/0370-2693(81)90365-8;%%
%255 citations counted in INSPIRE as of 03 Oct 2016

%\cite{Bonora:1980ar}
\bibitem{Bonora:1980ar}
  L.~Bonora, P.~Pasti and M.~Tonin,
{\it Geometric description of extended BRS symmetry in superfield formulation},\\
\textcolor{blue}{\href{http://link.springer.com/article/10.1007\%2FBF02772516}
{Nuovo Cim.\ A {\bf 63} (1981) 353}}.
%%CITATION = doi:10.1007/BF02772516;%%
%74 citations counted in INSPIRE as of 03 Oct 2016


%\cite{Bonora:1981rw}
\bibitem{Bonora:1981rw}
  L.~Bonora, P.~Pasti and M.~Tonin,
{\it Extended BRS symmetry in nonabelian gauge theories},\\
\textcolor{blue}{\href{http://link.springer.com/article/10.1007\%2FBF02812376}
{Nuovo Cim.\ A {\bf 64} (1981) 307}}.
  %%CITATION = doi:10.1007/BF02812376;%%
  %45 citations counted in INSPIRE as of 03 Oct 2016

\bibitem{BV1}
I.~A.~Batalin and G.~A.~Vilkovisky,
{\it Gauge algebra and quantization},\\
\textcolor{blue}{\href{http://www.sciencedirect.com/science/article/pii/0370269381902057}
{Phys.\ Lett.\ B {\bf 102} (1981) 27}}.
%%CITATION = PHLTA,B102,27;%%

\bibitem{BV2}
I.~A.~Batalin and G.~A.~Vilkovisky,
{\it Quantization of gauge theories with linearly dependent generators},\\
\textcolor{blue}{\href{http://journals.aps.org/prd/abstract/10.1103/PhysRevD.28.2567}
{Phys.\ Rev.\ D {\bf 28} (1983) 2567}}
(\textcolor{blue}{\href{http://journals.aps.org/prd/abstract/10.1103/PhysRevD.30.508}
{Erratum-ibid.\ D {\bf 30} (1984) 508)}}.
%%CITATION = PHRVA,D28,2567;%%

%\cite{Gomis:1994he}
\bibitem{Gomis:1994he}
  J.~Gomis, J.~Paris and S.~Samuel,
{\it Antibracket, antifields and gauge theory quantization},\\
\textcolor{blue}{\href{http://www.sciencedirect.com/science/article/pii/037015739400112G}
{Phys.\ Rept.\  {\bf 259} (1995) 1}}
[\textcolor{blue}{\href{https://arxiv.org/abs/hep-th/9412228}
{\sffamily hep-th/9412228}}].
  %%CITATION = doi:10.1016/0370-1573(94)00112-G;%%
  %286 citations counted in INSPIRE as of 03 Oct 2016

\bibitem{Kapustin:2013qsa}
A.~Kapustin and R.~Thorngren,
{\it Topological Field Theory on a Lattice, Discrete Theta-Angles and Confinement},\\
\textcolor{blue}{\href{http://www.intlpress.com/site/pub/pages/journals/items/atmp/content/vols/0018/0005/a004/}
{Adv.\ Theor.\ Math.\ Phys.\  {\bf 18} (2014) no.5 1233}}
[\textcolor{blue}{\href{https://arxiv.org/abs/1308.2926}{\sffamily arXiv:1308.2926 [hep-th]}}].
%%CITATION = doi:10.4310/ATMP.2014.v18.n5.a4;%%

\bibitem{Kapustin:2013uxa}
A.~Kapustin and R.~Thorngren,
{\it Higher symmetry and gapped phases of gauge theories},\\
\textcolor{blue}{\href{https://arxiv.org/abs/1309.4721}{\sffamily arXiv:1309.4721 [hep-th]}}.
  %%CITATION = ARXIV:1309.4721;%%

\bibitem{Bullivant:2016clk}
A.~Bullivant, M.~Calçada, Z.~Kádár, P.~Martin and J.~F.~Martins,
{\it Topological phases from higher gauge symmetry in 3+1D},\\
\textcolor{blue}{\href{https://arxiv.org/abs/1606.06639}{\sffamily arXiv:1606.06639 [cond-mat.str-el]}}.
%%CITATION = ARXIV:1606.06639;%%

\bibitem{Bullivant:2017sjz}
A.~Bullivant, M.~Calcada, Z.~Kádár, J.~F.~Martins and P.~Martin,
{\it Higher lattices, discrete two-dimensional holonomy and topological phases in (3+1) D with higher gauge symmetry},\\
\textcolor{blue}{\href{https://arxiv.org/abs/1702.00868}{\sffamily arXiv:1702.00868 [math-ph]}}.
%%CITATION = ARXIV:1702.00868;%%

\bibitem{Alexandrov:1995kv}
M.~Alexandrov, M.~Kontsevich, A.~Schwartz, and O.~Zaboronsky,
{\it The geometry of the master equation and topological quantum field theory},\\
\textcolor{blue}{\href{http://dx.doi.org/10.1142/S0217751X97001031}
{Int. J. Mod. Phys. A {\bf 12} (1997) 1405}} [\textcolor{blue}{
  \href{http://www.arxiv.org/abs/hep-th/9502010}{\sffamily hep-th/9502010}}].

\bibitem{Ikeda:2012pv} 
  N.~Ikeda,
 {\it Lectures on AKSZ Sigma Models for Physicists}, \\
\textcolor{blue}{\href{https://arxiv.org/abs/1204.3714}{\sffamily 
arXiv:1204.3714 [hep-th]}}.

\bibitem{Cattaneo:2010re} 
A.~S.~Cattaneo and F.~Schaetz,
{\it Introduction to supergeometry},\\
\textcolor{blue}{\href{http://dx.doi.org/10.1142/S0129055X11004400}
{Rev.\ Math.\ Phys.\  {\bf 23} 669 (2011)}}
[\textcolor{blue}{\href{https://arxiv.org/abs/1011.3401}{\sffamily 
arXiv:1011.3401 [math-ph]}}].
  %%CITATION = doi:10.1142/S0129055X11004400;%%
  %16 citations counted in INSPIRE as of 07 Nov 2016

\bibitem{zemmour;2016}
P.~Iglesias-Zemmour,
{\it An introduction to diffeology},\\
\textcolor{blue}{\href{https://ercpqg-espace.sciencesconf.org/resource/page/id/1}{New Spaces in Mathematics and
Physics, IHP, Sept. 28 – Oct. 2, 2015}}, proceedings, to appear. 
%\textcolor{blue}{\href{http://math.huji.ac.il/~piz/Site/The\%20Articles/12FB24BB-40D9-425F-A58B-03035426232C.html}{PIZ's website}}

\bibitem{Schwarz:1992nx}
A.~S.~Schwarz,
{\it Geometry of Batalin-Vilkovisky quantization},\\
\textcolor{blue}{\href{http://dx.doi.org/10.1007/BF02097392}{Commun. Math. Phys. {\bf 155} 249 (1993)}}
[\textcolor{blue}{\href{http://www.arxiv.org/abs/hep-th/9205088}{\sffamily hep-th/9205088}}].
%%CITATION = HEP-TH/9205088;%%

\bibitem{Severa:2006aa}
P.~\v Severa,
{\it $L_\infty$-algebras as 1-jets of simplicial manifolds (and a bit beyond)},\\
\textcolor{blue}{\href{http://www.arxiv.org/abs/math.DG/0612349}{\sffamily math.DG/0612349}}.

\bibitem{Ikeda:1993fh}
N.~Ikeda,
{\it Two-dimensional gravity and nonlinear gauge theory},\\
\textcolor{blue}{\href{http://www.sciencedirect.com/science/article/pii/S0003491684711043?via\%3Dihub}
{Annals Phys.\  {\bf 235} (1994) 435 }} 
[\textcolor{blue}{\href{https://arxiv.org/abs/hep-th/9312059}{\sffamily hep-th/9312059}}].
%%CITATION = doi:10.1006/aphy.1994.1104;%%
  

\bibitem{Schaller:1994es}
  P.~Schaller and T.~Strobl,
{\it Poisson structure induced (topological) field theories}, \\
\textcolor{blue}{\href{http://www.worldscientific.com/doi/abs/10.1142/S0217732394002951}{Mod.\ Phys.\ Lett.\ A {\bf 9} (1994) 3129}}
  [\textcolor{blue}{\href{https://arxiv.org/abs/hep-th/9405110}{\sffamily hep-th/9405110}}].
  %%CITATION = doi:10.1142/S0217732394002951;%%
  %257 citations counted in INSPIRE as of 31 Jan 2017

\bibitem{Cattaneo:1999fm}
  A.~S.~Cattaneo and G.~Felder,
{\it A Path integral approach to the Kontsevich quantization formula},\\
\textcolor{blue}{\href{http://link.springer.com/article/10.1007\%2Fs002200000229}{Commun.\ Math.\ Phys.\  {\bf 212} (2000) 591 }}
  [\textcolor{blue}{\href{https://arxiv.org/abs/math/9902090}{\sffamily math/9902090}}].
  %%CITATION = doi:10.1007/s002200000229;%%

\bibitem{Roytenberg:2006qz} 
  D.~Roytenberg,
{\it AKSZ-BV formalism and Courant algebroid-induced topological field theories},\\
\textcolor{blue}{\href{http://dx.doi.org/10.1007/s11005-006-0134-y}
{Lett.\ Math.\ Phys.\  {\bf 79} (2007) 143}}
[\textcolor{blue}{\href{https://arxiv.org/abs/hep-th/0608150}
{\sffamily hep-th/0608150}}].
  %%CITATION = doi:10.1007/s11005-006-0134-y;%%
  %54 citations counted in INSPIRE as of 08 Nov 2016
  
\end{thebibliography}
\end{document}